\documentclass[a4paper]{Frontiers}

\usepackage{url,hyperref,lineno,microtype}
\usepackage[onehalfspacing]{setspace}
\usepackage{amsmath,amssymb,mathrsfs,mathptmx,bm}
\usepackage{caption,graphicx,multirow,tikz,floatrow}
\usepackage{bibentry}

\newcommand{\hl}[1]{\bgroup \hskip0pt{#1}\egroup}
\newcommand{\subsubsec}[1]{\noindent\textbf{#1}:~}
\newtheorem{definition}{Definition}
\newcommand{\dfref}[1]{Definition~\ref{#1}}

\def\keyFont{\fontsize{8}{11}\helveticabold}
\def\firstAuthorLast{\copyright\ P. Loskot}
\def\Authors{Pavel Loskot$^{*}$}
\def\Address{ZJU-UIUC Institute, Haining, Zhejiang, China}
\def\corrAuthor{Pavel Loskot}
\def\corrEmail{pavelloskot@intl.zju.edu.cn}

\usepackage{changes}
\makeatletter
\@namedef{Changes@AuthorColor}{red}
\makeatother

\newcommand{\A}[1]{#1}

\DeclareSymbolFont{newfont}{OML}{cmm}{m}{it}
\DeclareMathSymbol{\Epsilon}{3}{newfont}{15}



\newcommand{\s}[1]{{\small\textsf{#1}}} 

\newcommand{\xrot}[1]{\rotatebox[origin=l]{90}{#1\ }}
\newcommand{\xrott}[2]{\xrot{#1}\xrot{#2}}

\newcommand{\scref}[1]{Section~\ref{#1}} 
\newcommand{\eref}[1]{(\ref{#1})} 
\newcommand{\tref}[1]{Table~\ref{#1}} 
\newcommand{\fref}[1]{Figure~\ref{#1}} 

\newcommand{\bibtex}{{\normalfont B\kern-.05em{\scshape i\kern-.025em b}%
    \kern-.08em \TeX}}

\newcommand{\Prob}[1]{\operatorname{Pr}\!\left(#1\right)}

\newcommand{\vy}{\bm{y}}
\newcommand{\ve}{\bm{e}}
\newcommand{\vs}{\bm{s}}
\newcommand{\BRN}{\mathrm{BRN}}
\newcommand{\hP}{\hat{P}}
\newcommand{\VR}{V_{\mathrm{R}}}
\newcommand{\NR}{{N_{\mathrm{R}}}}
\newcommand{\Qc}{\bar{Q}}
\newcommand{\Qa}{Q^\ast}
\newcommand{\nmin}{n_{\min}}
  
\begin{document}
\onecolumn\firstpage{1}

\title{A Query-Response Causal Analysis of Reaction Events in Biochemical
  Reaction Networks}

\author{\Authors}\address{}\correspondance{}

\extraAuth{ZJU-UIUC Institute, 718 East Haizhou Road Haining, Zhejiang 314400,
  China, pavelloskot@intl.zju.edu.cn}

\maketitle

\begin{abstract}
  The stochastic kinetics of biochemical reaction networks are described by a
  chemical master equation (CME) and the underlying laws of mass action. The
  CME must be usually solved numerically by generating enough traces of random
  reaction events. The resulting event-time series can be evaluated
  statistically to identify, for example, the reaction clusters, rare reaction
  events, and the periods of increased or steady-state activity. The aim of
  this paper is to newly exploit the empirical statistics of the reaction
  events in order to obtain causally and anti-causally related sub-sequences of
  reactions. This allows discovering some of the causal dynamics of the
  reaction networks as well as uncovering their more deterministic behaviors.
  In particular, it is proposed that the reaction sub-sequences that are
  conditionally nearly certain or nearly uncertain can be considered as being
  causally related or unrelated, respectively. Moreover, since time-ordering of
  reactions is locally irrelevant, the reaction sub-sequences can be
  transformed into the reaction event sets or multi-sets. The appropriately
  defined distance metrics can be then used to define equivalences between the
  reaction sub-sequences. The proposed framework for identifying causally
  associated reaction sub-sequences has been implemented as a computationally
  efficient query-response mechanism. The framework was evaluated assuming five
  selected models of genetic reaction networks in seven defined numerical
  experiments. The models were simulated in BioNetGen using the open-source
  stochastic simulator NFsim, which had to be modified to allow recording of
  the traces of reaction events. The generated event time-series were analyzed
  by Python and Matlab scripts. The whole process of data generation, analysis
  and visualization has been nearly fully automated using shell scripts. This
  also demonstrates the opportunities for substantially increasing the research
  productivity by creating automated data generation and processing pipelines.

  \keyFont{ \section{Keywords:} biochemical reaction network, causal inference,
    dynamic system, event time-series, NFsim, query-response, state-space,
    stochastic simulation}

\end{abstract}

\newpage

\section{Introduction}\label{sc:intro}

Biochemical reaction networks (BRNs) represent the systems of chemical species
that are interacting through chemical reactions. The deterministic or
stochastic models of these systems must be often analyzed numerically
\cite{gillespie2007, wolf2010, warne2019, loskot2019}. The objective of these
analyses is to generate the time trajectories of the system state represented
by the copy counts of chemical species, or by the sequences of reaction events.
The statistics of these trajectories then reveal the properties of a dynamic
system including its stability and the transition to a steady-state, if it
exists.

In addition to the properties inferred from the statistical observations,
uncovering causality can yield additional important information about the
system dynamics by relating causes to effects, and effects to causes. For
instance, a procedure for establishing the causal relationships as the
dependency between reactions in protein interaction networks was defined in
\cite{dang2015}. Causal ordering of reactions is crucial in reconstructing the
reaction pathways, and in inferring a topology of BRN \cite{villaverde2013,
  lowe2022}. In \cite{dang2015}, the causality between reactions is defined as
their dependency. The causal ordering can be achieved by comparing the
correlation peaks, exploiting asymmetry of the conditional correlation matrix,
and by examining the correlation network in a series of statistical
independence tests \cite{villaverde2013}. Moreover, many inverse problems such
as estimating the reaction rates implicitly involve the causal inferences
\cite{loskot2019}. In \cite{park2021}, the genes causing a disease were
identified by constructing a linear structural causal model (SCM), which also
accounts for the confounding and differential effects.

It can be argued that it is much easier to causally relate the sub-sequences of
reaction events than to causally relate changes in observed copy counts or
concentrations of chemical species. A rigorous framework for deriving the
reaction trajectories by solving the chemical master equation (CME) was
presented in \cite{sunkara2009, sunkara2019}. The traces of reactions events
were recorded, and subsequently used in analysis in \cite{gilbert2019} and in
\cite{connolly2022}. Moreover, as pointed out in \cite{gilbert2019}, the
reaction traces uniquely define changes in the molecular copy counts. However,
the opposite may not necessarily be true, since the copy counts are usually
only recorded at predefined time intervals.

The reaction events are categorical random variables. The causal inference
involving categorical data can be performed by a series of conditional
independence tests \cite{runge2019}, or by using information-theoretic methods
\cite{hlavackova2007}. The reaction events obtained from simulating the BRN
models represent time-series data. A comprehensive review of causal discovery
in time-series data is provided in \cite{moraffah2021} including the metrics
for evaluating the causal inferences. In \cite{soo2018}, it is argued that
causality in time-series must account for temporal trends and dependencies in
order to reach a valid conclusion.

In general, causality in multivariate time-series is usually defined as Granger
causality \cite{hlavackova2007}. Other notions of causality in time-series
include intervention causality and structural causality \cite{eichler2011,
  runge2019}. Granger causality was re-formulated in terms of the conditional
probability distributions in \cite{chikahara2018}. In particular, provided that
conditioning on observations changes the probability distribution, the
corresponding random variables can be assumed to be causally related. The
commonly agreed requirement for causality is that cause must precede the
effect, and cause must contain unique information about the effect, which is
not available from elsewhere \cite{hlavackova2007}. Alternatively, a change in
cause can be detected by its effect \cite{eichler2011}. The causality in
non-linear dynamical systems can be detected from observations assuming
state-space representations \cite{zhang2017, runge2019}. Different methods for
determining the direction and the strength of direct linear and non-linear
causal effects were compared by simulation in \cite{papana2013}. A supervised
learning for determining the causal direction between random variables has been
studied in \cite{lopez2015}. The stationary distribution obtained from solving
a CME was converted into a SCM in \cite{ness2019}. Recently, the causality
between categorical random variables was investigated in our paper
\cite{loskot2022}.

Alternatively, the reaction event sub-sequences can be related assuming various
similarity or distance metrics. The definitions of distance metrics between
data sequences can be found, for example, in \cite{vlachos2003, cassisi2012,
  batista2013}. A distance measure suitable for categorical variables such as
nucleotide bases was proposed in \cite{zielezinski2017}. These metrics can be
used to classify data sub-sequences, to identify shapelets (i.e., frequently
reappearing sequence patterns) \cite{mueen2011}, and to perform a matrix
profile analysis of longitudinal data \cite{yeh2018}.

Mathematical models of BRNs and the corresponding methods of statistical
inferences are comprehensively surveyed in \cite{loskot2019}. The algorithms
for simulating BRNs are listed in \cite{warne2019}. The software tools for
simulating and analyzing BRNs include, for example, \s{COPASI}
\cite{bergmann2017}, \s{CERENA} \cite{kazeroonian2016}, and \s{BioNetGen}
\cite{bionetgen}. The latter provides a rule-based modeling to compactly
describe BRNs involving molecules with multiple binding and modification sites.
These BRN models can be effectively and exactly simulated at the level of
reaction rules by tracking the corresponding copy counts of molecular complexes
without the need to fully extract all chemical reactions, and to track the
counts of individual species \cite{faeder2009, sneddon2011}. A tool for
visualizing the causal reaction pathways has been reported, for example, in
\cite{dang2015}. The network-based and network-free simulations of BRNs are
compared in \cite{gupta2018}. The time-scale aspects of the rule-based models
of BRNs are investigated in \cite{klinke2012}.

In this paper, the causal associations between reaction event sub-sequences in
well-stirred stochastic kinetic models of BRNs are investigated. The objective
is to identify sub-sequences that can be considered to be causally or
anti-causally related. Since the reactions are occurring at random with varying
probabilities, their associations cannot be inferred directly from the
structure of BRNs, but they must be estimated from the observed traces of
reaction events. Consequently, it is proposed to define causality in terms of
the empirical conditional probabilities. In particular, it is claimed that, if
one reaction sub-sequence precedes another, and if the corresponding empirical
conditional probability is close to 1, then these reaction sub-sequences are
nearly certain, and thus, they can be considered to be causally related. On the
other hand, if the empirical conditional probability is close to 0, then the
corresponding reaction events are conditionally nearly uncertain, so they can
be considered to be causally unrelated or even independent.

The proposed strategy of causal inference is implemented by a computationally
efficient mechanism of causal and anti-causal queries and the corresponding
responses. Moreover, since the exact ordering of reaction events is locally
irrelevant, distance metrics between event sub-sequences can be considered by
assuming the event sets or multi-sets instead of sub-sequences. The event
sub-sequences with zero mutual distance are then assumed to be equivalent. The
distance metrics can be also used to discover common patterns in the reaction
event time-series by computing the relevant matrix profile.

Numerical examples were produced in \s{NFsim}, an open-source stochastic
simulator of BRNs written in \s{C++}. This software admits the rule-based
kinetic models, and generates trajectories of chemical species counts. However,
this software had to be modified in order to allow recording of the reaction
event histories \cite{nfsimurl1}. The generated sequences of molecule copy
counts were discarded in our analysis of causal associations. The event
time-series were processed and visualized by \s{Python} and \s{Matlab} scripts.
The complete pipeline of data generation, processing and visualization was
nearly fully automated using \s{Bash} shell-scripts. This enables generating
extensive amount of numerical results with the minimum required manual
interventions within relatively short times. The complete results for the five
selected BRN models produced in the seven defined numerical experiments are
summarized in Supplementary; in the main text, only the selected results for
Model A are presented and discussed for illustration in Section 3.

\section{Methods}\label{sc:methods}

\begin{figure}[h!]
  \includegraphics[scale=1.0]{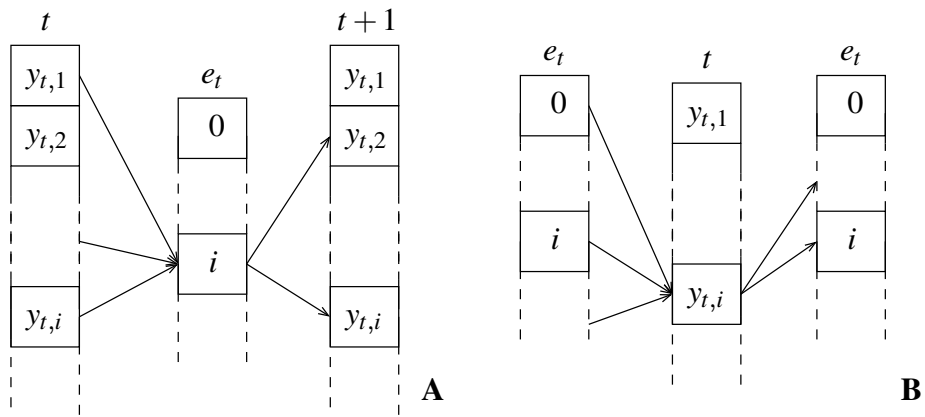}
  \caption{The causal associations between chemical species and chemical
    reactions in a BRN; a reaction event-centric model (A), and a chemical
    species count-centric model (B).\label{fg:1}}
\end{figure}

A stochastic kinetic model of a BRN is normally defined by the set of $R$
chemical reactions including their reaction rates, and the set of $S$
corresponding molecular species including their initial counts. It constitutes
a stochastic dynamic system, which is described by the vector state,
$\vy_t\in\{0,1,2,\ldots\}^S$, of molecule counts at time $t$. The states
undergo random transitions due to periodically reoccurring reaction events,
$e_t$, i.e.,
\begin{equation}\label{eq:10}
  \vy_t = \BRN(\vy_{t-1},e_t).
\end{equation}
The reaction event $e_t$ is selected at random from $R$ chemical reactions with
the probability dependent on the current state, $\vy_{t-1}$. This makes the
sequence of stochastic events, $\{e_t\}_t$, to be Markovian. The sequential
model \eref{eq:10} describes the dynamics of BRN exactly, i.e., it is an exact
solution of the CME \cite{gillespie2007}. Every defined chemical reaction binds
some reactants with certain products; the reactants are consumed, and the
molecule counts of products increase. Since often, $R\gg S$, chemical species
are involved in more than one reaction as depicted in \fref{fg:1}. The states
$\vy_t$ are non-negative integer vectors, although its large components can be
approximated as non-negative real numbers. Over any finite time interval, the
changes in molecule counts are always finite. These constraints impose
dependency between successive reaction events, $e_t$, representing categorical
or nominal random variables.

\begin{figure}[h!]
  \includegraphics[scale=1.4]{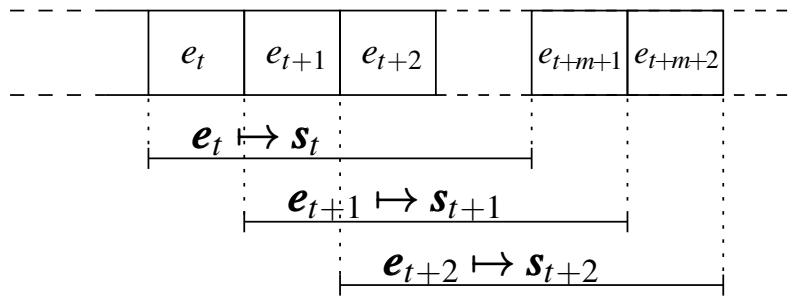}  
  \caption{The sliding-window event sub-sequences, $\ve_t$, of $(m+1)$ events
    mapped into event multi-sets or ordinary sets, $\vs_t$.\label{fg:2}}
\end{figure}

A classical analysis of BRNs assumes descriptive or inferential statistics of
the state trajectory, $\{\vy_t\}_t$. This paper departs from analyzing the
molecule counts, and instead only considers the sequences of reaction events,
$\{e_t\}_t$, and completely ignores the states, $\vy_t$. The sequence of
$(m+1)$ reaction events, $\ve_t=(e_t,e_{t+1},\ldots,e_{t+m})$, transforming the
state $\vy_t$ at time $t$ into a state $\vy_{t+m}$ at time $(t+m)$ can be
arbitrarily reordered (in time) without changing the end-state, $\vy_{t+m}$.
However, such reordering of reactions may temporarily violate the
non-negativity constraint or other limits imposed on the copy counts, $\vy_t$.
Consequently, the sub-sequences of events, $\ve_t$, can be assumed to be
multi-sets (the same events can appear multiple times, but their ordering is
irrelevant), or they can be converted into ordinary sets (repeated events are
discarded, and only the unique reactions are considered). These sets or
multi-sets are denoted as, $\vs_t$. The event sub-sequences can be created by a
sliding-window partitioning of the original event time-series as indicated in
\fref{fg:2}.

The sub-sequences $\ve_t$ or $\vs_t$ of categorical random variables can be
studied by assuming their probability distributions, and also by defining
various distance metrics. The former approach will be used to identify the
causal relationships among event sub-sequences, whereas the latter approach
enables the matrix profile analysis of the event time-series.

\bigskip\subsection{Causal Associations of Event Sub-Sequences}

In general, given two event sub-sequences $\ve_1$ and $\ve_2$ of reaction
events, the objective is to determine whether these sequences could be causally
related. For causal learning, given a cause $\ve_1$, the task is to determine,
if $\ve_2$ can be its effect. For anti-causal learning, given the effect
$\ve_2$, the task is to determine the corresponding cause $\ve_1$. The causal
specificity implies that a single cause leads to a single effect, or a single
effect has exactly one cause. If the cause is sufficient, then it is enough to
cause or to prevent an effect, whereas a necessary cause appears in every
sufficient cause \cite{peters2017}.

The first obvious condition is that, for both causal and anti-causal learning,
the events in $\ve_1$ representing a cause must precede all events in a
possible effect $\ve_2$, i.e., it is sufficient that the intersection,
$\ve_1\cap \ve_2$, is an empty set. The second condition is that, if the
sequences $\ve_1$ and $\ve_2$ are statistically independent, they cannot be
causally associated. The independence can be formulated as a null-hypothesis,
and then tested using, for example, chi-square independence tests. Since the
statistical tests are always evaluated with a certain level of significance,
the independence tests may yield false positives as well as false negatives.
Moreover, in some scenarios, the independence must be tested conditioned on
other random variables such as confounders.

Different strategies were considered in the literature to determine causal
models of time-series data as discussed in \scref{sc:intro}. Fundamentally,
provided that the causal effect is identifiable, it can be transformed into a
probability expression containing only the observed variables \cite{pearl2009}.
However, causal effects in multiple time-series cannot be identified, provided
that their SCM contains instantaneous effects \cite{peters2017}, i.e., causal
networks cannot be inferred from steady-state data. Moreover, the unknown
causal dependencies in SCM can be replaced with conditional distributions.

Consider now event time-series generated by model \eref{eq:10}. The probability
of events $\ve_2$ is conditionally dependent on the preceding events $\ve_1$
\emph{and} the sequence of states $\vy_{1-2}$, which have occurred between
$\ve_1$ and $\ve_2$, at times $t_1$ and $t_2$, respectively. The conditional
probability of interest is,
\begin{equation}\label{eq:20}
  \Prob{ \ve_2 | \ve_1 } = \sum_{\vy_{1-2}} \Prob{ \ve_2 | \ve_1 , \vy_{1-2} }
  \Prob{ \vy_{1-2} }.
\end{equation}
Removing the dependency on states $\vy_{1-2}$ in \eref{eq:20} by averaging
occurs naturally, when the conditional probability between the reaction events
is estimated empirically as a relative frequency of occurrence of events
$\ve_1$ and $\ve_2$ over sufficiently long sequences of observed reaction
events.

The conditional probability \eref{eq:20} is the likelihood that the events
$\ve_1$ have occurred, provided that the events $\ve_2$ were observed. In order
to causally relate $\ve_1$ to $\ve_2$, i.e., to claim that $\ve_1$ is a cause
of $\ve_2$, or $\ve_2$ is an effect of $\ve_1$, the Do-calculus framework of
\cite{pearl2009} requires to enforce $\ve_1$ to occur; then, the conditional
probability, $\Prob{\ve_2| \mathrm{Do}(\ve_1)}$, can be assumed instead. It is
the probability that specific $\ve_2$ is observed, provided that $\ve_1$ has
occurred. In this paper, it is claimed that the limiting cases of conditional
probability \eref{eq:20}, when it is either close to 1 or close to 0, can be
interpreted as interventions within the Do-calculus framework.

\begin{definition}\label{df:1}
  A sufficient condition for the event sub-sequences $\ve_1$ and $\ve_2$, such
  that all events in $\ve_1$ have occurred before any event in $\ve_2$, to have
  a cause-effect relationship is that their conditional probability,
  $\Prob{\ve_2|\ve_1}\to 1$. In such a case, $\ve_1$ is said to be the cause of
  $\ve_2$, and equivalently, $\ve_2$ is the effect of $\ve_1$. For anti-causal
  association, i.e., finding a cause for given effect, it is sufficient that,
  $\Prob{\ve_1|\ve_2}\to 1$.
\end{definition}

\begin{definition}\label{df:2}
  For the event sub-sequences $\ve_1$ and $\ve_2$ not to be causally related,
  it is sufficient that either one or more events in $\ve_1$ have occurred
  after any event in $\ve_2$, or that their conditional probability,
  $\Prob{\ve_2|\ve_1}\to 0$, and also, $\Prob{\ve_1|\ve_2}\to 0$, in order to
  also rule out their anti-causal association.
\end{definition}

However, there are also cases when $\ve_1$ causes $\ve_2$, and at the same
time, $\ve_2$ can cause $\ve_1$. Moreover, causal relationships are generally
asymmetric, and normally, $\Prob{\ve_2|\ve_1} \neq \Prob{\ve_1|\ve_2}$. The
conditional probability $\Prob{\ve_2|\ve_1}$ assumed in \dfref{df:1} as being
close to 1 indicates that, given $\ve_1$, there are only a few possible event
sub-sequences $\ve_2$ following $\ve_1$. More importantly, \dfref{df:1} cannot
identify all event sub-sequences $\ve_2$ that are caused by $\ve_1$, nor all
event sub-sequences $\ve_1$ that are cause of $\ve_2$.

In order to practically identify the pairs of event sub-sequences, which are
causally related by \dfref{df:1}, the event time-series are first partitioned
into sliding-window sub-sequences $\ve_t$ of $N$ events, such that the time
index, $t=1,2,\ldots$, indicates the first event in the sub-sequence, $\ve_t$.
The sub-sequences $\ve_t$ are then split into two disjoint segments $\ve_1$ and
$\ve_2$ (omitting the time index, $t$, for brevity) of $N_1$ and $N_2=N-N_1$
events, respectively. These segments are referred to as the left and the right
event sub-sequence, respectively.

The number of identified causally related event sequences can be increased by
assuming smaller values of $N$, and by introducing a notion of equivalent
sub-sequences. In particular, denote as $\vs_i$ and $\vs_j$ the sets or
multi-sets corresponding to the event sub-sequences $\ve_i$ and $\ve_j$,
respectively. Denote as $d_0$ the Hamming distance between $\ve_i$ and $\ve_j$.
Assuming the equal-length sequences $\ve_i$ and $\ve_j$, their distance can be
defined in multiple ways as follows.
\begin{subequations}
  \begin{eqnarray}
    d(\ve_i,\ve_j) &=& d_0 - | \vs_i \cup \vs_j | \label{eq:40a} \\
    d(\ve_i,\ve_j) &=& d_0 - | \vs_i \cap \vs_j | \\
    d(\ve_i,\ve_j) &=& d_0 - (|\vs_i| + |\vs_j|) \\
    d(\ve_i,\ve_j) &=& d_0 - \max(|\vs_i|,|\vs_j|) \\
    d(\ve_i,\ve_j) &=& \max(|\vs_i|,|\vs_j|) - \min(|\vs_i|,|\vs_j|) \\
    d(\ve_i,\ve_j) &=& \min(|\vs_i\setminus \,\vs_j|, |\vs_j\setminus
    \,\vs_i|). \label{eq:40f}
  \end{eqnarray}
\end{subequations}
Hence, always, $d(\ve_i,\ve_j)\geq 0$ (non-negativity),
$d(\ve_i,\ve_j)=d(\ve_j,\ve_i)$ (symmetry), and $|\vs_i|\leq |\ve_i|$, where
$|\cdot|$ denotes the cardinality of a set or the sequence length.

\begin{figure}[h!]
  \includegraphics[scale=1.0]{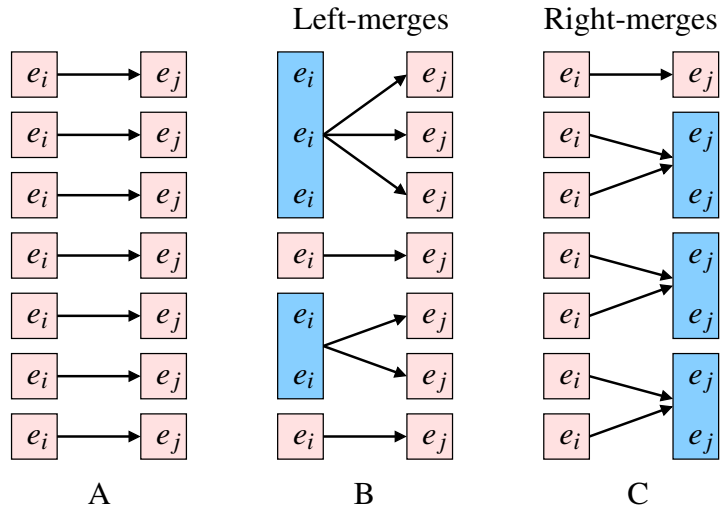}
  \caption{The original pairs of unique event sub-sequences (A), merging
    equivalent left event sub-sequences as potential causes (B), and merging
    equivalent right event sub-sequences as potential effects (C). \A{The event
      sub-sequences are equivalent, if their suitably defined distance is zero.}
    \label{fg:3}}
\end{figure}

The metrics \eref{eq:40a}--\eref{eq:40f} allow considering the distances
between sub-sequences of categorical or discrete-valued variables. The distance
metrics can be used to cluster sub-sequences of reaction events, which affects
the frequency of occurrence estimates of the conditional probabilities
considered in \dfref{df:1}, and as illustrated in \fref{fg:3}. The distance
metrics also enable matrix profile analysis of observed event time-series. The
equivalence of event sub-sequences is defined as follows.

\begin{definition}\label{df:3}
  The event sub-sequences $\ve_i$ and $\ve_j$ are said to be equivalent,
  provided that their distance, $d(\ve_i,\ve_j)=0$.
\end{definition}

\bigskip\subsection{Frequency Analysis of Reaction Events}

The most basic analysis assumes the frequency of occurrence of individual
reaction events. Denote as $\hP_i$ the estimated probability, or equivalently,
the relative frequency of occurrence of reaction $i$ in a very long sequence of
observed reaction events. For simplicity of discussion, the reaction events are
not separated into steady-state and a transition to steady-state. The reaction
events can be then sorted and also clustered by values $\hP_i$. In particular,
let $\hP_{(1)}> \hP_{(2)}> \hP_{(3)} > \ldots$. The disjoint clusters
$C_1,C_2,\ldots$ of reaction events can be defined as,
\begin{equation}\label{eq:50}
  C_j = \{ i:\min \hP_{(i)} \gg \hP_{(i^\prime)}, i^\prime\in C_{j+1}\cup
  C_{j+2} \cup \ldots \}. 
\end{equation}
In other words, the reaction clusters are defined, so that there is a
relatively large decrease in the reaction probability between the subsequent
clusters of reaction events $C_j$ and $C_{j+1}$. In all BRNs investigated in
this paper, the number of such clusters was equal to three, and in some cases,
equal to two.

Furthermore, assuming again the reaction ordering,
$\hP_{(1)}> \hP_{(2)}> \hP_{(3)} > \ldots$, it may be of interest to evaluate
if there are a few frequently occurring reactions dominating the reaction
dynamics in a BRN, or whether the reactions are more evenly distributed. Such a
spread of reactions over the observed event sequence can be measured by a
sample variance. It is computed for the event sub-sequence, $\ve_t$, as,
\begin{equation}\label{eq:60}
  \VR = \sum_{(i)=1}^\NR \left( (i) - 
    \sum_{(i)=1}^\NR (i) \hP_{(i)} \right)^2 \hP_{(i)},\quad
  \sum_{(i)=1}^\NR \hP_{(i)} = 1
\end{equation}
where $\NR$ denotes the total number of reactions defined in the BRN.
Alternatively, we can consider the random fraction of reaction types that
appear in the event set $\vs_t$, i.e.,
\begin{equation}\label{eq:70}
  v_t = |\vs_t|/\NR,\ t=1,2,\ldots.
\end{equation}
The spread of such a random variable can be measured as the median, and the
quartiles Q1 and Q3, respectively.

\bigskip\subsection{Query-Response Causal Analysis}

The discovery of causally related event sub-sequences by \dfref{df:1} is
performed in two steps. First, the distance metric is chosen in order to group
equivalent sub-sequences of events by \dfref{df:3}. Then, a histogram of event
sub-sequences is obtained to estimate the relevant conditional probabilities as
a relative frequency of occurrence. Although this is a valid strategy, it is
numerically very expensive, and it restricts the length $N_1=|\ve_1|$ and
$N_2=\ve_2|$ of the event sub-sequence that can be considered to relatively
smaller values. In order to enable exploring a possibly very large number of
event sub-sequences of length $N\gg 1$, the following two query-response
procedures are proposed for finding the causally and anti-causally related
event sub-sequences. Both procedures start from creating sliding window event
sub-sequences of equal lengths $N$. Each sub-sequence is then split into the
first $N_1$ and the last $N_2=N-N_1$ events. The procedures are depicted in
\fref{fg:4}, and they consist of the following steps.

\noindent\textbf{Discovery of causal event sub-sequences}
\begin{enumerate}
\item The input query defines criteria for selecting the left sub-sequences
  $\ve_1$. Let $Q$ be such a set of initial time indexes of sub-sequences
  $\ve_1$ satisfying the query, and let $\Qc$ be the complement of $Q$, i.e.,
  $Q\cup\Qc$ is the set of time indexes $t$ of all sliding-window sub-sequences
  $\ve_t$ considered.
  \item The response is formed by exploring the right event sub-sequences
  $\ve_2$ corresponding to the time indexes in the set $Q$. The task is to find
  a common property or feature shared by as many selected sub-sequences $\ve_2$
  as possible, while such property or feature must not be observed among any of
  the sub-sequences $\ve_2$ with the time indexes in the complement set $\Qc$.
  \item Denote such a set of time indexes of event sub-sequences, for which the
  property or feature identified in Step 2 is satisfied as $\Qa$, so that
  $\Qa\subseteq Q$ and $\Qa \cap \Qc=\emptyset$ (empty set). The response to
  the query defined in Step 1 is then the identified property or feature, and
  the set $\Qa$. The sub-sequences $\ve_1$ and $\ve_2$ associated with the set
  $\Qa$ can be then assumed to be causally related.
\end{enumerate}

\noindent\textbf{Discovery of anti-causal event sub-sequences}
\begin{enumerate}
  \item The input query defines criteria for selecting the right sub-sequences
  $\ve_2$. Let $Q$ be such a set of initial time indexes of sub-sequences
  $\ve_2$ satisfying the query, and let $\Qc$ be the complement of $Q$, i.e.,
  $Q\cup\Qc$ is the set of the time indexes $t$ of all sliding-window
  sub-sequences $\ve_t$ considered.
  \item The response is formed by exploring the left event sub-sequences
  $\ve_1$ corresponding to the time indexes in the set $Q$. The task is to find
  a common property or feature shared by as many selected sub-sequences $\ve_1$
  as possible, while such a property or feature must not be observed among any
  of the sub-sequences $\ve_1$ with the time indexes in the complement set
  $\Qc$.
  \item Denote such a set of time indexes of event sub-sequences, for which the
  property or feature identified in Step 2 is satisfied as $\Qa$, so that
  $\Qa\subseteq Q$ and $\Qa \cap \Qc=\emptyset$ (empty set). The response to
  the query defined in Step 1 is then the identified property or feature, and
  the set $\Qa$. The sub-sequences $\ve_1$ and $\ve_2$ associated with the set
  $\Qa$ can be then assumed to be anti-causally related.
\end{enumerate}

Numerical examples considered in the next section assume the queries formulated
as the minimum number of reaction events from a given reaction cluster that
have occurred within the left or right event sub-sequence, respectively. The
corresponding responses assume the intersection of all the left or the right
event sub-sets $\vs_1$ or $\vs_2$, respectively.

\bigskip\subsection{Matrix Profile of Event Time-Series}

The canonical matrix profile effectively shows the minimum distances between
the constant length sub-sequences in time-series data \cite{mueen2011}. In
particular, for every sub-sequence, the matrix profile indicates the value of
the smallest distance to any other sub-sequence, and possibly, also the
location of such sub-sequence having the minimum distance. The calculation of
matrix profile is greatly optimized in order to avoid many unnecessary
computations, so it can be used for processing very long sequences of data. The
small values of the mutual distance in matrix profile indicate common patterns
(motifs), and large values point at the locations of rare patterns (discords).
The matrix profile can be also used to identify the time instances where the
distance-based sequence statistics have changed.

The choice of distance metric and the sub-sequence length are the two most
important considerations in matrix profile analysis. In addition, instead of
searching for the minimum distance values, the maximum distances may reveal the
level of dissimilarity between different parts of time-series data. It is also
useful to count multiplicities of the distance values to understand how they
are distributed. However, interpreting matrix profile for categorical random
variables is beyond the scope of this paper, so the corresponding numerical
results are not reported.

\bigskip\subsection{Numerical Implementation and Software Tools}

The devised causal framework is investigated assuming stochastic kinetic models
\eref{eq:10} of five BRNs. The open-source \s{BioNetGen} software was chosen to
perform the simulations; version 2.7.0 of this program was downloaded from
\cite{bionetgenurl}. The BRN models for \s{BioNetGen} are described in an
intuitive \s{BNGL} language \cite{blinov2004, bionetgen}. These models are
stored as ordinary text files. The \s{BNGL} model description includes the
lists of chemical species, reaction rates and other parameters, reaction rules,
quantities to be recorded as observations, and the description of other model
components. The \s{BNGL} file is processed by a \s{Perl} script distributed
with \s{BioNetGen} to generate the model description file in a System Biology
Markup Language (SBML) format. The SBML file can be then directly simulated in
\s{NFsim} \cite{sneddon2011}.

The stochastic simulation algorithm (SSA) in network-free simulations is
performed exactly and efficiently by assuming reaction rules while tracking the
copy counts of molecular objects with multiple binding and modification sites
\cite{faeder2009}. The reaction rules can be fully extracted into a
combinatorially large number of reactions with corresponding chemical species
to perform traditional network-based simulations \cite{gupta2018}. More
importantly, the linear kinetic model \eref{eq:10} and the causal framework
defined earlier in this section are still valid, provided that the reaction
rules and molecular complexes are considered instead of the actual reactions
involving individual chemical species.

The \s{NFsim} version 1.11 was downloaded from \cite{nfsimurl}, and the
modified version 1.20 of this software was uploaded to \cite{nfsimurl1}. There
are several issues with version 1.11 of \s{NFsim}. First, it cannot record the
generated reaction events, since the dynamics of BRNs so far have always been
elucidated from the observed copy counts rather than from the reaction events.
Therefore, a new software feature was added to \s{NFsim} to allow recording the
generated reaction events into a separate file. More specifically, since the
reaction rules are not extracted during the network-free simulations, the
recorded traces of reaction events are the reaction rules selected by the SSA.
Consequently, the causal analysis is performed for sub-sequences of reaction
rules rather than sub-sequences of individual reactions. Second, the original
source code cannot be recompiled incrementally when only some source files have
been modified, which slows down the code development substantially. Therefore,
a new global \s{make}-file was created to allow the incremental compilings and
to speed-up the code development. Third, the version 1.11 of \s{NFsim} was
released a decade ago (in 2012), so it does not compile under more recent
versions of \s{C++} language. Hence, the original source code was further
refactored to remove compiling errors under the more recent version of the
\s{gcc} compiler. The updated version 1.20 of \s{NFsim} is now freely available
for download from \cite{nfsimurl1}.

The generated event time-series were processed and visualized by the custom
scripts written in \s{Matlab} and \s{Python}. The canonical matrix profile can
be calculated using a \s{Python} library \s{stumpy} \cite{stumpy}. The overall
process of performing the simulations, processing the recorded event
time-series, producing the plots, and generating the \LaTeX\ source code for
inserting figures and tables into the supplementary file was largely automated
using \s{Bash} scripts.

The \s{BNGL} model files are provided in \s{models} folder, the \s{Matlab} and
\s{Python} scripts are available in \s{scripts} folder, and the samples of
generated event history files are stored in \s{data} folder of the \s{NFsim}
public repository uploaded to \s{Github} \cite{nfsimurl1}.

\bigskip\section{Results}\label{sc:results}

Numerical results were produced for five selected rule-based models of BRNs in
seven defined numerical experiments. The models and experiments are detailed
below. The models are labeled by letters A--E as well as by their acronyms. The
experiments are labeled by integers 1--7.

Each model was simulated once for 100s of simulation time, and the sequence of
generated random reaction events was recorded into a comma-separated file (CSV)
by \s{NFsim}. The field with reaction times was removed from the CSV file,
since it is not relevant for the subsequent analysis. The event time-series
were then processed and analyzed by multiple \s{Matlab} and \s{Python} scripts.
The script names are listed in the Supplementary file including the
corresponding figures and tables that were generated by these scripts.

It should be noted that, in general, finding the reliable SBML files for
kinetic simulations of BRNs is challenging. These files are often incomplete,
i.e., they do not define all initial molecule counts, reaction rates and other
parameters required to faithfully and unambiguously represent the selected
biochemical system, or, these files contain many errors of various kind, which
are very difficult to remove, especially for larger models.

\bigskip\subsection{BRN Models}

\subsubsec{Model A (Multi-states)} This is a rule-based model considered in
\cite{blinov2004}, \cite{colvin2009} and \cite{colvin2010}. The model
represents an idealized multivalent ligand–receptor binding with multiple
ligand-induced receptor aggregates. This is an example of a model with lower
complexity consisting of only 5 species and 6 reaction rules.

\subsubsec{Model B (Chemotaxis)} This rule-based model has been proposed in
\cite{hansen2008} to study chemical signaling of chemotaxis in Escherichia
coli, and to understand signaling adaptation to changes in chemical
concentrations. The receptor enzymes involved in signaling act on small
assistance neighborhoods (AN) of 5 to 7 receptor homodimers. The model contains
12 species and 41 reaction rules.

\subsubsec{Model C (Chemotaxis-ext)} This rule-based model is an extended
version of previous Model B. It has the updated receptor topology with 37
species, but only 32 reaction rules.

\subsubsec{Model D (TLBR)} This rule-based model is a kinetic version of the
Goldstein-Perelson model of trivalent ligand-bivalent receptor (TLBR)
\cite{yang2011, sneddon2011}. The model allows predicting the distribution of
ligand-receptor aggregate sizes and configurations. This is the smallest model
considered consisting of only 4 species and 6 reaction rules.

\subsubsec{Model E (Multi-sites)} This is a rule-based model that was assumed
in \cite{colvin2009} and further analyzed in \cite{colvin2010} to test and
compare the network-based simulations of BRNs. The model represents an
idealized autophosphorylation of receptor tyrosine kinase (RTK) having multiple
receptor binding sites. The model rules can be extracted into 68 species and
290 reactions.

\bigskip\subsection{Numerical Experiments}

The numerical results presented here were selected to support the key
observations. The complete numerical results that were generated for the five
BRN models considered across all the experiments can be found in the
Supplementary file.

\subsubsec{Experiment 1}\label{ex:1}

The objective is to evaluate the model complexity in terms of the number of
functions, species, parameters, reaction rules and the molecule types
(\tref{tab1}). The models A and E defined by the \s{BNGL} file can be fully
extracted over several iterations in order to enumerate all the combinations of
reactions specified by the reaction rules. However, the reaction network
extraction failed for the models B, C and D (investigating the reasons why the
extraction algorithm failed to terminate is beyond the scope of this paper).
The total number of groups, species and reactions in the extracted models A and
E are listed in \tref{tab1}.

The wall-clock times to simulate the BRN models over 100s of simulation time,
the number of randomly generated reaction events, and the number of unique
reaction rules are compared in \tref{tab2}. These values demonstrate the great
effectiveness of the network-free modeling and simulations. The reaction events
are produced the fastest for model E, and at the slowest for model D.

The recorded event history files were processed to find unique sliding-window
$N$-tuples of reaction events and their counts (i.e., histograms). The
resulting $N$-tuple statistics can be found in supplementary Tables S1--S5 for
the sliding window sub-sequences of $N=1$, $3$ and $5$ reaction events,
respectively. Moreover, the event time-series were split into 100 blocks of one
second duration each. For models B--E (Chemotaxis, Chemotaxis-ext, Multi-sites,
and TLBR), the probabilities of reaction $N$-tuples are nearly constant within
the blocks. On the other hand, model A (Multi-states) appears to experience a
transition into a steady-state where the reaction probabilities become
constant. Furthermore, the event $N$-tuples can be naturally clustered by
assuming their frequencies of occurrence as indicated in Tables S1--S5.

\begin{table}[!t]
  \centering
  \caption{Summary of the BRN model complexities.}\label{tab1}
  \begin{tabular}{lccccc} \hline
    Model & A & B & C & D & E \\ \hline
    functions  & -  & 17  & 11  & - & - \\
    parameters  & 11  & 55  & 53  & 22 & 11 \\
    groups  & 6  & -  & -  & - & 5 \\    
    molecule types  & 5  & -  & -  & 4 & 5 \\
    species  & 5  & 12  & 37  & 4 & 5 \\
    ext. species  & 11  & -  & -  & - & 68 \\
    reaction rules  & 6  & 41  & 32  & 6 & 17 \\    
    ext. reactions  & 20  & -  & -  & - & 290 \\
    ext. iterations & 5 & -  & -  & - & 10 \\
    \hline
  \end{tabular}
\end{table}

\begin{table}[!t]
  \centering
  \caption{The simulations of BRNs over 100s of simulation time.}\label{tab2}
  \begin{tabular}{lccccc}\hline
    Model & A & B & C & D & E \\ \hline
    reaction events & 133,513 & 57,621 & 53,457 & 17,238 & 2,353,366 \\
    reaction types & 6 & 22 & 16 & 24 & 18 \\
    walk-clock time [s] & 0.1 & 2.8 & 1.5 & 1.5 & 1.6 \\
    \hline
  \end{tabular}
\end{table}

\subsubsec{Experiment 2}\label{ex:2}

\begin{figure}[!t]
  \includegraphics[scale=0.85]{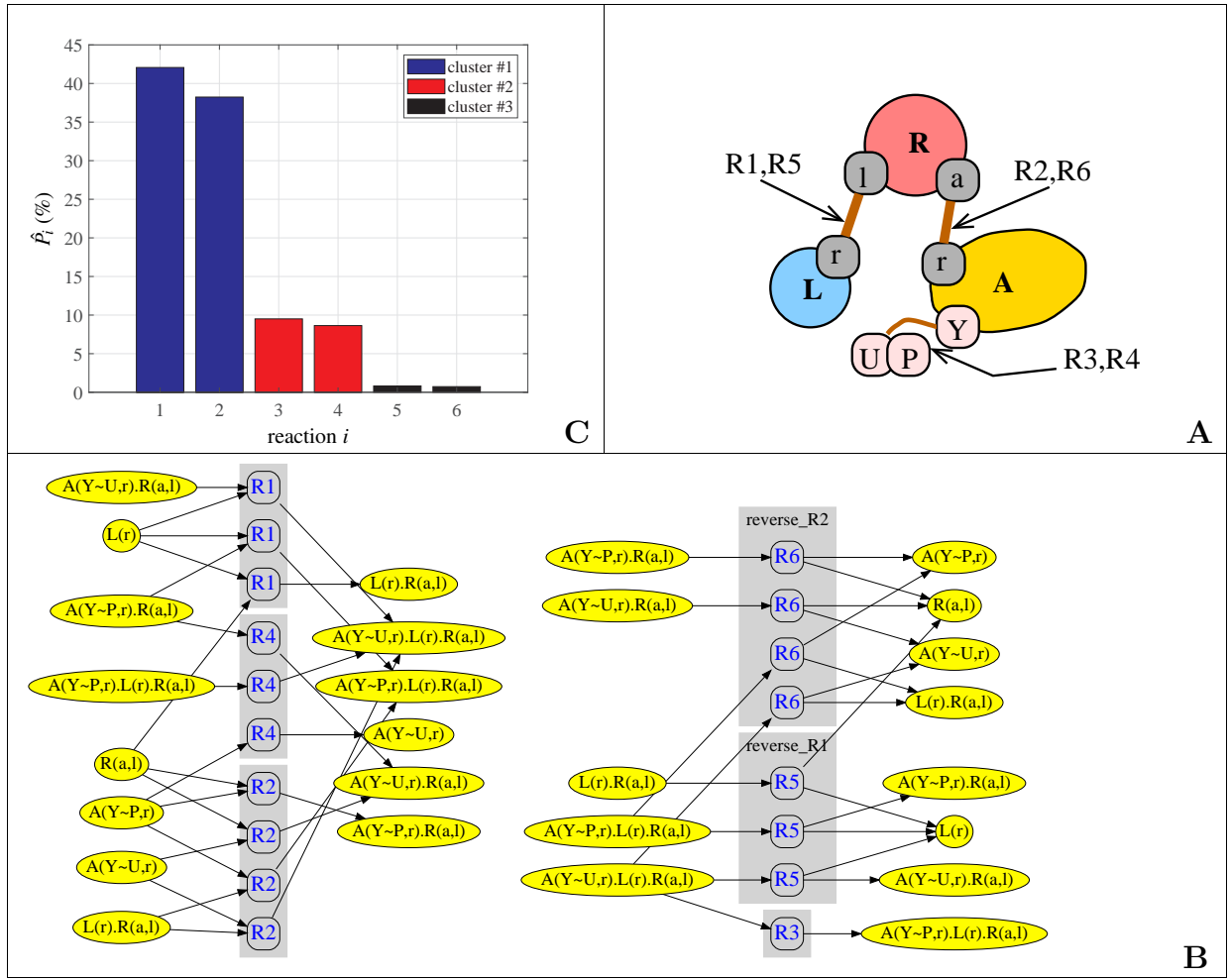}
  \caption{\A{Model A is an example of a reaction network with multi-site
      chemical species (panel A). It models ligand-receptor (L-R) binding in
      the presence of adapter protein (A). The adapter protein contains
      tyrosine (Y), which can be in phosporylated (P) or unphosporylated (U)
      state. The expanded reaction rules R1--R6 are shown in (panel B).} The
    long-term probability of individual reactions (panel C). The reactions can
    be naturally clustered into 3 groups.\label{fg:5}}
\end{figure}

The objective of this experiment is to compare the frequency of occurrence of
individual reaction rules. The long-term probabilities of individual reactions
for Model A are shown in \fref{fg:5}A, and for other models, they are shown in
Figures S2--S5. These probabilities can be used to naturally cluster the
reaction rules. In particular, 3 reaction clusters can be defined for Models A,
B and C, but only 2 such clusters are defined for Models D and E.

\subsubsec{Experiment 3}\label{ex:3}

\begin{figure}[!t]
  \includegraphics[scale=0.75]{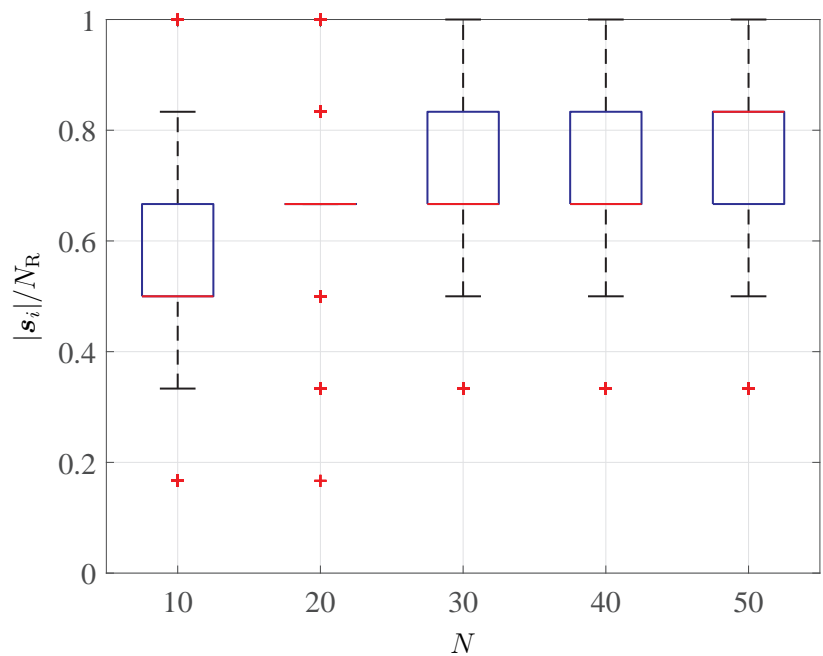}
  \caption{The box plot of relative frequency occurrences of unique reactions
    in Model A showing the median, quartiles Q1 and Q3, and the outliers
    outside the $1.5\times|Q3-Q1|$ range. \A{The reaction event sub-sequences,
      $\ve_i$, were converted first into multi-sets, $\vs_i$.}\label{fg:6}}
\end{figure}

This experiment evaluates how many unique reaction events are occurring within
the sub-sequences $\ve_t$ of $N$ reaction events. The median, the quartiles Q1
and Q3 as well as the outliers of the random variables, $|\ve_t|/\NR$, are
shown as box plots in \fref{fg:6}, for Model A, and in Figures S7--S10, for
Models B--E. In general, it can be observed that the median is non-linearly and
nearly monotonically increasing with the length of the reaction event
$N$-tuples until it eventually levels-off. Interestingly, for some values $N$,
the inter-quartile range $|Q3-Q1|$ collapses, and the median and the quartiles
Q1 and Q3 coincide.

\subsubsec{Experiment 4}\label{ex:4} 

This experiment implements the proposed query-response method to determine
causal relationships between the neighboring pairs of the reaction event
sub-sequences. In particular, the query here is formed by the minimum number of
reactions from a selected reaction cluster that must have occurred within the
left sub-sequences of $N_1$ reaction events. The right sub-sequences of $N_2$
reaction events are first converted into the sets of unique reaction events.
These sets can be then divided into three groups. The first group contains the
reaction rules from the right sub-sequences of $N_2$ events corresponding to
the left sub-sequences of $N_1$ events, which were selected by the query, and
that do not appear in the reaction event sub-sequences outside (i.e., not
selected by) the query. The second group are the reaction sets that only appear
in the right event sub-sequences outside (i.e., not selected by) the query. The
third group represents the reaction sets that are shared between the right
sub-sequences within the query-selected as well as the query-not selected right
sub-sequences.

\begin{figure}[h!]
  \includegraphics[scale=1.0]{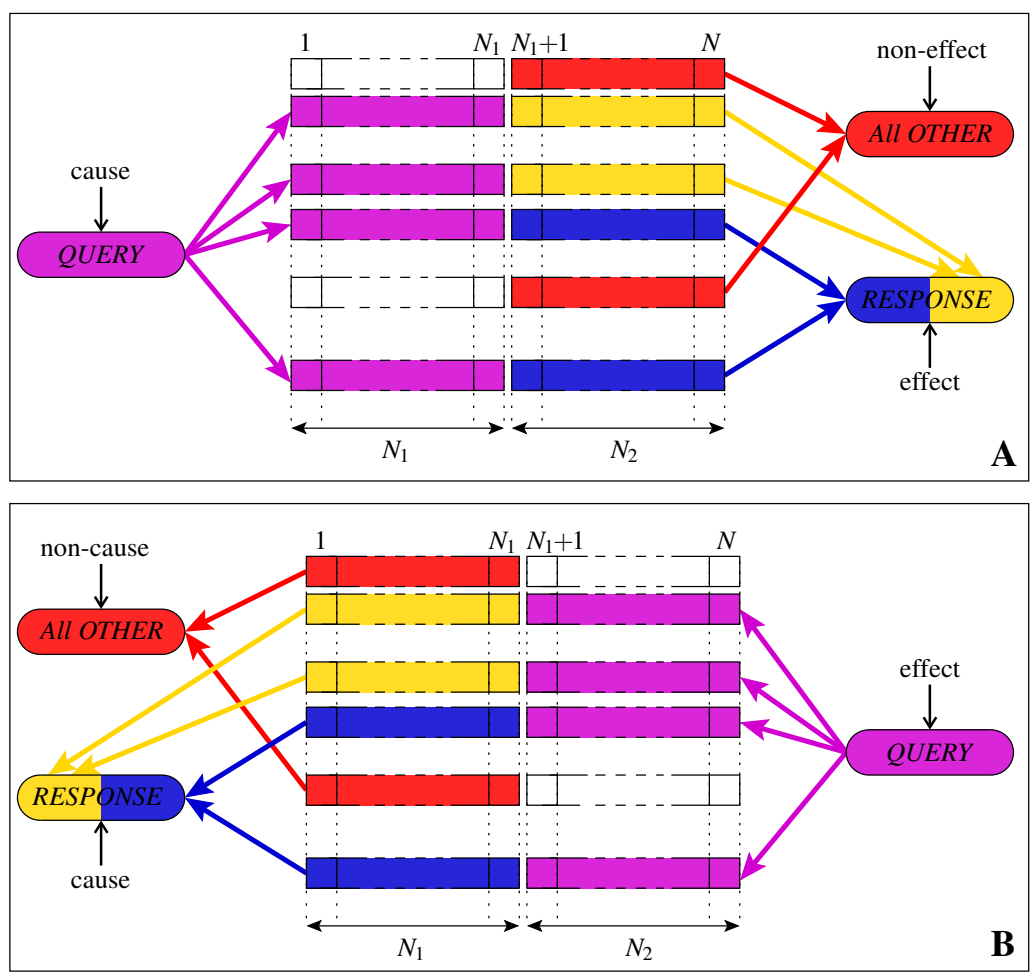}
  \caption{\A{The causal (anti-causal) inference over reaction event
      sub-sequences is performed by formulating a query to select the event
      sub-sequences representing a cause (effect). The response to the query is
      a collection of event sub-sequences representing the effect (cause), for
      which some shared feature can be identified. This feature is then used to
      formulate a causal or anti-causal statement about a biochemical reaction
      network.} A query-response causal (A) and anti-causal (B) discovery of
    cause-effect relationships between the left and right event sub-sequences
    of $N_1$ and $N_2=(N-N_1)$ reaction events, respectively. The responses
    corresponding to the queried sub-sequences (magenta boxes) are divided into
    sub-sequences that can be described by some defined shared property or
    feature (blue boxes) and those that cannot (yellow boxes). Ideally, the
    number of yellow boxes is very small, or even zero. \label{fg:4}}
\end{figure}

\begin{figure}[!t]
  \includegraphics[scale=1.1]{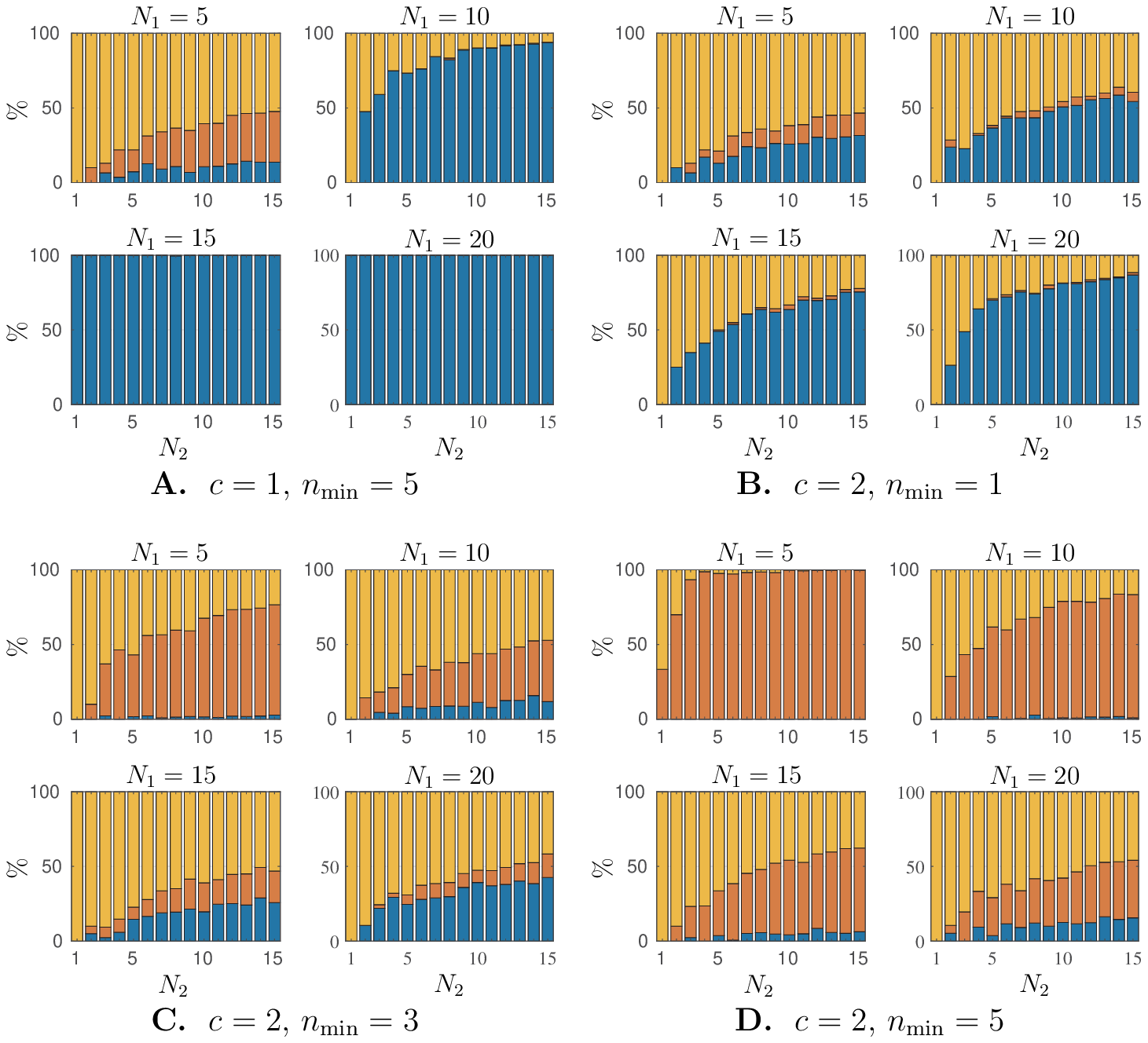}
  \caption{\A{The bar plots for Model A quantifying the responses to a given
      causal query defined as all sub-sequences of $N_1=5,10,15$ and $20$
      reaction events, respectively, containing at least $\nmin$ reaction
      events from the reaction group, $c\in\{1,2,3\}$. The length $N_2$ of the
      response sub-sequences (the effects) is varied between $1$ and $15$
      events. The blue bars are fractions of the response sub-sequences to the
      given causal query. The red bars are fractions of sub-sequences outside
      the causal query. The yellow bars are the response sub-sequences, which
      are shared between the response sub-sequences contained within and
      outside the given causal query, respectively. This coloring scheme
      corresponds to the one used in \fref{fg:4}.}\label{fg:7}}
\end{figure}

Denote as $r_1$, $r_2$ and $r_3$, respectively, the relative sizes of these
three groups, so that $r_1+r_2+r_3=100\%$ (see \fref{fg:4}A). The values $r_1$
(blue bars), $r_2$ (red bars) and $r_3$ (yellow bars) for Model A are shown as
stacked bar plots in \fref{fg:7}. The experiments are labeled as ``4-C-F'',
where $F\geq 1$ is the required minimum number of reactions from the reaction
cluster, $C\in\{1,2,3\}$, to have occurred in each left sub-sequence of $N_1$
reaction events. The complete experimental results for all models and $F=1,3$
and $5$ are provided in Figures S11--S88 in the Supplementary.

The plots in \fref{fg:7} suggest that both $C$ and $F$ are the important
parameters in formulating the query. Ideally, the size of the response (the
effects) matches the size of the query (the causes). This is the case when
$r_1\to 100\%$, i.e., the responses fully cover different queries, which is
indicated by the full blue bars in \fref{fg:7}. When $r_2\to 100\%$ (tall red
bars in \fref{fg:7}), none or only a few responses match the queries about the
effects. Such responses indicate that most of the right event sub-sequences are
causally unrelated to the query. Larger $r_3$ values (large yellow bars)
correspond to cases when none or only a few event sub-sequences in the response
can be considered to be either causally related or causally unrelated to the
query, so these cases have the smallest information insight.

In general, it can be concluded that larger values of $N_1$, which are used to
formulate the query about the effects as well as larger values of $N_2$, which
are used to identify a common property or feature among the right event
sub-sequences chosen by the query, shift the probability mass towards either
the causal or non-causal event sets.

\subsubsec{Experiment 5}\label{ex:5}

This experiment extends Experiment 4 in order to illustrate how to generate
causal statements about the BRN considered. In particular, consider again Model
A as an example. This model represents ligand-receptor (L-R) binding in the
presence of adapter protein (A). The adapter protein contains tyrosine (Y),
which can be in phosporylated (P) or unphosporylated (U) form
\cite{colvin2009}. The model is referred to as being multi-site, since all
species L, R and A contain dedicated binding sites (\fref{fg:5}B). There are
only six reactions defined in this model. Specifically, R1 and R5 denote
binding (forward reaction) and unbinding (reverse reaction) of ligand to
receptor, respectively, whereas R2 and R6 denote binding (forward reaction) and
unbinding (reverse reaction) of adapter to receptor, respectively. The
phosporylation and unphosporylation are labeled as reactions R3 and R4,
respectively. The completely expanded reaction rules are then shown in
\fref{fg:5}C.

The examples of generated causal statements are summarized in \tref{tab3}. In
particular, the query representing a cause is defined as the minimum number of
reactions from the given reaction cluster $C=1$, $2$ and $3$ that must have
occurred within the reaction sub-sequences of $N_1$ events. The query size in
\tref{tab3} indicates how many of such reaction sub-sequences satisfy this
condition, while only the unique sub-sequences are counted. The response
representing the effect is formed by all subsequent sub-sequences of $N_2$
reaction events; only the unique such sub-sequences are again counted in
\tref{tab3} to indicate the size of the response. In addition, the minimum and
the maximum numbers of all six reactions defined in Model A contained within
the response event sub-sequences are reported in \tref{tab3}. For instance, it
can be observed from \tref{tab3} that some reactions are guaranteed to occur
with a certainty (non-zero minimum counts) within the query response.

\begin{table}[!t]
  \centering
  \caption{Example causal statements generated from the simulated event-series
    for model A assuming the queries of $N_1=25$ reaction events, and the
    responses of $N_2=25$ reaction events.} \label{tab3}
  \begin{tabular}{|c|c|c|c|c|c|c|c|c|c|}\hline
    \multicolumn{2}{|c|}{query} & query & response &
    \multicolumn{6}{c|}{min--max observed reactions} \\
    \multicolumn{1}{|c}{$\nmin$} & cluster & size & size &
    \multicolumn{1}{c}{$R1$} &
    \multicolumn{1}{c}{$R2$} &
    \multicolumn{1}{c}{$R3$} &
    \multicolumn{1}{c}{$R4$} &
    \multicolumn{1}{c}{$R5$} &
    \multicolumn{1}{c|}{$R6$} \\
    \hline
    5 &	1 & 2670 & 919  & 2--24 & 0--20 & 0--11 & 0--8 & 0--3 &  0--3 \\
    15 & 1 & 2645 & 913 & 2--24 & 0--20 & 0--11 & 0--8 & 0--3 & 0--3 \\
    5 &	2 & 1267 & 593 & 3--21 & 0--17 & 0--9 & 0--7 & 0--2 & 0--3 \\
    10 & 2 & 46 & 45 & 5--18 & 1--14 & 0--8 & 0--5 & 0--1 & 0--1 \\
    1 &	3 & 824 & 469 & 3--19 & 2--17 & 0--7 & 0--8 & 0--2 & 0--3 \\
    3 &	3 & 17 & 17 & 6--10 & 10--11 & 1--3 & 2--5 & 0--1 & 0--1 \\
    \hline
  \end{tabular}
\end{table}

\subsubsec{Experiment 6}\label{ex:6}

This experiment is similar to Experiment 4, however, now the query about causes
is anti-causal as it selects the right reaction events as effects, and the
returned response are the corresponding left reaction sub-sequences to be
evaluated as possible reaction causes. The values $l_1$, $l_2$ and $l_3$,
$l_1+l_2+l_3 = 100\%$ can be again defined to be the fractions of sizes of the
left reaction event sets that are within the given query (blue bars), outside
the given query (red bars), or lie in the intersection of these two groups
(yellow bars) (cf. \fref{fg:4}B).

These experiments are labeled as ``5-C-F'' with the same definition for $C$ and
$F$ as in Experiment 5. Interestingly, given $C$ and $F$, the bar plots for
Experiment 4 and 5 are similar but not identical. However, this is a
consequence of how the queries are formulated, and how the responses with the
common property or feature are determined in our examples, rather than a
general rule.

\subsubsec{Experiment 7}\label{ex:7}

Similarly to Experiment 5, which has been used to produce the examples of
causal statements corresponding to Experiment 4, Experiment 7 generates
anti-causal statements for Experiment 6. Model A described in \fref{fg:5} is
again considered as example. In particular, the query representing the effects
is formed by the sub-sequences of $N_2$ reaction events containing at least the
given number of reactions from the reaction cluster $C=1$, $2$ and $3$,
respectively. The preceding sub-sequences of $N_1$ reaction events representing
the cause contain a certain minimum and maximum number of reactions as shown in
\tref{tab4}. The minimum counts are sometimes non-zero, so that the
corresponding reactions are guaranteed to occur within the event sub-sequences
corresponding to the cause.
  
\begin{table}[!t]
  \centering
  \caption{Example anti-causal statements generated from the simulated
    event-series for model A assuming the queries of $N_2=25$ reaction events,
    and the responses of $N_1=25$ reaction events.} \label{tab4}
  \begin{tabular}{|c|c|c|c|c|c|c|c|c|c|}\hline
    \multicolumn{2}{|c|}{query} & query & response &
    \multicolumn{6}{c|}{min--max observed reactions} \\
    \multicolumn{1}{|c}{$\nmin$} & cluster & size & size &
    \multicolumn{1}{c}{$R1$} &
    \multicolumn{1}{c}{$R2$} &
    \multicolumn{1}{c}{$R3$} &
    \multicolumn{1}{c}{$R4$} &
    \multicolumn{1}{c}{$R5$} &
    \multicolumn{1}{c|}{$R6$} \\
    \hline
    5 &	1 & 2670 & 887 & 3-24 & 0-18 & 0-12 & 0-8 &  0-3 &  0-3 \\
    15 & 1 & 2648 & 884 & 3-24 & 0-18 & 0-12 & 0-8 &  0-3 &  0-3 \\
    5 &	2 & 1264 & 590 & 3-21 & 0-18 & 0-12 & 0-8 &  0-2 &  0-3 \\
    10 & 2 & 39 & 38 & 6-15 & 4-11 & 0-6 &  2-6 &  0-1 &  0-2 \\
    1 &	3 & 879 & 455 & 3-16 & 3-17 & 0-8 &  0-8 &  0-2 &  0-3 \\
    3 &	3 & 16 & 16 & 7-14 & 7-13 & 0-4 &  0-2 &  0-2 & 0-2 \\
    \hline
  \end{tabular}
\end{table}

\bigskip\section{Discussion}

In this paper, BRNs were newly analyzed assuming the recorded traces of
reaction events. The traces were produced by network-free simulations, so they
represent the sequences of reaction rules rather than actual reactions. The
reaction rules can be used to perfectly reconstruct changes in the copy counts
of molecular objects having multiple binding and modification sites, so they
contain complete information about the system dynamics. It is different from
the previous strategy, where the dynamics of BRNs are deduced from traces of
the species copy counts. Working with event traces also requires much less
storage, since only a single reaction index needs to be recorded, whereas
storing the complete system state is usually only done at predefined time
instances in order to reduce the overall storage size.

The main objective of this paper is to identify causally or anti-causally
related and causally unrelated event sub-sequences rather than to explicitly
infer the system dynamics. The traditional approach for obtaining the SCM of a
BRN assumes a steady-state distribution of molecule copy counts, and then
defining the transformations between dependent random variables as demonstrated
in \cite{ness2019}. The causality in the present paper assumes the empirical
probabilities of sub-sequences of reaction rules over one or more reaction
traces obtained from stochastic simulations of a given BRN. This approach
allows inferring, with a very high probability, the causality between
sub-sequences of reaction events, and also formulating causal statements such
as ``if the observed sub-sequence of reaction events contains a defined
multiplicity of certain reactions, then, with near certainty, the subsequent
sub-sequence of reactions will contain at least or at most some predefined
number of other reactions''. The interesting problem for future research is how
to reconcile the SCM models capturing causality among the molecule copy counts
with the causal models defined for reaction event sub-sequences.

There are many definitions of causality in the literature. These definitions
usually assume either interventions, or they determine the independences
between random variables. For instance, causality between chemical reactions
has been associated with their dependency in \cite{dang2015}. General causality
in multivariate time-series has been studied extensively \cite{hlavackova2007,
  eichler2011, runge2019}. The basic consideration is whether the statistical
associations such as correlations or conditional dependencies may be sufficient
to claim causality under some conditions, and whether causality can be inferred
solely from the observations without any prior knowledge. A short answer is
that a causal discovery from data and from observed statistical associations
are critically dependent on the structure of data. The event time-series
described by linear model \eref{eq:10}, which was assumed in this paper,
involve categorical random variables.

The causal ordering of reaction event sub-sequences provides additional
qualitative information about the dynamics of BRNs. Once causally related
reaction sub-sequences have been identified, they can be interpreted as
guaranteed changes in the species concentrations, i.e., as causal changes in
the state of a BRN. This may allow identifying the components and modules of
BRNs with more predictable dynamics. However, the reaction events can only be
recorded in computer simulations. For in vivo and in vitro experiments, the
reaction events must be inferred from the observed molecular concentrations.
Moreover, it may not be completely known, which reactions are occurring in the
given BRN. In such a case, the causally related reaction sub-sequences can be
exploited as prior information on the multiplicities of possible reactions
occurring between the observations.

In this paper, it was proposed to identify the causally related and unrelated
reaction event sub-sequences by measuring their empirical conditional
probabilities, and then identifying the cases when these probabilities are
either very large or very small. The proposed method was implemented as a
computationally efficient query-response mechanism for selecting the reaction
event sub-sequences. Specifically, the sliding-window event sub-sequences are
further split into the left and the right event sub-sequences, which may be
then converted into sets or multi-sets. The causal query about the effects
identifies a set of left reaction events, and the corresponding right events
are then evaluated to identify their shared property or feature. It is crucial
that this property or feature is not present in any other right event
sub-sequences that were not selected by the query. The response to a given
query is the list of event sub-sequences having the identified common property
or feature. For anti-causal query about causes, the selected effects consisting
of the right event sub-sequences are matched with their left event
sub-sequences. The search is then performed to again identify a shared property
or feature, which can be returned as the cause in the response to a given
query. Furthermore, the reasoning in both causal and anti-causal
query-responses can be inverted in order to identify non-causal associations
instead.

The future work can consider more systematic approach to formulating complex
queries about the reaction events, and devising computationally effective
methods for searching shared properties or features among the event
sub-sequences forming the response. The causal discovery may also exploit both
the reaction events and the molecule copy counts. More importantly, once the
causally related or unrelated event sub-sequences have been identified, the
crucial question is how to interpret such information to better understand the
functions or dynamics of a BRN, especially in the context of the biological
system that this BRN represents. Moreover, causality could be also studied
assuming deterministic models of BRNs, which has not been considered in this
paper.

The distance metrics listed in Section 2 for the sequences, sets and multi-sets
of reaction events allow defining equivalences between the events, and to
perform the matrix profile analysis. The canonical matrix profile can be
modified to not only show the minimum distance values, but to also provide the
maxima and any other statistics of the distances including their
multiplicities. It resembles a feature extraction mechanism where the original
time-series are transformed into another time-series representation.
Unfortunately, at present, the libraries implementing the canonical matrix
profile are only optimized for the minimum Euclidean distance calculations. It
would be extremely useful to provide a highly optimized implementation of the
matrix profile algorithm allowing for arbitrary distance metrics and other
statistics. There is also a need to have a general and fast computing framework
for sliding-window processing of multivariate time-series data in \s{Python}
and in other languages.

The \s{NFsim} software can be further modified to remove unused or rarely used
features from the source code, and to refactor the source code for better
readability, speed and efficiency. Lastly, as demonstrated in this paper, the
data processing and visualization can be readily automated; however, also
automating the interpretation of the produced results is much harder, but also
much more interesting problem to consider in the future \cite{loskot2018}.

\bigskip\section{Conclusion}

The event time-series generated by the stochastic simulations of BRNs were used
to discover causally related event sub-sequences. The causality was defined in
terms of conditionally certain and conditionally uncertain reaction events. The
causal, non-causal and anti-causal associations were determined by the
computationally efficient query-response mechanism. In addition, since the
ordering of reaction events is locally irrelevant, the event sub-sequences were
transformed into the event sets or multi-sets. This can be exploited in
defining various distance metrics between the event sub-sequences. The event
sub-sequences having the zero distance are then assumed to be equivalent. The
sequence equivalence increases the likelihood of identifying the causally
related events. The defined distance metrics can be also used in the matrix
profile analysis of the event time-series.

Producing the numerical results required to modify and update the popular
stochastic simulator of BRNs \s{NFsim}. The reaction event time-series from the
simulations of five models of BRNs were processed by \s{Python} and \s{Matlab}
scripts within the seven defined numerical experiments. All tables and figures
automatically generated by the scripts are summarized in Supplementary, whereas
only the selected results are presented in the main paper.

\bigskip\section*{Acknowledgment}

This work was supported by a research grant from Zhejiang University.

\bigskip\section*{Figures}

\clearpage

\label{sc:refer}
\bibliographystyle{FrontiersStyle}
\bibliography{xrefer,xrefer1}

\clearpage


\renewcommand{\thetable}{S\arabic{table}}
\newcommand{\sscite}[1]{{\footnotesize\cite{#1}}}
\newcommand{\model}[1]{\noindent\textbf{#1}}
\newcommand{\code}[1]{{\small\texttt{#1}}}
\newlength{\phonyoutput}
\settowidth{\phonyoutput}{\emph{Output}}

\onecolumn\firstpage{1}

\title{{\huge\helveticaitalic{\textbf{Supplementary Material}}}}
\bigskip

The numerical results were generated for 5 biochemical models labeled as A--E
in 7 experiments labeled as 1--7.\bigskip

The files are available at \s{GitHub} repository\
{\small \url{https://github.com/ploskot/nfsim/tree/update}}:

\begin{tabular}{rl}
  \s{data/} & sample history files produced from a single run of stochastic
  simulation \\  
  \s{models/} & model definitions as \s{BNGL} files\\
  \s{scripts/} & \s{Python} and \s{Matlab} scripts for experiments
\end{tabular}\bigskip

\section*{Biochemical models}

\bigskip\model{Model A:} a multi-states system defined in
\code{models/modelA.bngl}

\begingroup
\renewcommand{\clearpage}{} 
\renewcommand{\section}[2]{}

\endgroup

\bigskip\model{Model B:} a model of chemotaxis defined in
\code{models/modelB.bngl}

\begingroup
\renewcommand{\clearpage}{} 
\renewcommand{\section}[2]{}

\endgroup

\bigskip\model{Model C:} an extended model of chemotaxis defined in
\code{models/modelC.bngl}

\begingroup
\renewcommand{\clearpage}{} 
\renewcommand{\section}[2]{}

\endgroup

\bigskip\model{Model D:} a model of trivalent ligand-bivalent receptor defined
in \code{models/modelD.bngl}

\begingroup
\renewcommand{\clearpage}{} 
\renewcommand{\section}[2]{}

\endgroup

\newpage

\bigskip\model{Model E:} a multi-sites system defined in
\code{models/modelE.bngl}

\begingroup
\renewcommand{\clearpage}{} 
\renewcommand{\section}[2]{}

\endgroup

\section*{Experiments}

\bigskip\model{Experiment 1:} \s{Python} scripts \code{script1.py} and
\code{script2.py}

\emph{Objective:} find unique reaction $N$-tuples and their counts, so they can
be clustered by their frequency of occurrence; in addition, the $N$-tuple
statistics assume partitioning of events into 100 blocks of 1 second duration,
and the minimum and maximum values across blocks are reported

\emph{Output:} Tables \ref{tabS1}--\ref{tabS5}

\bigskip\model{Experiment 2:} \s{Matlab} script \code{mscript1.m}

\emph{Objective:} sort and cluster reactions by their frequency of occurrence

\begin{figure}[p]
 \centering
  \begin{floatrow}
    \ffigbox[\FBwidth]{\includegraphics[scale=1.0]{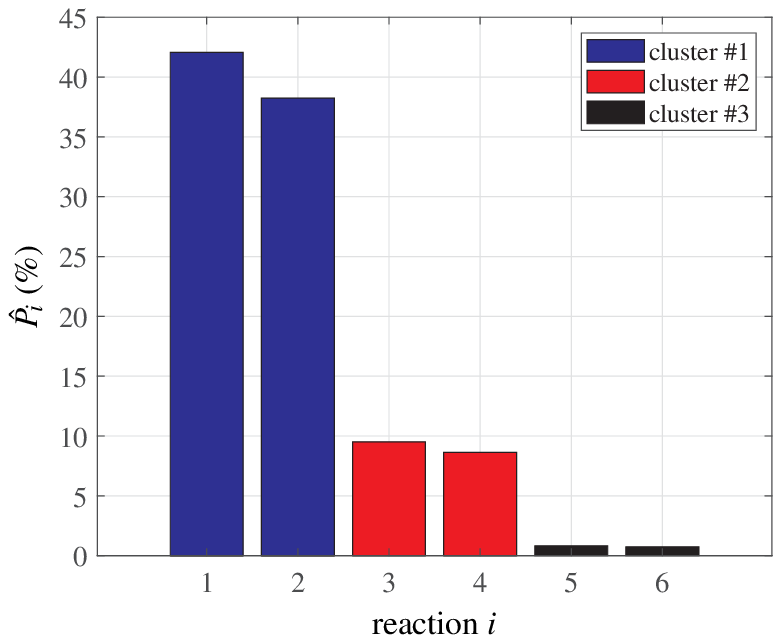}}
    {\caption{Model A, reaction frequency}\label{rxnfrq1}}
    \ffigbox[\FBwidth]{\includegraphics[scale=1.0]{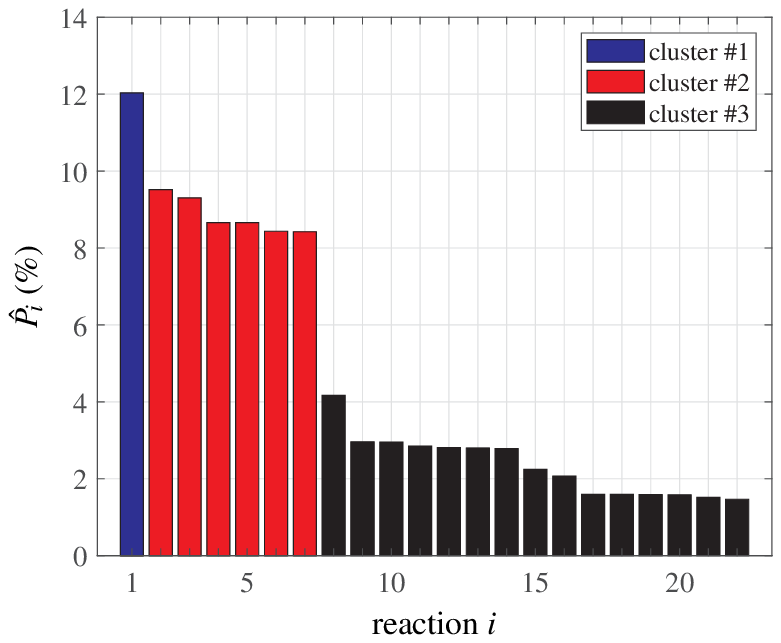}}
    {\caption{Model B, reaction frequency}\label{rxnfrq2}}
  \end{floatrow}
\end{figure}

\begin{figure}[p]
 \centering
  \begin{floatrow}
    \ffigbox[\FBwidth]{\includegraphics[scale=1.0]{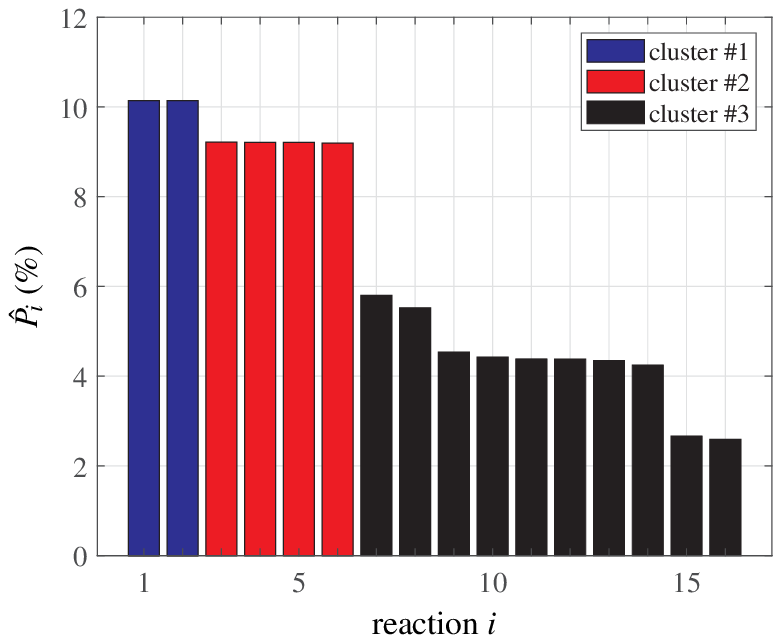}}
    {\caption{Model C, reaction frequency}\label{rxnfrq3}}
    \ffigbox[\FBwidth]{\includegraphics[scale=1.0]{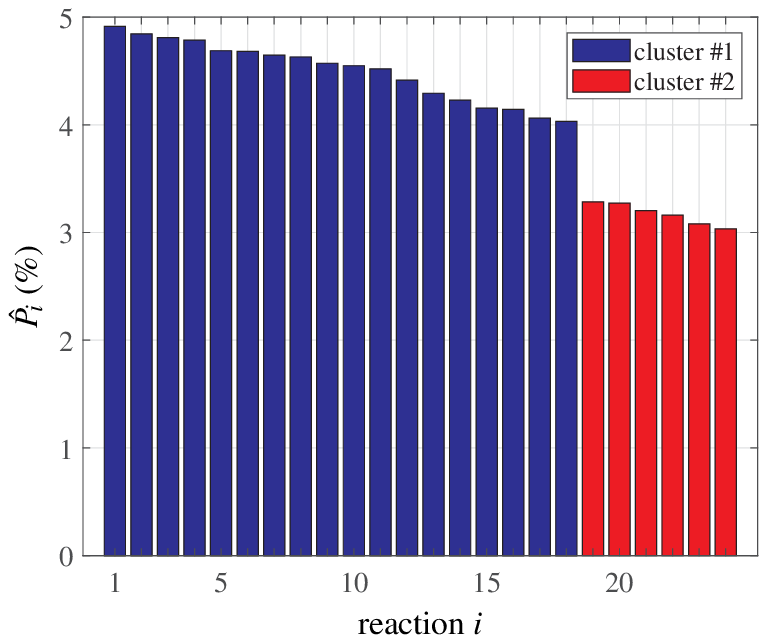}}
    {\caption{Model D, reaction frequency}\label{rxnfrq4}}
  \end{floatrow}
\end{figure}

\begin{figure}[p]
  \centering
  \begin{floatrow}
    \ffigbox[\FBwidth]{\includegraphics[scale=1.0]{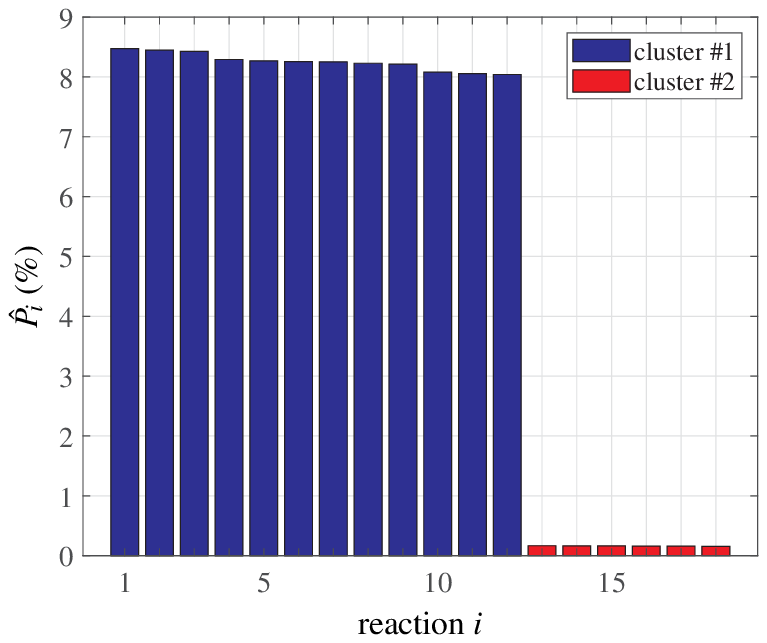}}
    {\caption{Model E, reaction frequency}\label{rxnfrq5}}
\end{floatrow}
\end{figure}

\emph{Output:} Figures \ref{rxnfrq1}--\ref{rxnfrq5}

\bigskip\model{Experiment 3:} \s{Matlab} script \code{mscript2.m}

\emph{Objective:} the box plots of unique reactions within sliding-window event
$N$-tuples to visualize their statistics; it is assumed that reactions are
sorted by their frequency of occurrence as in Figures
\ref{rxnfrq1}--\ref{rxnfrq5}

\begin{figure}[p]
 \centering
  \begin{floatrow}
    \ffigbox[\FBwidth]{\includegraphics[scale=0.5]{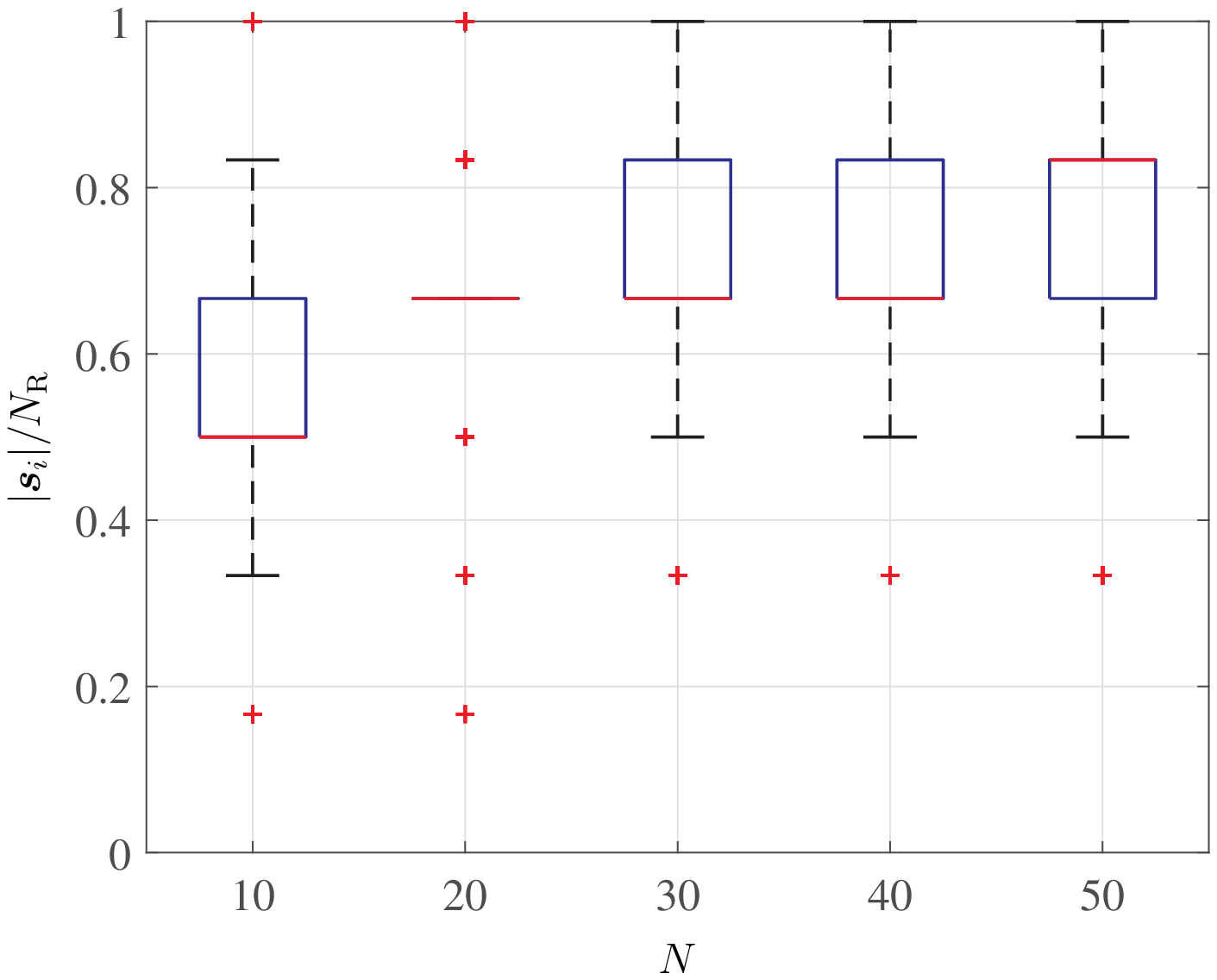}}
    {\caption{Model A, reaction statistics}\label{rxnvars1}}
    \ffigbox[\FBwidth]{\includegraphics[scale=0.5]{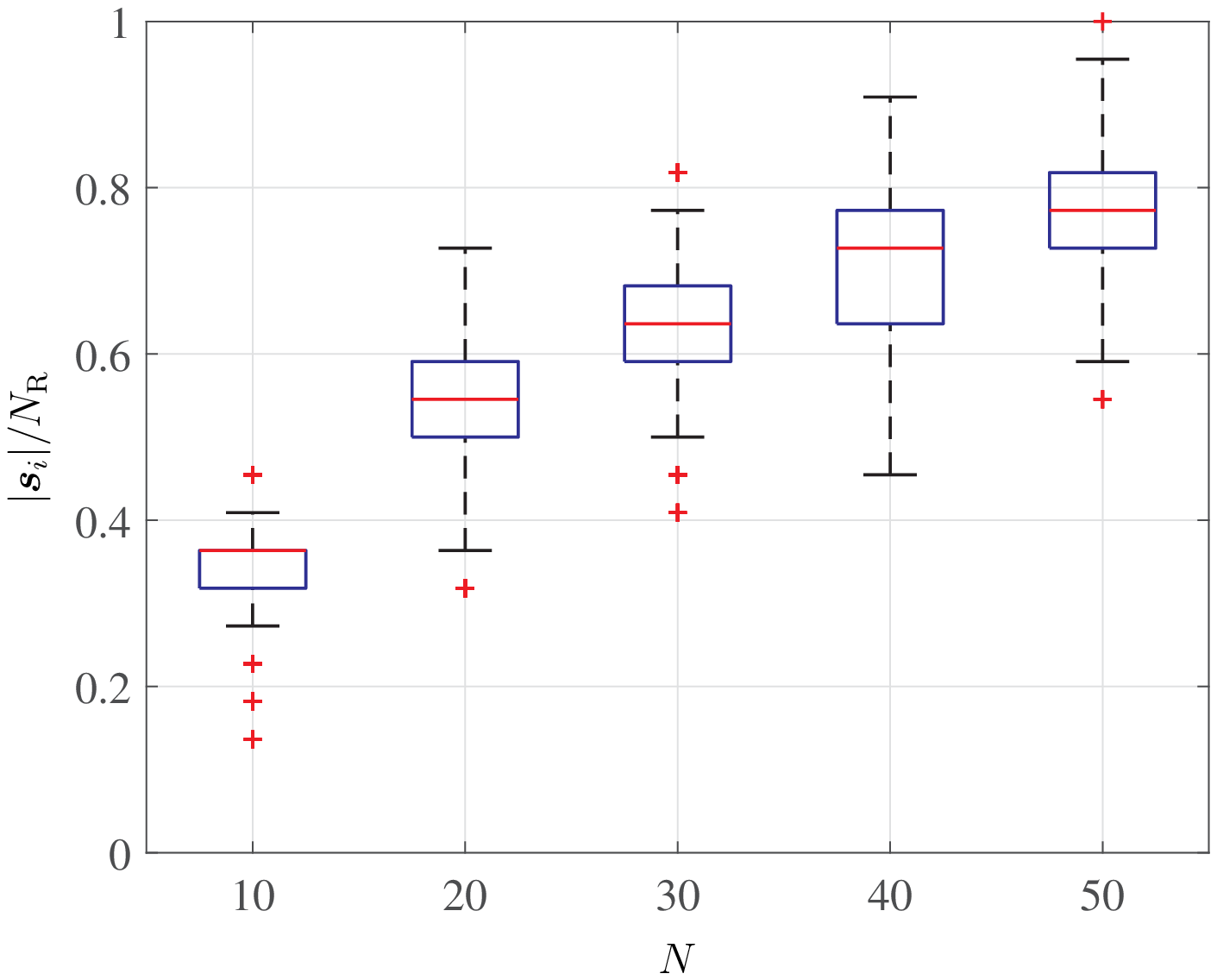}}
    {\caption{Model B, reaction statistics}\label{rxnvars2}}
  \end{floatrow}
\end{figure}

\begin{figure}[p]
 \centering
  \begin{floatrow}
    \ffigbox[\FBwidth]{\includegraphics[scale=0.5]{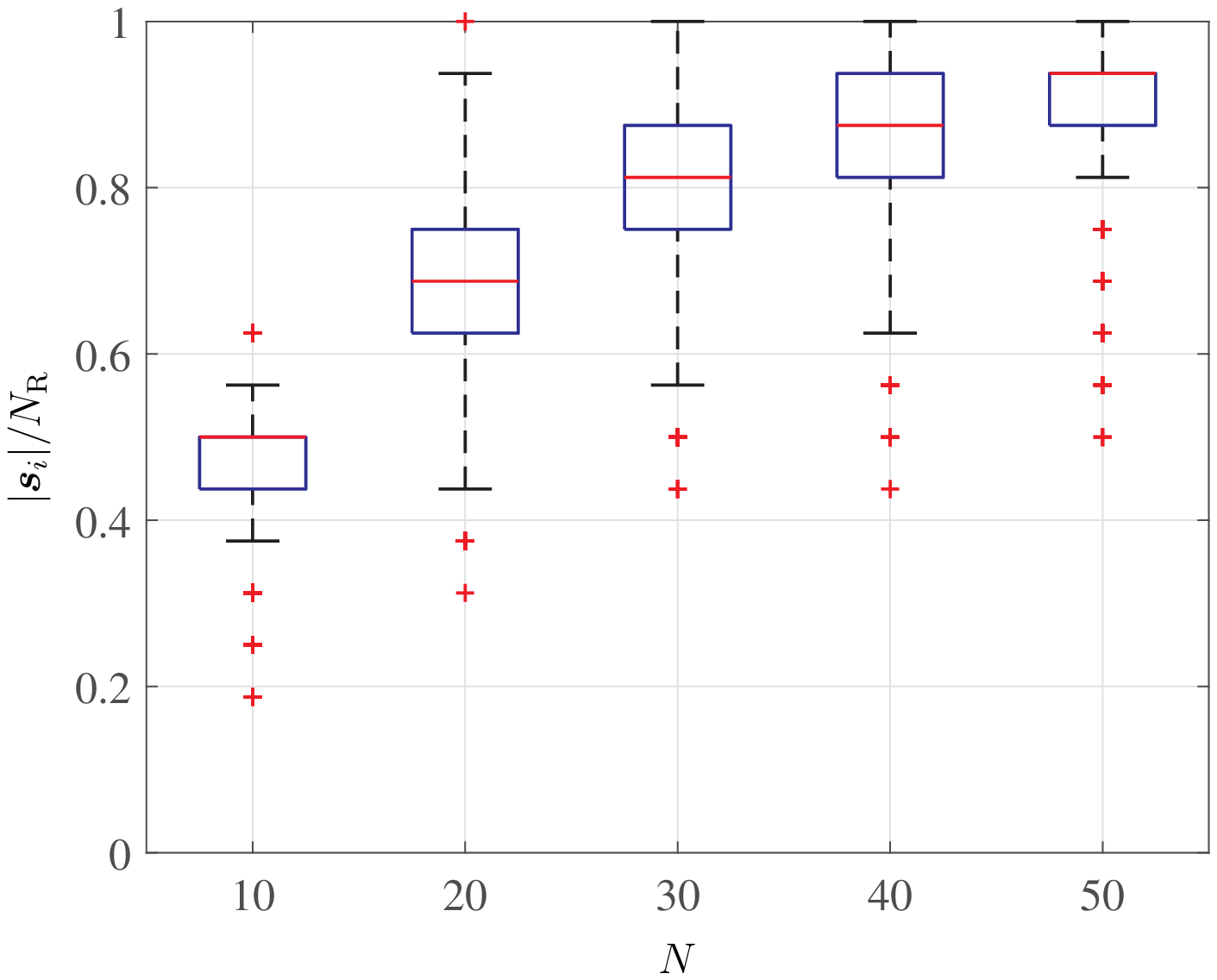}}
    {\caption{Model C, reaction statistics}\label{rxnvars3}}
    \ffigbox[\FBwidth]{\includegraphics[scale=0.5]{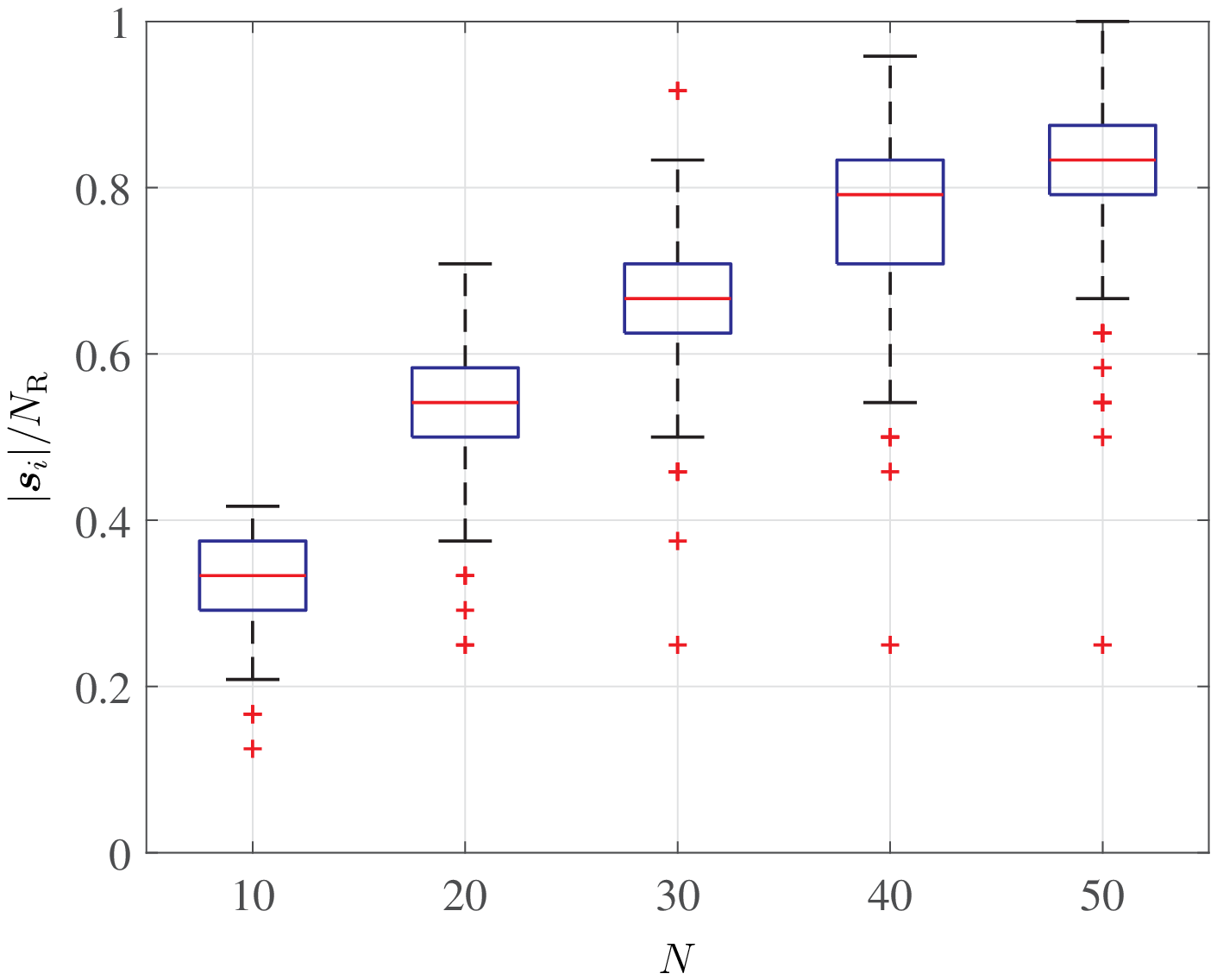}}
    {\caption{Model D, reaction statistics}\label{rxnvars4}}
  \end{floatrow}
\end{figure}

\begin{figure}[p]
  \centering
  \begin{floatrow}
    \ffigbox[\FBwidth]{\includegraphics[scale=0.5]{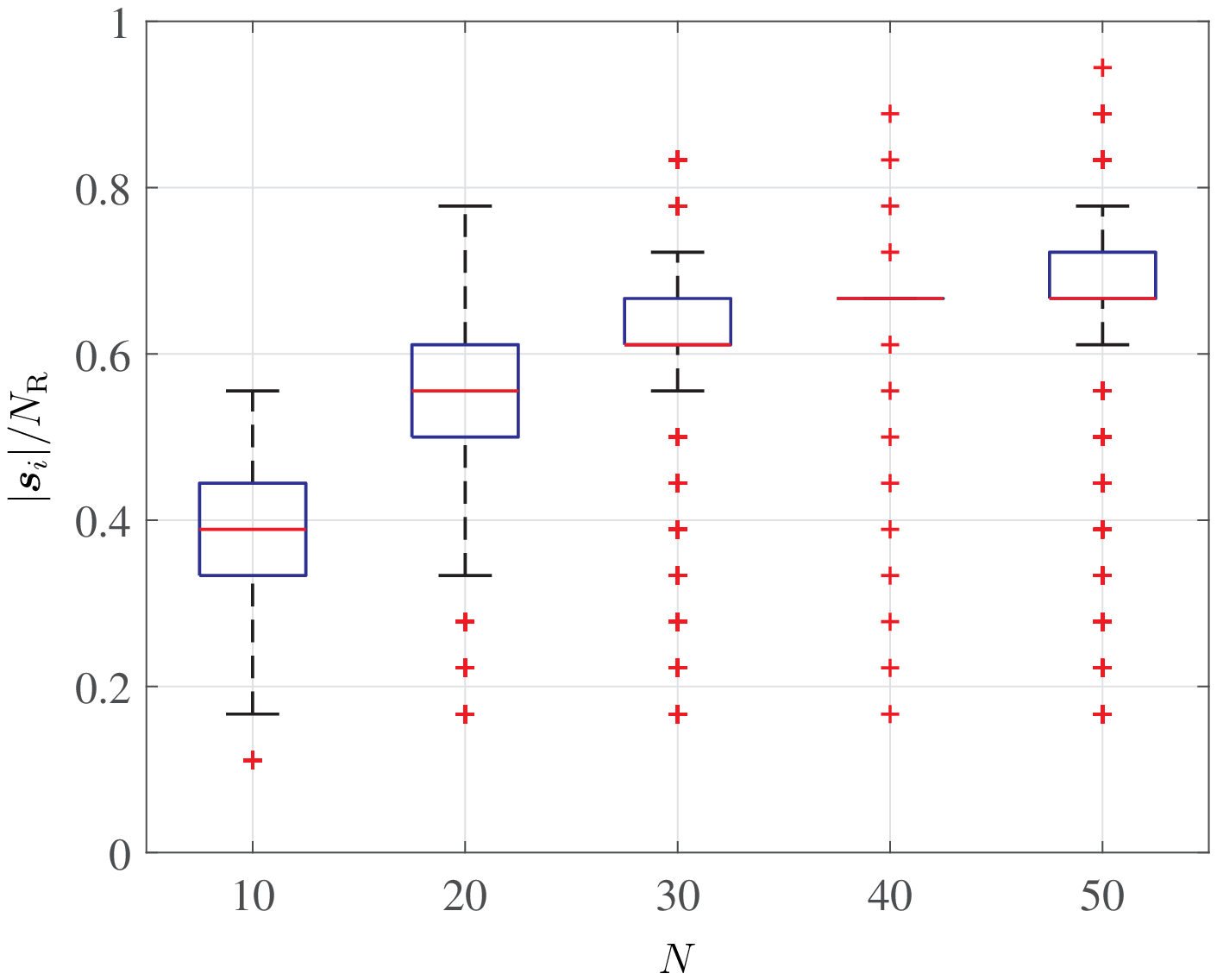}}
    {\caption{Model E, reaction statistics}\label{rxnvars5}}
\end{floatrow}
\end{figure}

\emph{Output:} Figures \ref{rxnvars1}--\ref{rxnvars5}

\bigskip\model{Experiment 4:} \s{Matlab} script \code{mscript3.m}

\emph{Objective:} assessment of reaction $N_2$-tuples as effects for the query
about $N_1$-tuple as causes; the experiments and figures are labeled by the
query, i.e., ``$4-C-F$'', where $F\geq 1$ is the required minimum number of
reactions from the reaction cluster $C\in\{1,2,\ldots\}$ occurring within
reaction $N_1$-tuples

\begin{figure}[p]
 \centering
  \begin{floatrow}
    \ffigbox[\FBwidth]{\includegraphics[scale=1.0]{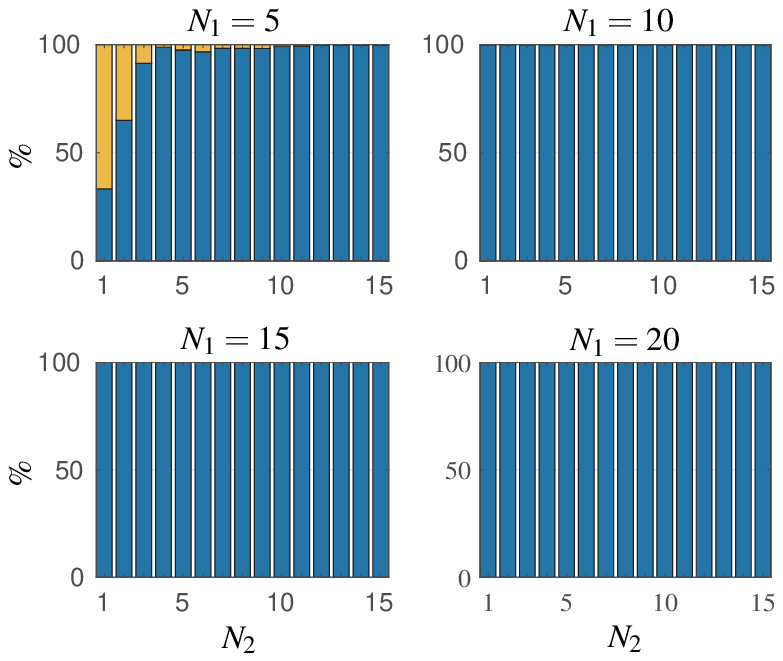}}
    {\caption{Model A, Experiment 4-1-1}\label{pict11-11}}
    \ffigbox[\FBwidth]{\includegraphics[scale=1.0]{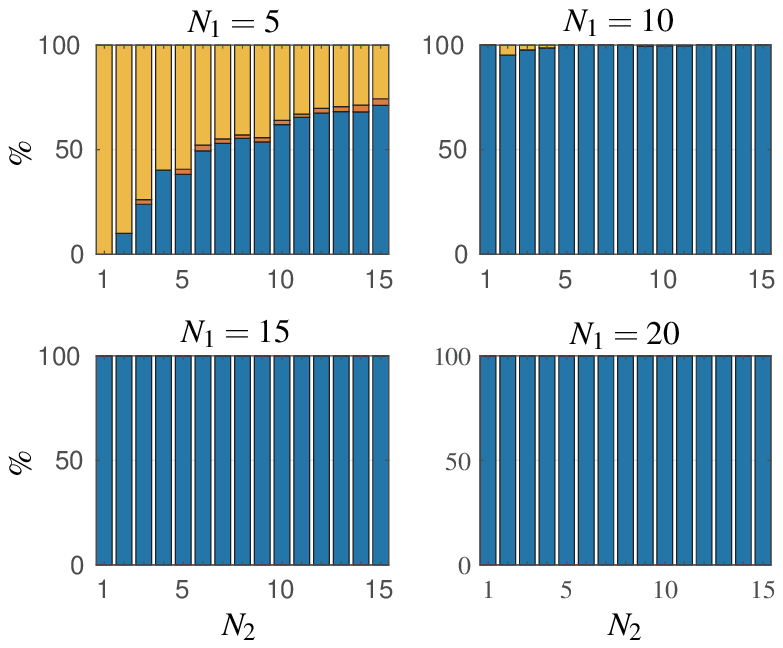}}
    {\caption{Model A, Experiment 4-1-3}\label{pict11-13}}
  \end{floatrow}
\end{figure}

\begin{figure}[p]
 \centering
  \begin{floatrow}
    \ffigbox[\FBwidth]{\includegraphics[scale=1.0]{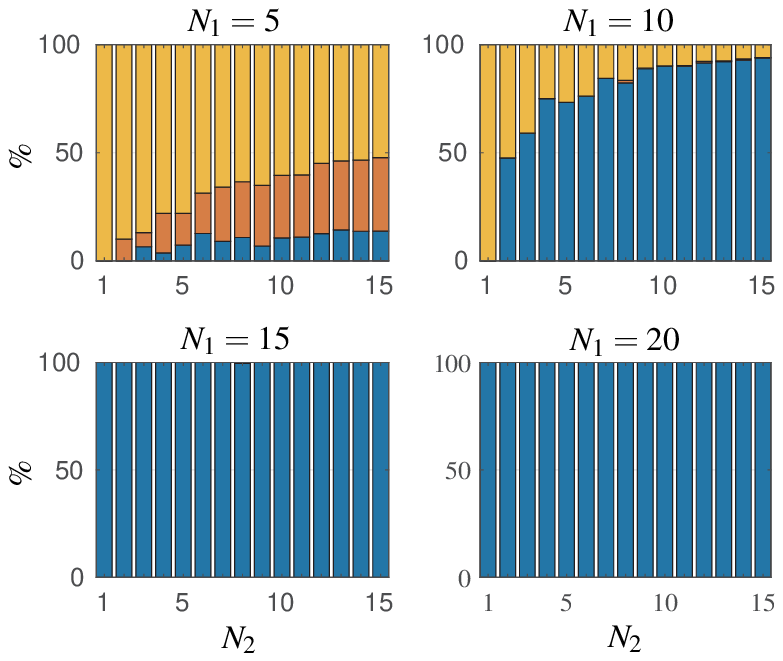}}
    {\caption{Model A, Experiment 4-1-5}\label{pict11-15}}
    \ffigbox[\FBwidth]{\includegraphics[scale=1.0]{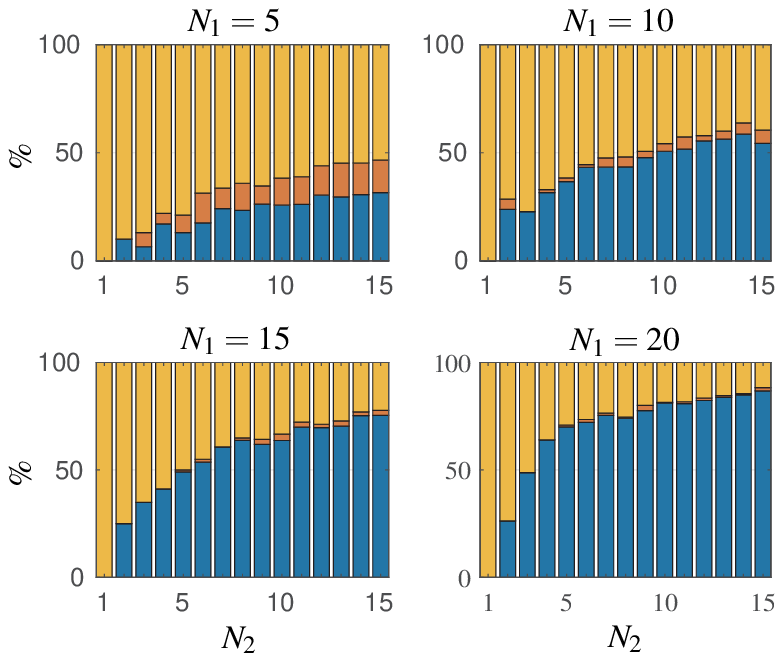}}
    {\caption{Model A, Experiment 4-2-1}\label{pict11-21}}
  \end{floatrow}
\end{figure}

\begin{figure}[p]
 \centering
  \begin{floatrow}
    \ffigbox[\FBwidth]{\includegraphics[scale=1.0]{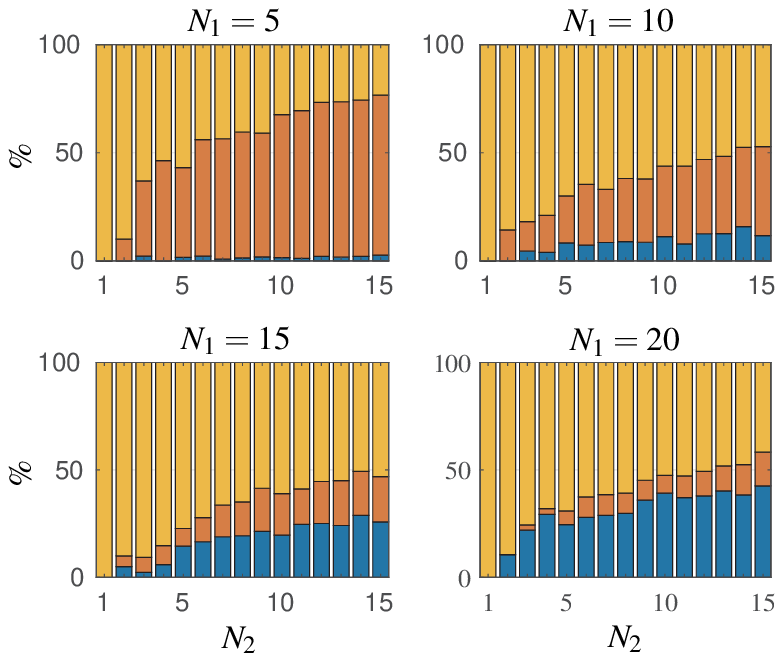}}
    {\caption{Model A, Experiment 4-2-3}\label{pict11-23}}
    \ffigbox[\FBwidth]{\includegraphics[scale=1.0]{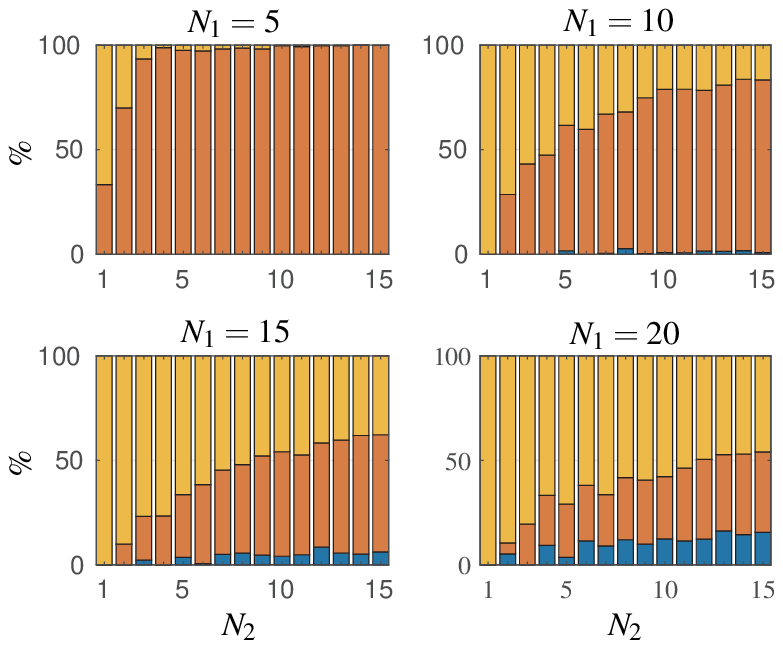}}
    {\caption{Model A, Experiment 4-2-5}\label{pict11-25}}
  \end{floatrow}
\end{figure}

\begin{figure}[p]
 \centering
  \begin{floatrow}
    \ffigbox[\FBwidth]{\includegraphics[scale=1.0]{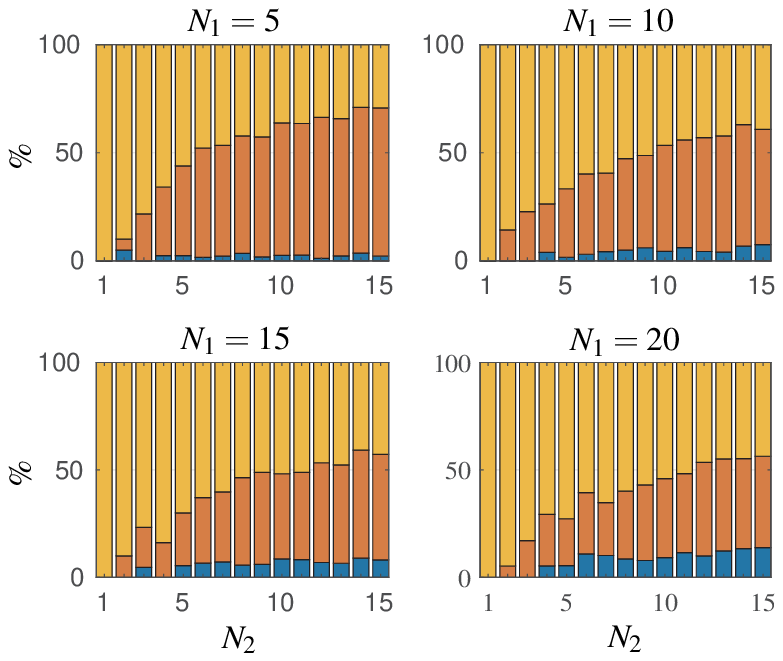}}
    {\caption{Model A, Experiment 4-3-1}\label{pict11-31}}
    \ffigbox[\FBwidth]{\includegraphics[scale=1.0]{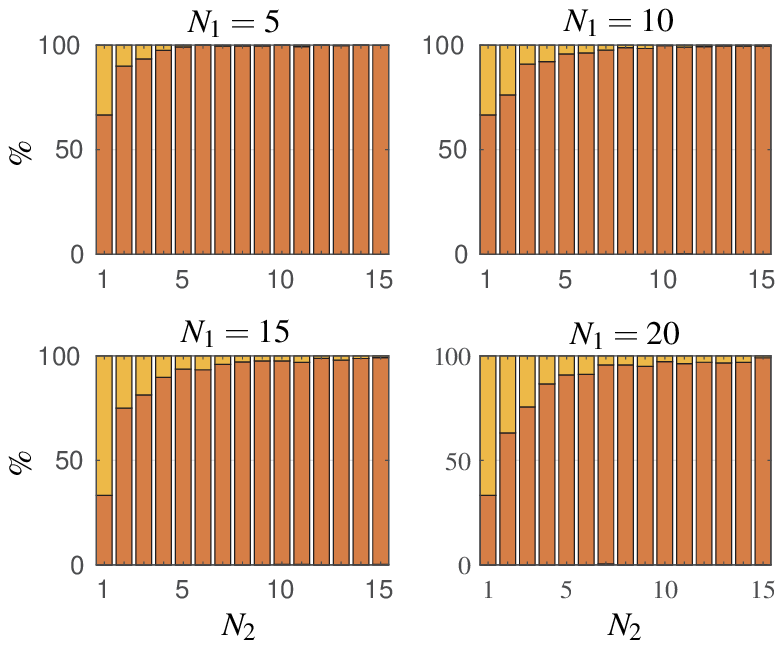}}
    {\caption{Model A, Experiment 4-3-3}\label{pict11-33}}
  \end{floatrow}
\end{figure}

\begin{figure}[p]
  \centering
  \begin{floatrow}
    \ffigbox[\FBwidth]{\includegraphics[scale=1.0]{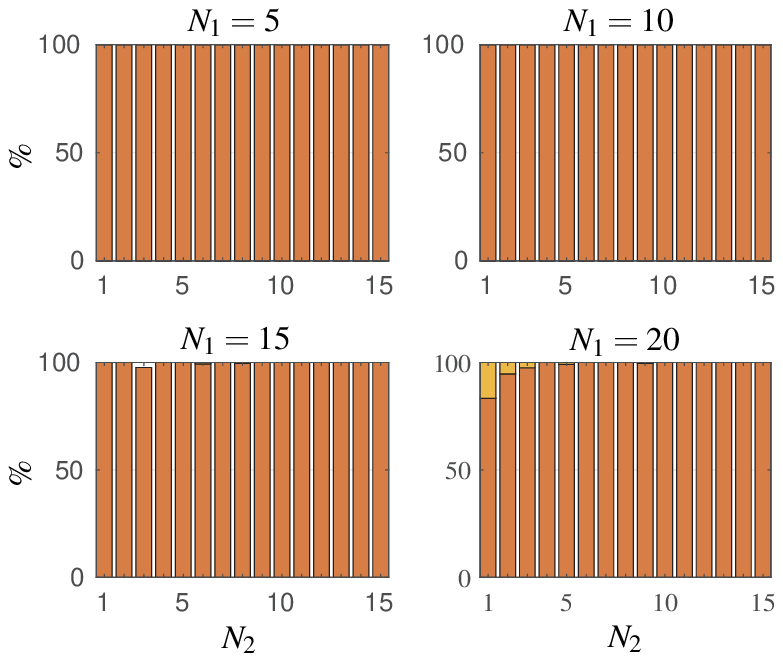}}
    {\caption{Model A, Experiment 4-3-5}\label{pict11-35}}
\end{floatrow}
\end{figure}

\begin{figure}[p]
 \centering
  \begin{floatrow}
    \ffigbox[\FBwidth]{\includegraphics[scale=1.0]{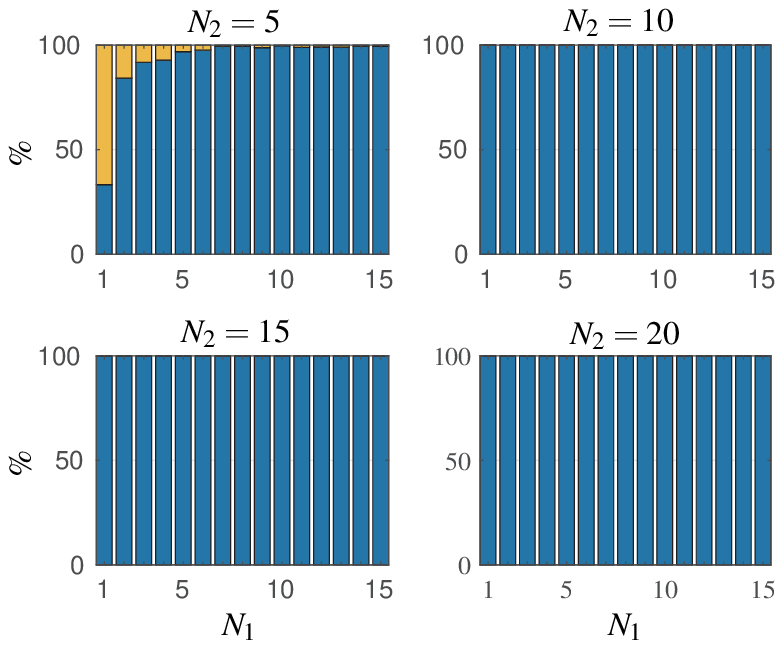}}
    {\caption{Model A, Experiment 5-1-1}\label{pict21-11}}
    \ffigbox[\FBwidth]{\includegraphics[scale=1.0]{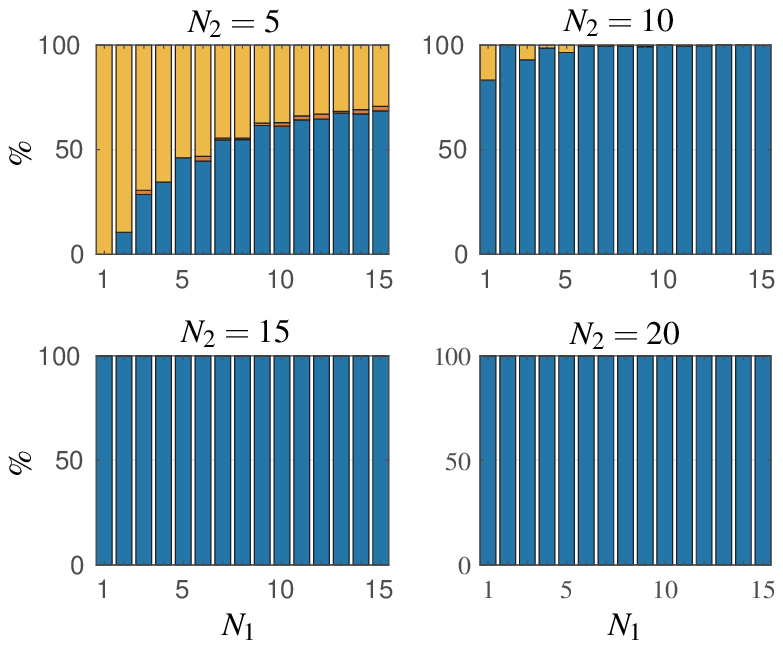}}
    {\caption{Model A, Experiment 5-1-3}\label{pict21-13}}
  \end{floatrow}
\end{figure}

\begin{figure}[p]
 \centering
  \begin{floatrow}
    \ffigbox[\FBwidth]{\includegraphics[scale=1.0]{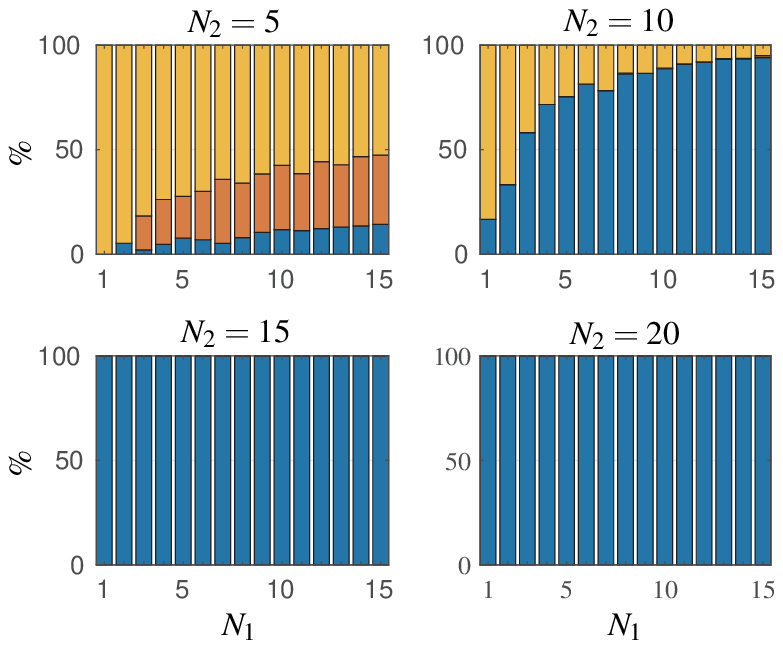}}
    {\caption{Model A, Experiment 5-1-5}\label{pict21-15}}
    \ffigbox[\FBwidth]{\includegraphics[scale=1.0]{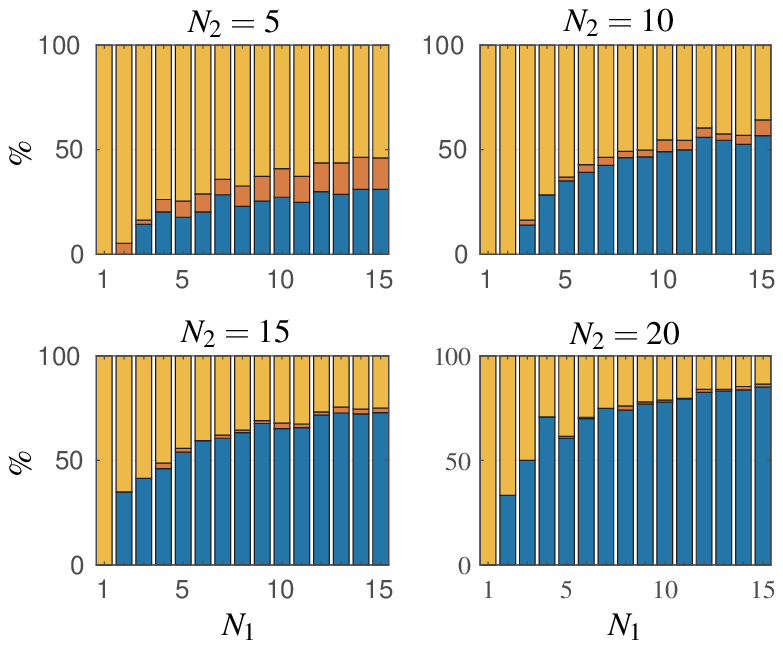}}
    {\caption{Model A, Experiment 5-2-1}\label{pict21-21}}
  \end{floatrow}
\end{figure}

\begin{figure}[p]
 \centering
  \begin{floatrow}
    \ffigbox[\FBwidth]{\includegraphics[scale=1.0]{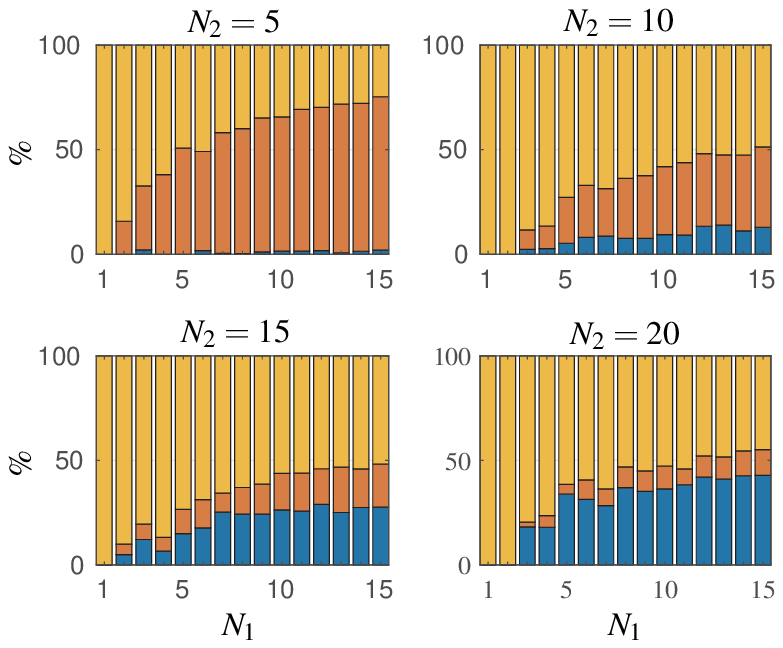}}
    {\caption{Model A, Experiment 5-2-3}\label{pict21-23}}
    \ffigbox[\FBwidth]{\includegraphics[scale=1.0]{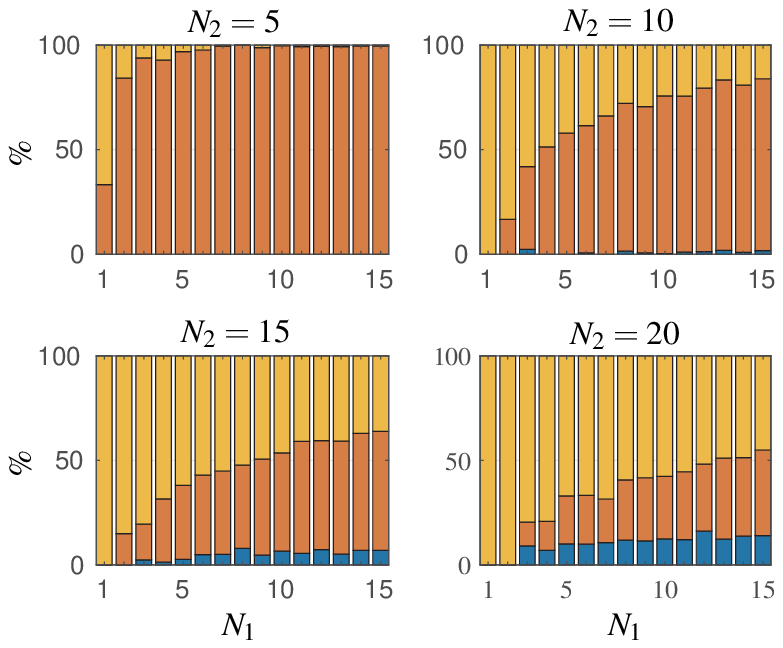}}
    {\caption{Model A, Experiment 5-2-5}\label{pict21-25}}
  \end{floatrow}
\end{figure}

\begin{figure}[p]
 \centering
  \begin{floatrow}
    \ffigbox[\FBwidth]{\includegraphics[scale=1.0]{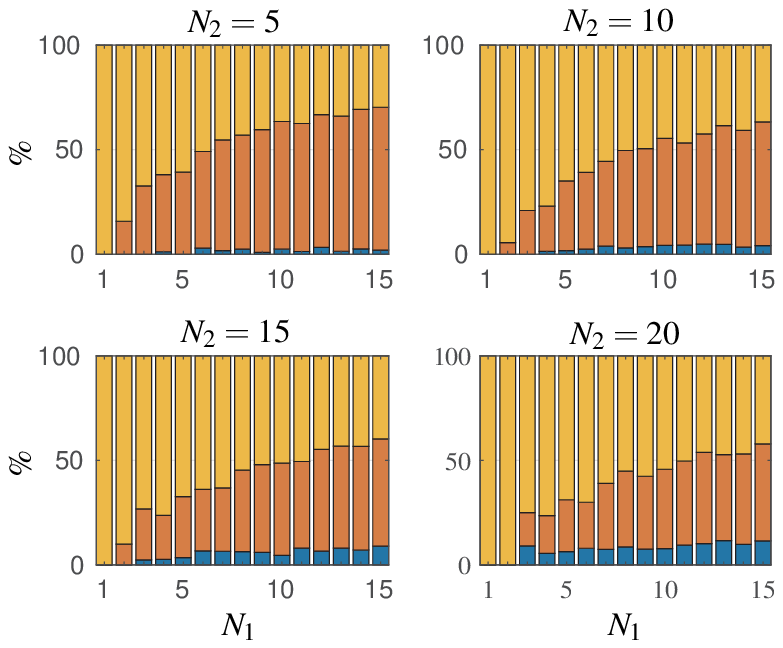}}
    {\caption{Model A, Experiment 5-3-1}\label{pict21-31}}
    \ffigbox[\FBwidth]{\includegraphics[scale=1.0]{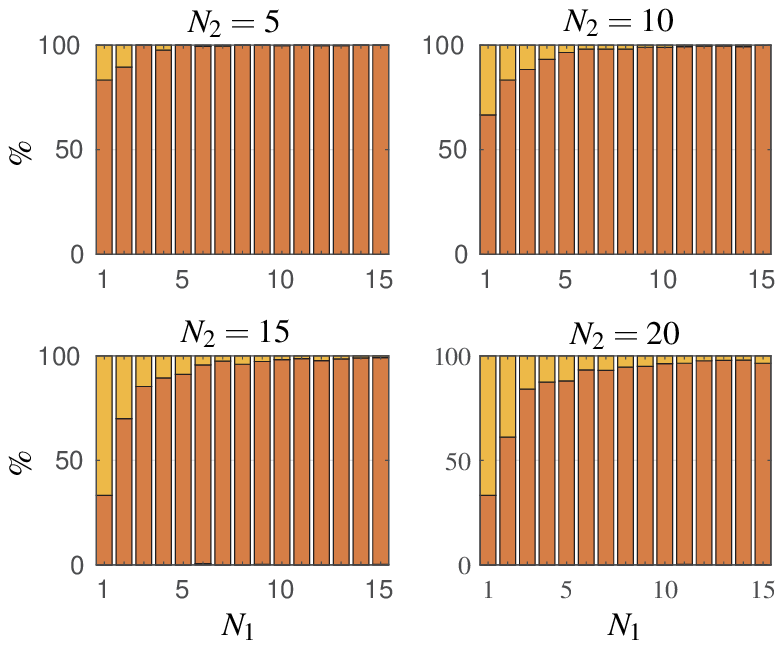}}
    {\caption{Model A, Experiment 5-3-3}\label{pict21-33}}
  \end{floatrow}
\end{figure}

\begin{figure}[p]
  \centering
  \begin{floatrow}
    \ffigbox[\FBwidth]{\includegraphics[scale=1.0]{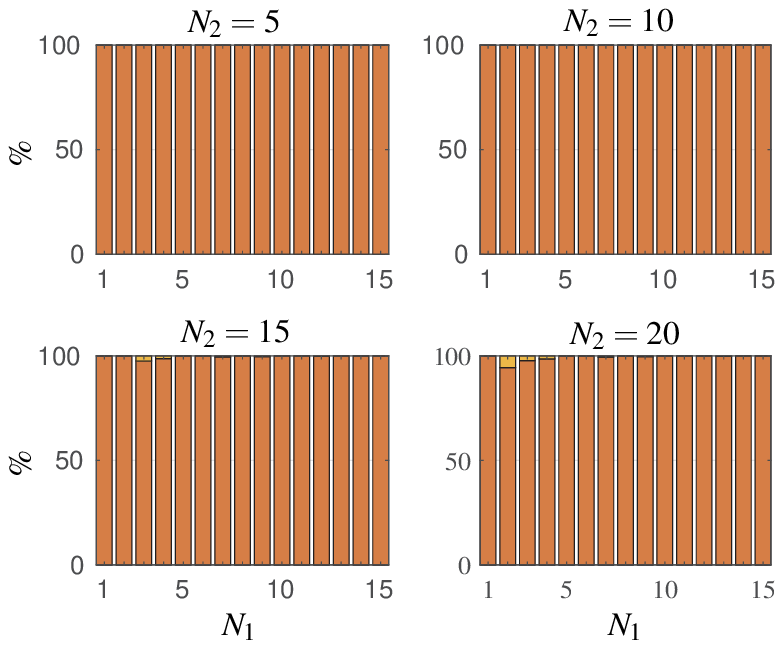}}
    {\caption{Model A, Experiment 5-3-5}\label{pict21-35}}
\end{floatrow}
\end{figure}

\begin{figure}[p]
 \centering
  \begin{floatrow}
    \ffigbox[\FBwidth]{\includegraphics[scale=1.0]{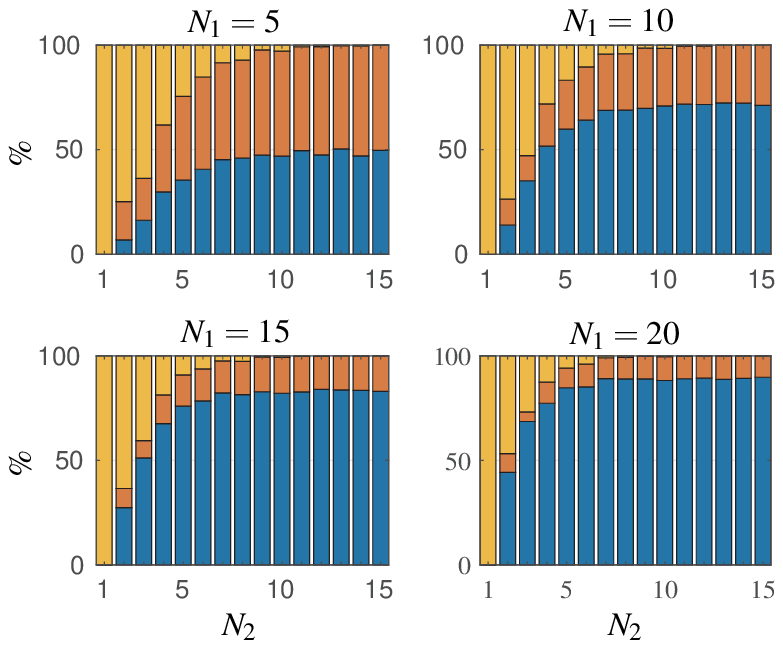}}
    {\caption{Model B, Experiment 4-1-1}\label{pict12-11}}
    \ffigbox[\FBwidth]{\includegraphics[scale=1.0]{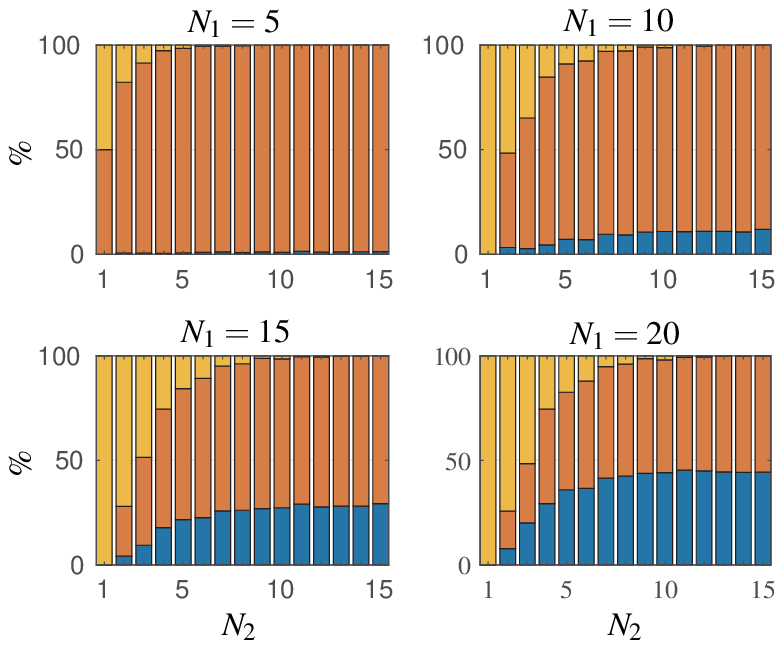}}
    {\caption{Model B, Experiment 4-1-3}\label{pict12-13}}
  \end{floatrow}
\end{figure}

\begin{figure}[p]
 \centering
  \begin{floatrow}
    \ffigbox[\FBwidth]{\includegraphics[scale=1.0]{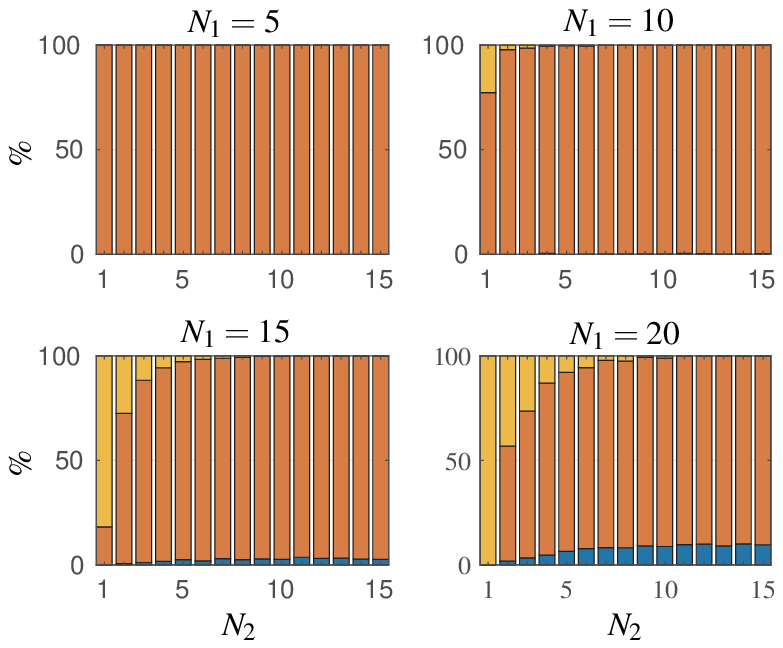}}
    {\caption{Model B, Experiment 4-1-5}\label{pict12-15}}
    \ffigbox[\FBwidth]{\includegraphics[scale=1.0]{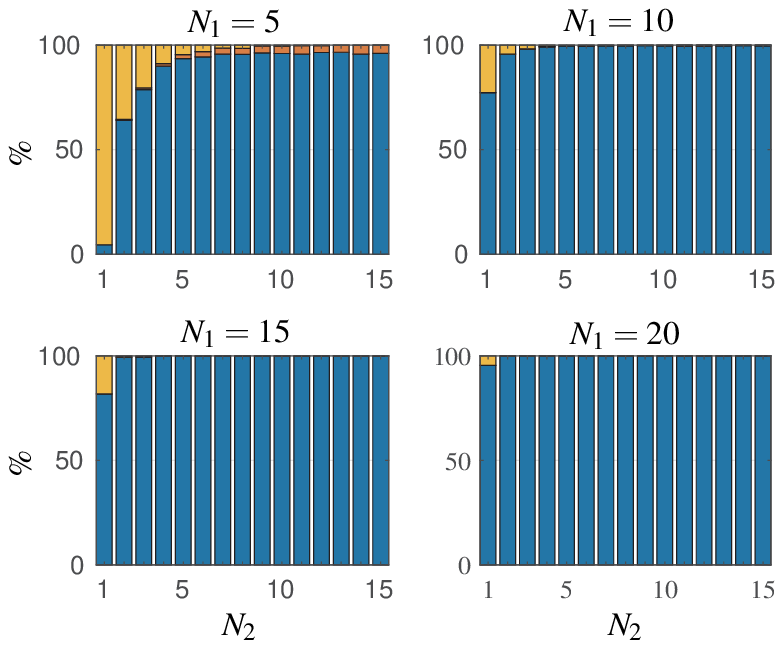}}
    {\caption{Model B, Experiment 4-2-1}\label{pict12-21}}
  \end{floatrow}
\end{figure}

\begin{figure}[p]
 \centering
  \begin{floatrow}
    \ffigbox[\FBwidth]{\includegraphics[scale=1.0]{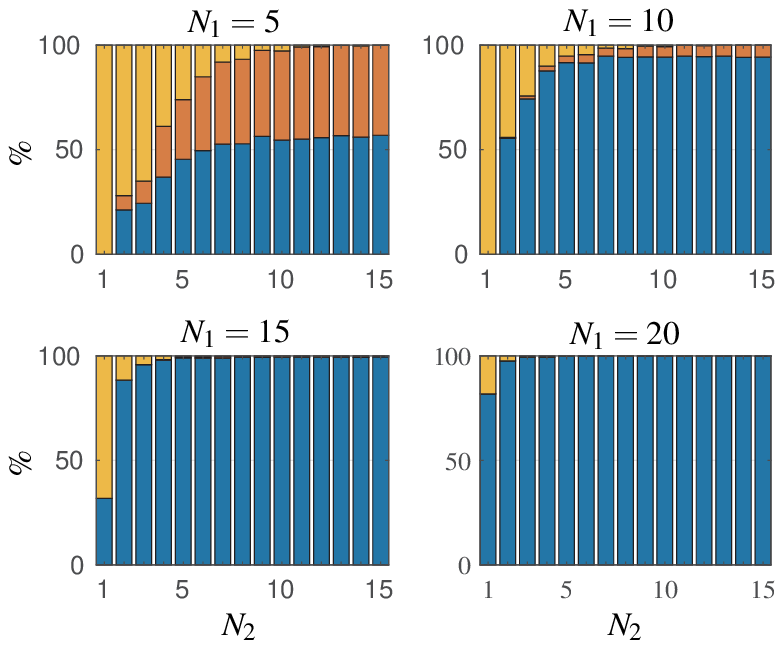}}
    {\caption{Model B, Experiment 4-2-3}\label{pict12-23}}
    \ffigbox[\FBwidth]{\includegraphics[scale=1.0]{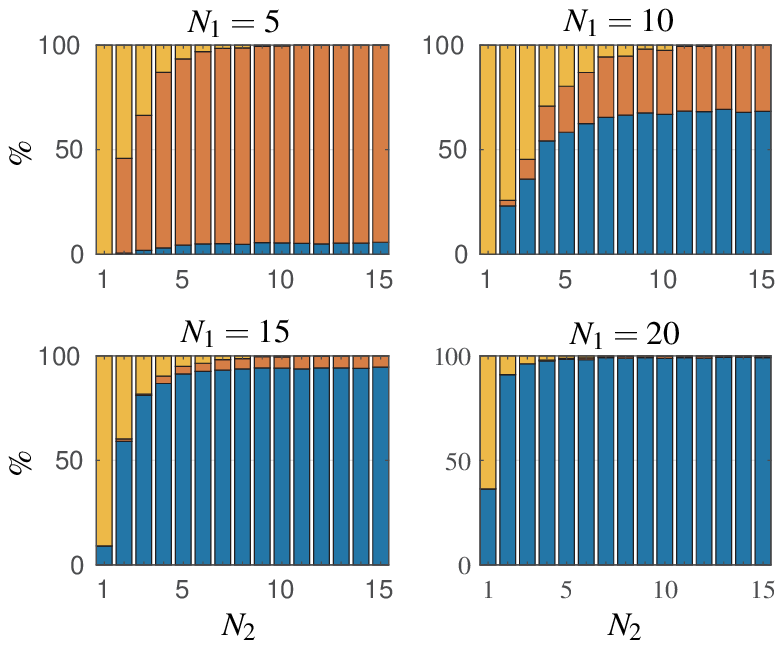}}
    {\caption{Model B, Experiment 4-2-5}\label{pict12-25}}
  \end{floatrow}
\end{figure}

\begin{figure}[p]
 \centering
  \begin{floatrow}
    \ffigbox[\FBwidth]{\includegraphics[scale=1.0]{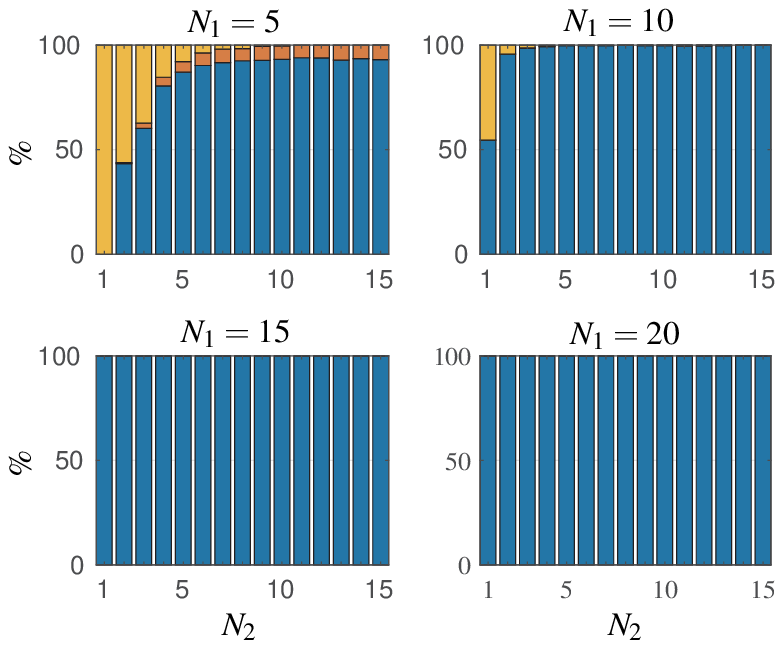}}
    {\caption{Model B, Experiment 4-3-1}\label{pict12-31}}
    \ffigbox[\FBwidth]{\includegraphics[scale=1.0]{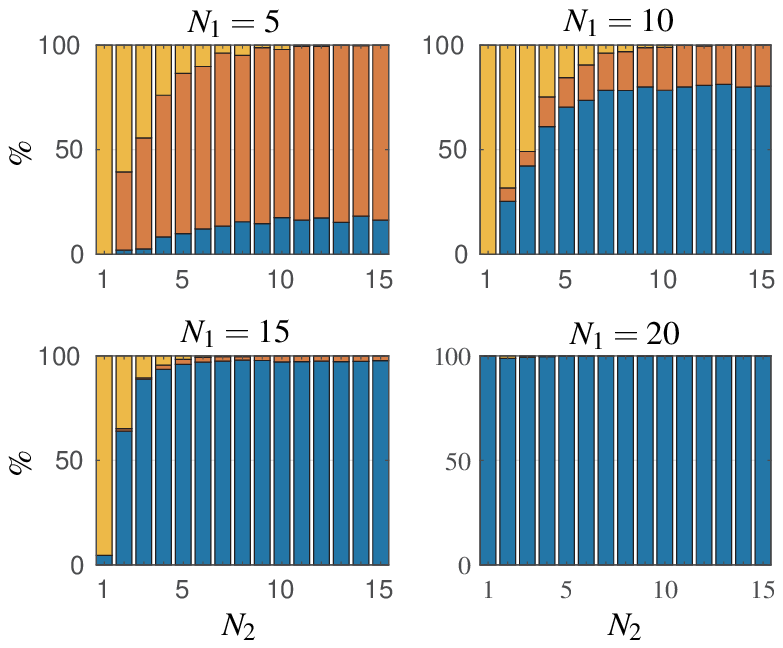}}
    {\caption{Model B, Experiment 4-3-3}\label{pict12-33}}
  \end{floatrow}
\end{figure}

\begin{figure}[p]
  \centering
  \begin{floatrow}
    \ffigbox[\FBwidth]{\includegraphics[scale=1.0]{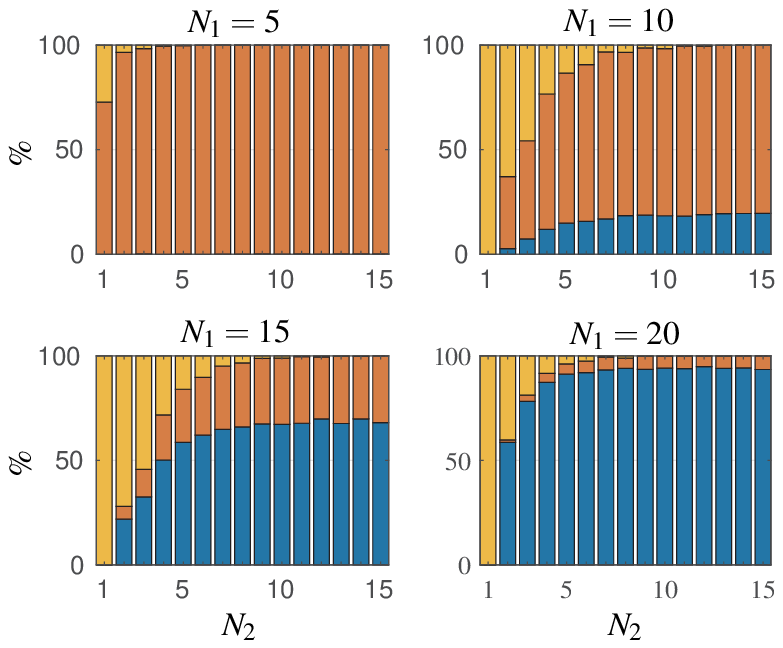}}
    {\caption{Model B, Experiment 4-3-5}\label{pict12-35}}
\end{floatrow}
\end{figure}

\begin{figure}[p]
 \centering
  \begin{floatrow}
    \ffigbox[\FBwidth]{\includegraphics[scale=1.0]{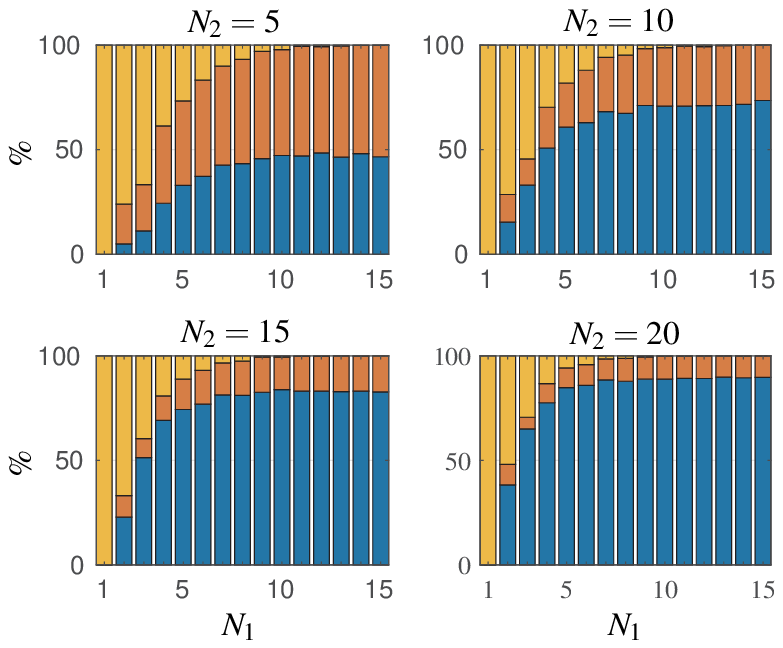}}
    {\caption{Model B, Experiment 5-1-1}\label{pict22-11}}
    \ffigbox[\FBwidth]{\includegraphics[scale=1.0]{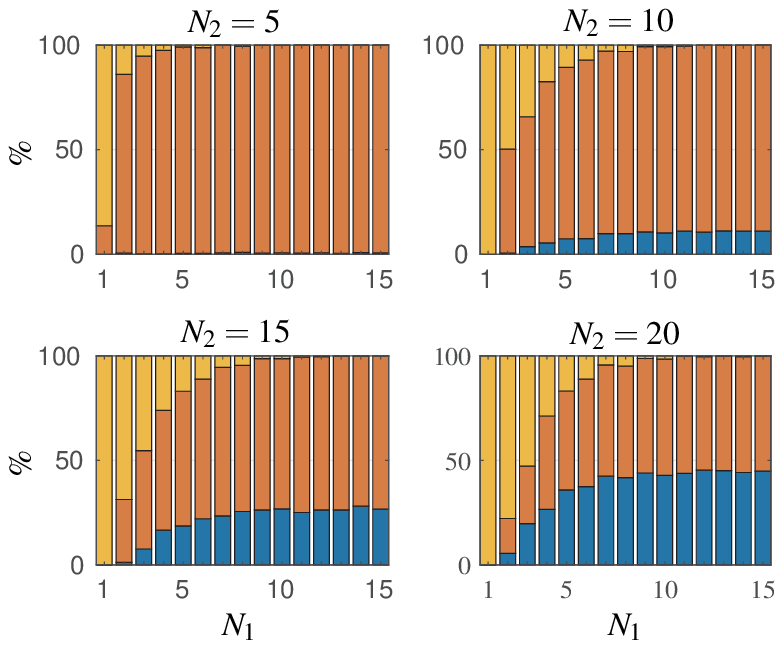}}
    {\caption{Model B, Experiment 5-1-3}\label{pict22-13}}
  \end{floatrow}
\end{figure}

\begin{figure}[p]
 \centering
  \begin{floatrow}
    \ffigbox[\FBwidth]{\includegraphics[scale=1.0]{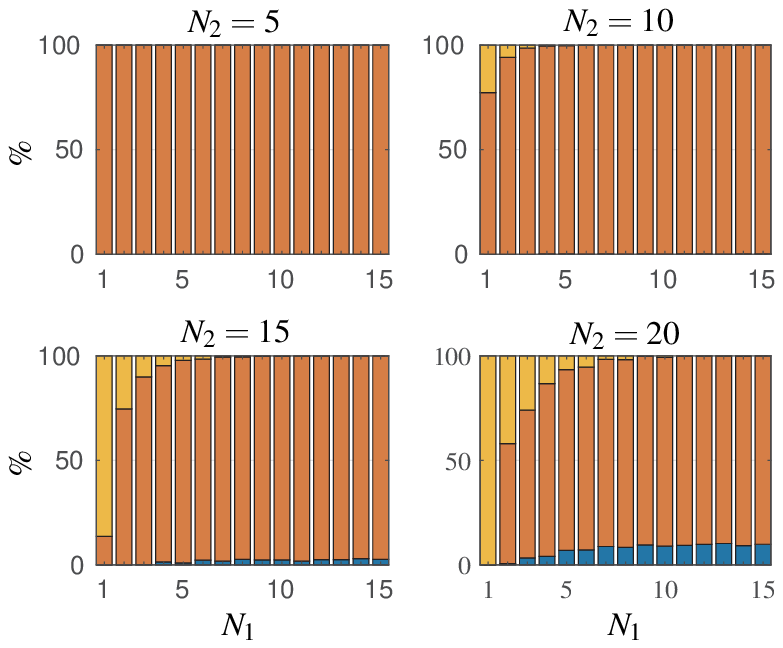}}
    {\caption{Model B, Experiment 5-1-5}\label{pict22-15}}
    \ffigbox[\FBwidth]{\includegraphics[scale=1.0]{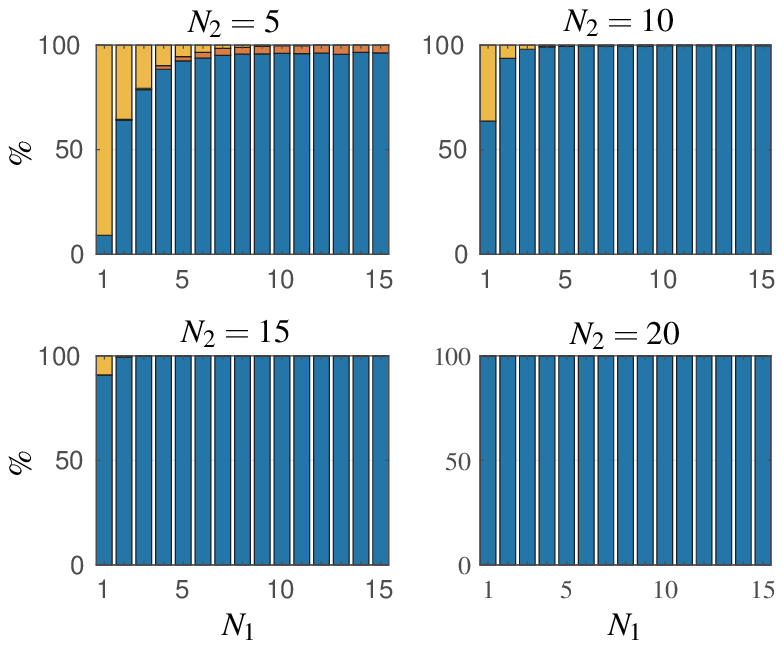}}
    {\caption{Model B, Experiment 5-2-1}\label{pict22-21}}
  \end{floatrow}
\end{figure}

\begin{figure}[p]
 \centering
  \begin{floatrow}
    \ffigbox[\FBwidth]{\includegraphics[scale=1.0]{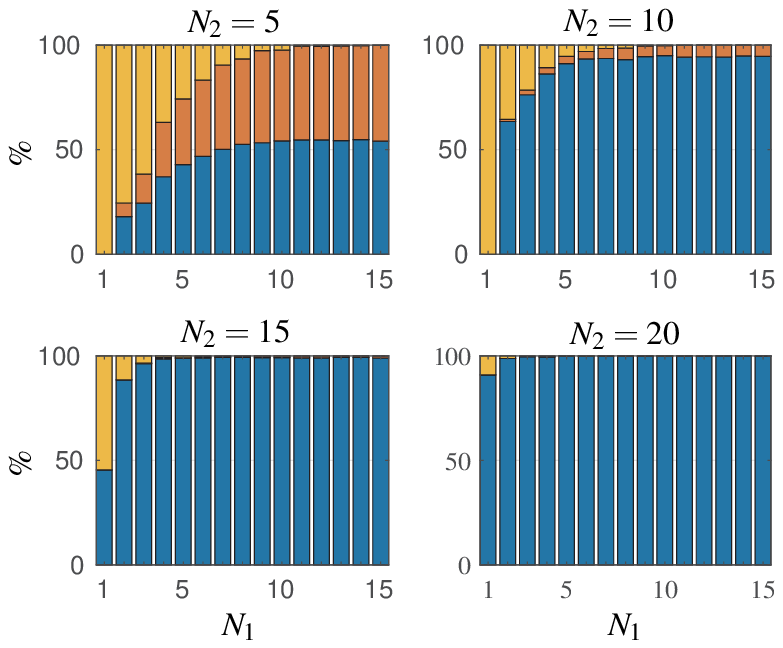}}
    {\caption{Model B, Experiment 5-2-3}\label{pict22-23}}
    \ffigbox[\FBwidth]{\includegraphics[scale=1.0]{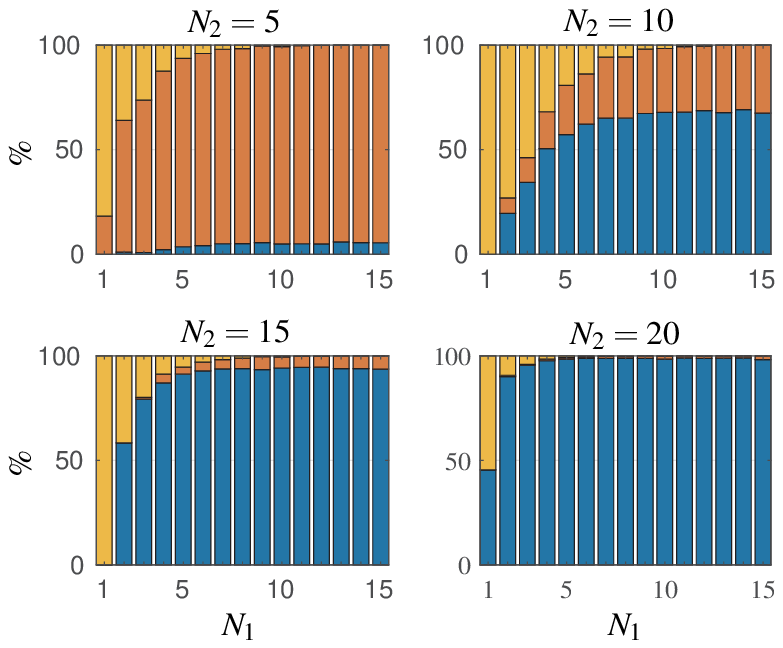}}
    {\caption{Model B, Experiment 5-2-5}\label{pict22-25}}
  \end{floatrow}
\end{figure}

\begin{figure}[p]
 \centering
  \begin{floatrow}
    \ffigbox[\FBwidth]{\includegraphics[scale=1.0]{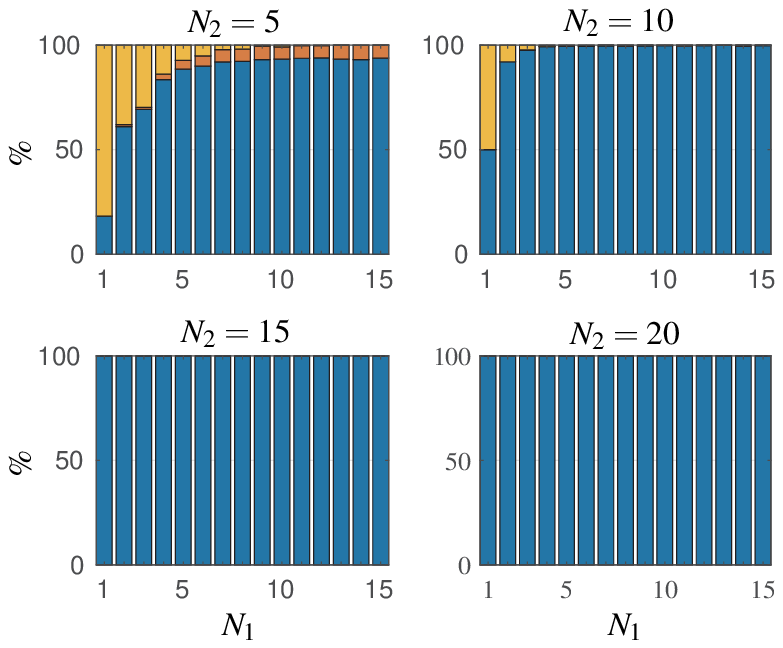}}
    {\caption{Model B, Experiment 5-3-1}\label{pict22-31}}
    \ffigbox[\FBwidth]{\includegraphics[scale=1.0]{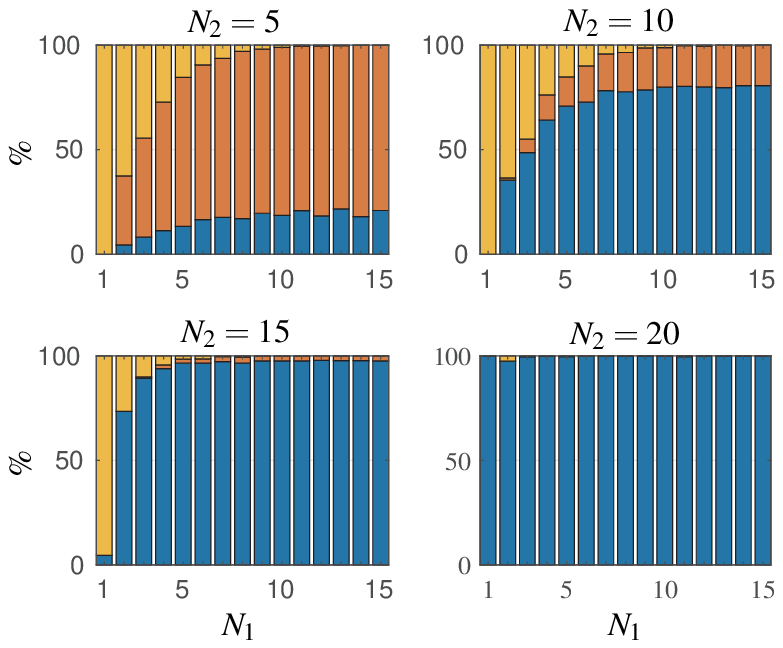}}
    {\caption{Model B, Experiment 5-3-3}\label{pict22-33}}
  \end{floatrow}
\end{figure}

\begin{figure}[p]
  \centering
  \begin{floatrow}
    \ffigbox[\FBwidth]{\includegraphics[scale=1.0]{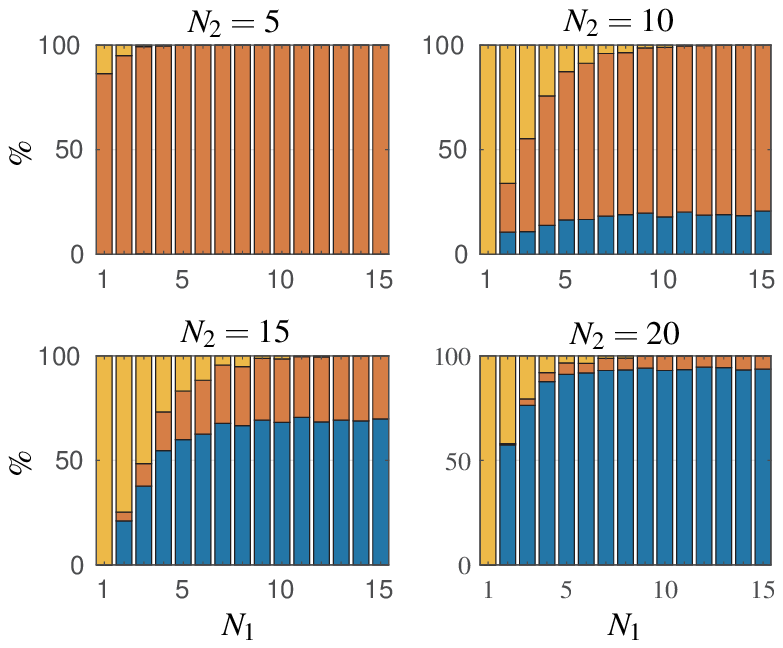}}
    {\caption{Model B, Experiment 5-3-5}\label{pict22-35}}
\end{floatrow}
\end{figure}

\begin{figure}[p]
 \centering
  \begin{floatrow}
    \ffigbox[\FBwidth]{\includegraphics[scale=1.0]{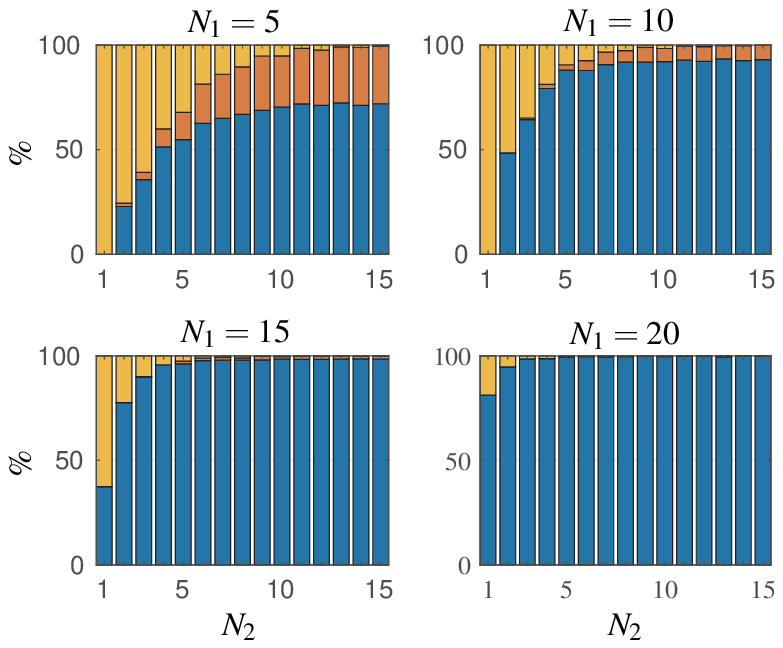}}
    {\caption{Model C, Experiment 4-1-1}\label{pict13-11}}
    \ffigbox[\FBwidth]{\includegraphics[scale=1.0]{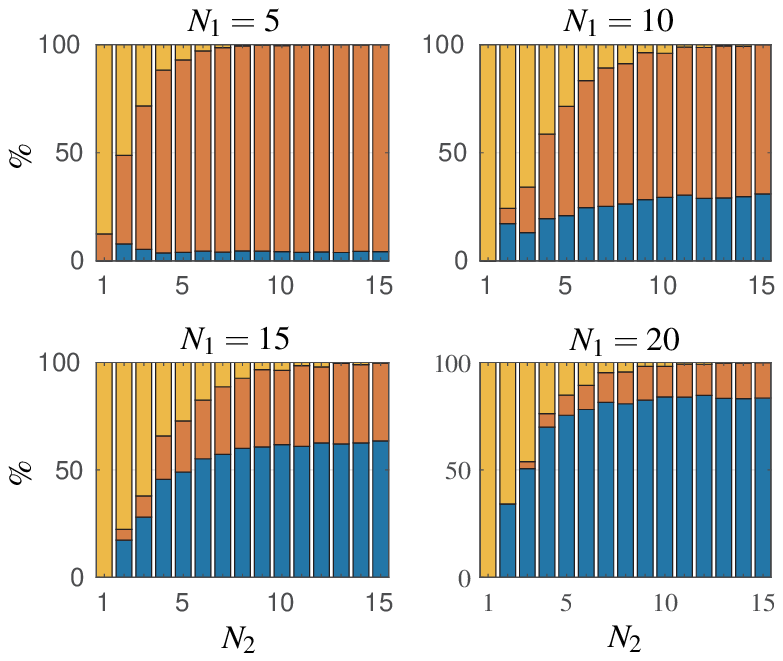}}
    {\caption{Model C, Experiment 4-1-3}\label{pict13-13}}
  \end{floatrow}
\end{figure}

\begin{figure}[p]
 \centering
  \begin{floatrow}
    \ffigbox[\FBwidth]{\includegraphics[scale=1.0]{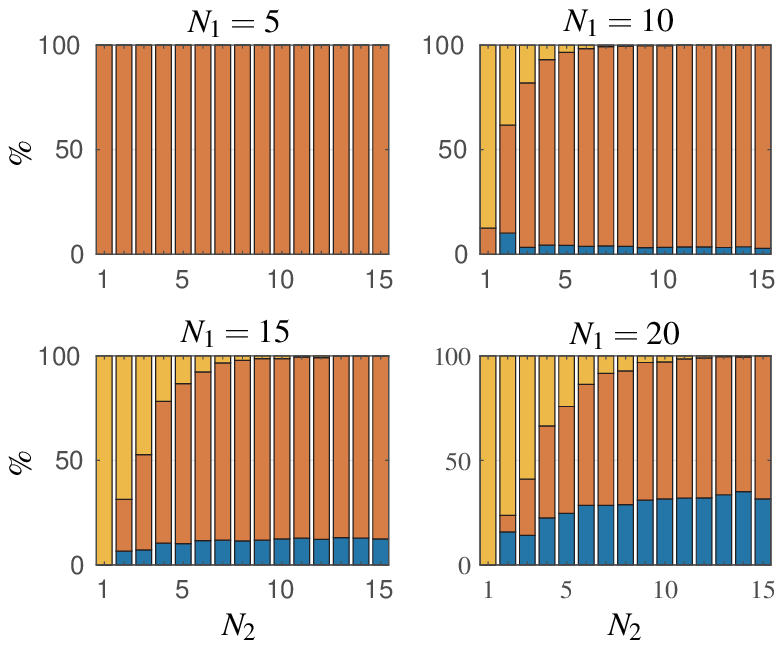}}
    {\caption{Model C, Experiment 4-1-5}\label{pict13-15}}
    \ffigbox[\FBwidth]{\includegraphics[scale=1.0]{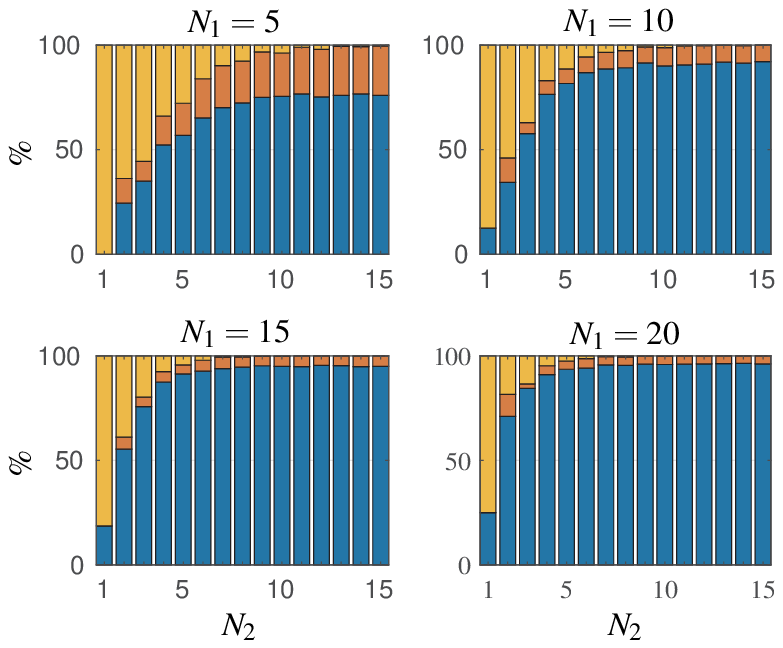}}
    {\caption{Model C, Experiment 4-2-1}\label{pict13-21}}
  \end{floatrow}
\end{figure}

\begin{figure}[p]
 \centering
  \begin{floatrow}
    \ffigbox[\FBwidth]{\includegraphics[scale=1.0]{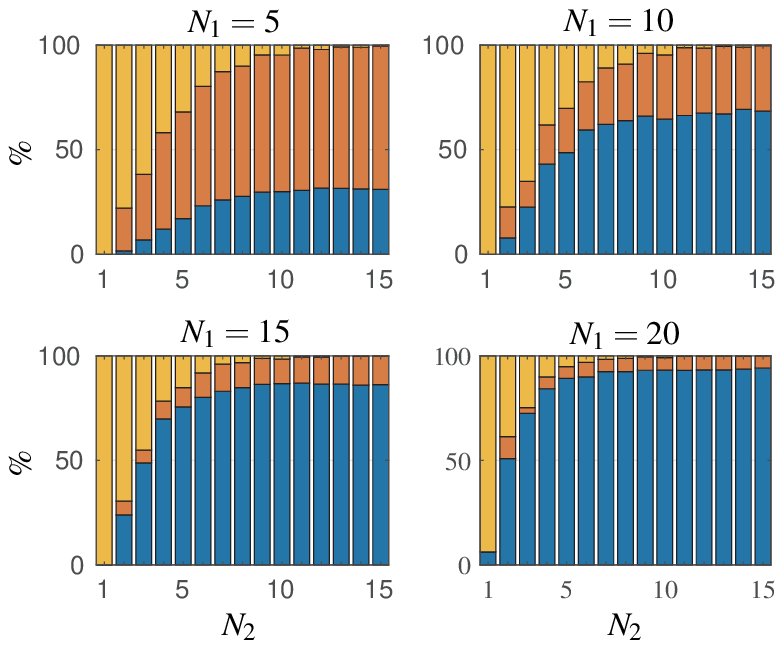}}
    {\caption{Model C, Experiment 4-2-3}\label{pict13-23}}
    \ffigbox[\FBwidth]{\includegraphics[scale=1.0]{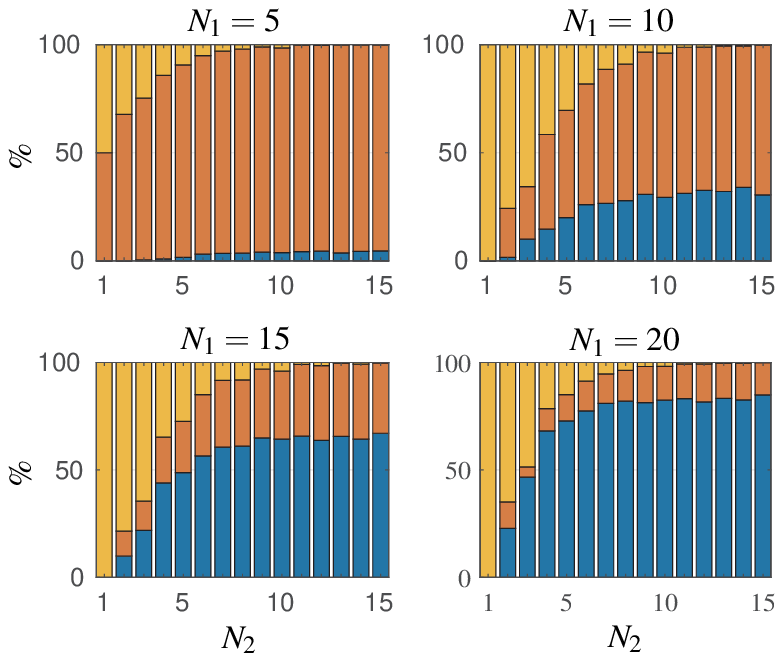}}
    {\caption{Model C, Experiment 4-2-5}\label{pict13-25}}
  \end{floatrow}
\end{figure}

\begin{figure}[p]
 \centering
  \begin{floatrow}
    \ffigbox[\FBwidth]{\includegraphics[scale=1.0]{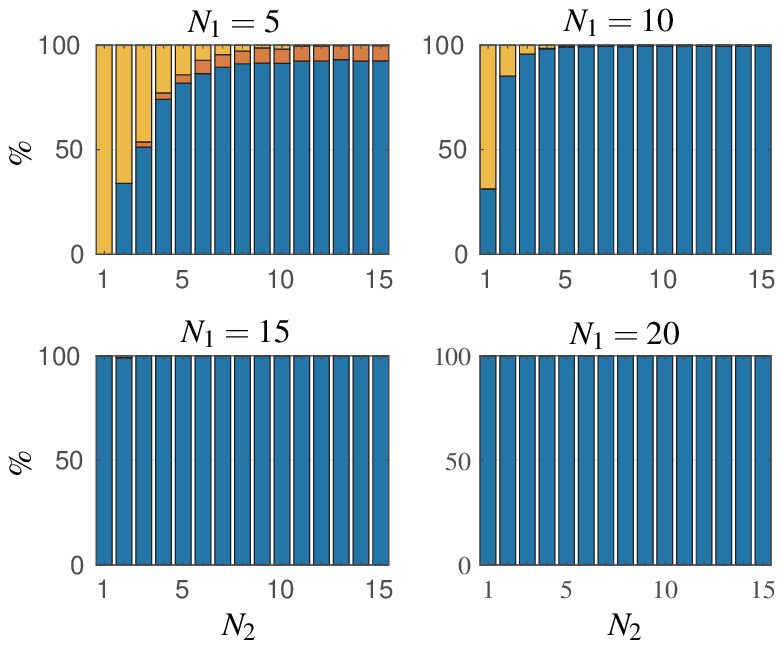}}
    {\caption{Model C, Experiment 4-3-1}\label{pict13-31}}
    \ffigbox[\FBwidth]{\includegraphics[scale=1.0]{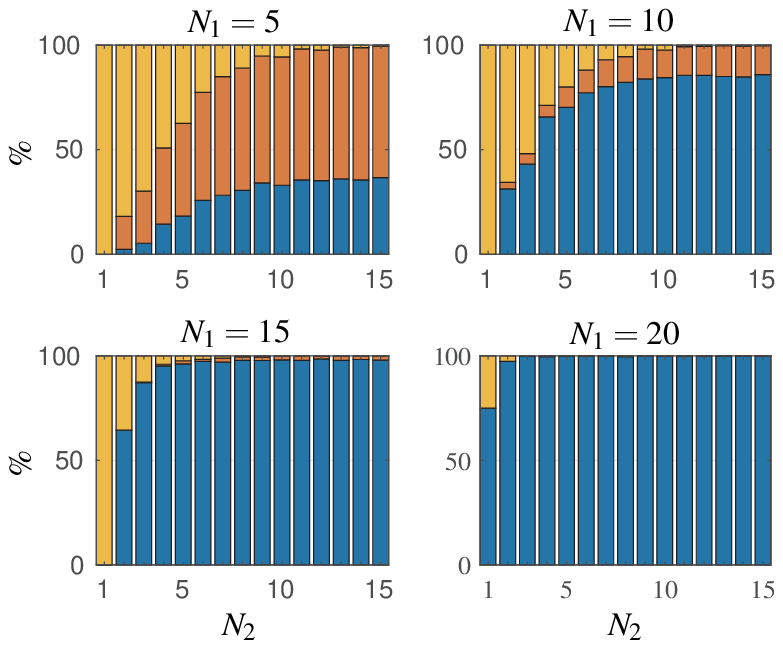}}
    {\caption{Model C, Experiment 4-3-3}\label{pict13-33}}
  \end{floatrow}
\end{figure}

\begin{figure}[p]
  \centering
  \begin{floatrow}
    \ffigbox[\FBwidth]{\includegraphics[scale=1.0]{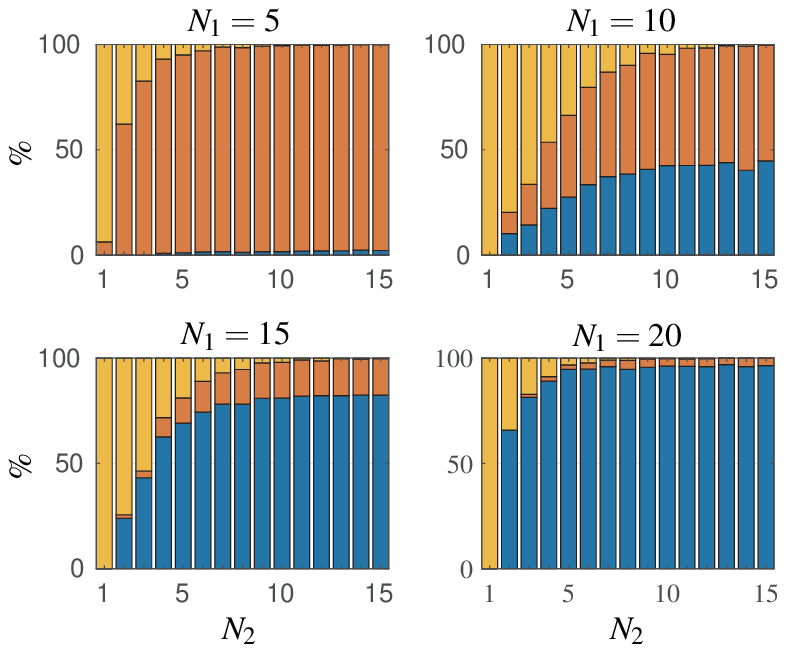}}
    {\caption{Model C, Experiment 4-3-5}\label{pict13-35}}
\end{floatrow}
\end{figure}

\begin{figure}[p]
 \centering
  \begin{floatrow}
    \ffigbox[\FBwidth]{\includegraphics[scale=1.0]{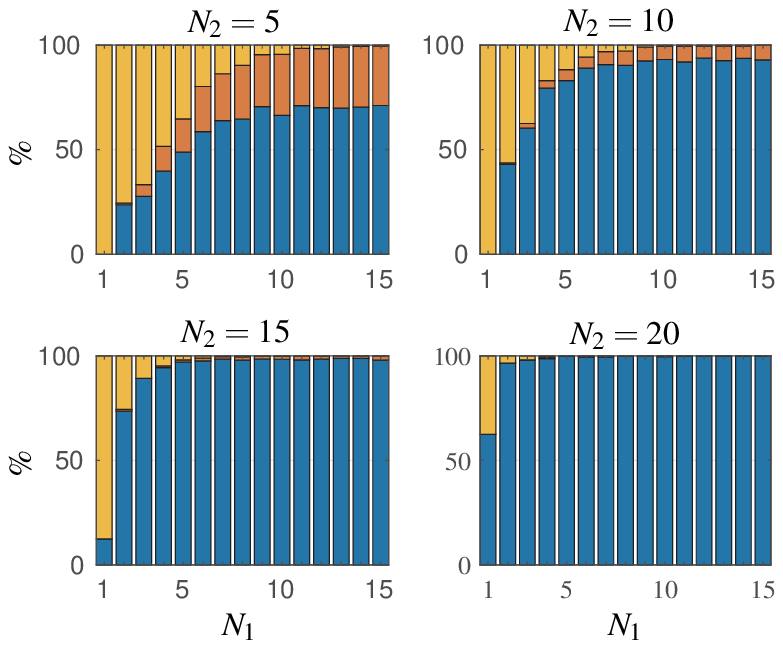}}
    {\caption{Model C, Experiment 5-1-1}\label{pict23-11}}
    \ffigbox[\FBwidth]{\includegraphics[scale=1.0]{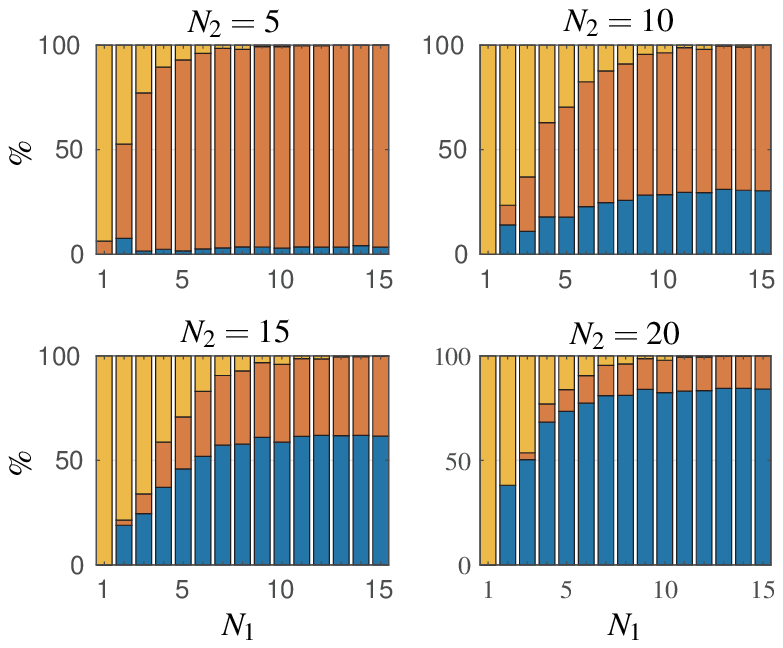}}
    {\caption{Model C, Experiment 5-1-3}\label{pict23-13}}
  \end{floatrow}
\end{figure}

\begin{figure}[p]
 \centering
  \begin{floatrow}
    \ffigbox[\FBwidth]{\includegraphics[scale=1.0]{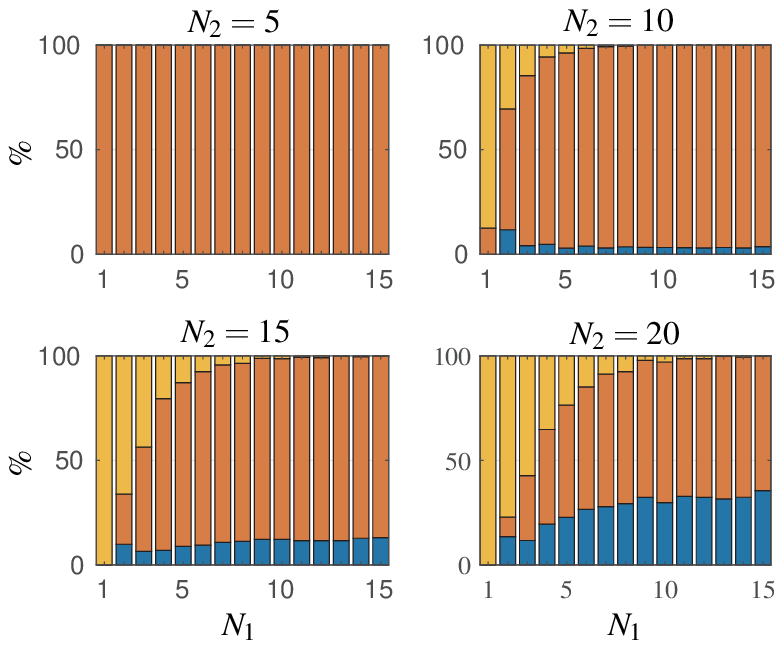}}
    {\caption{Model C, Experiment 5-1-5}\label{pict23-15}}
    \ffigbox[\FBwidth]{\includegraphics[scale=1.0]{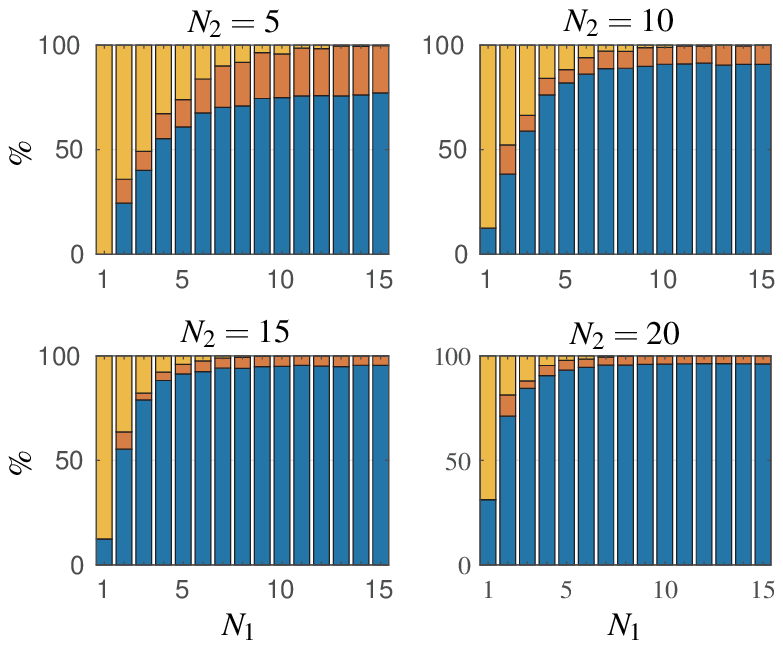}}
    {\caption{Model C, Experiment 5-2-1}\label{pict23-21}}
  \end{floatrow}
\end{figure}

\begin{figure}[p]
 \centering
  \begin{floatrow}
    \ffigbox[\FBwidth]{\includegraphics[scale=1.0]{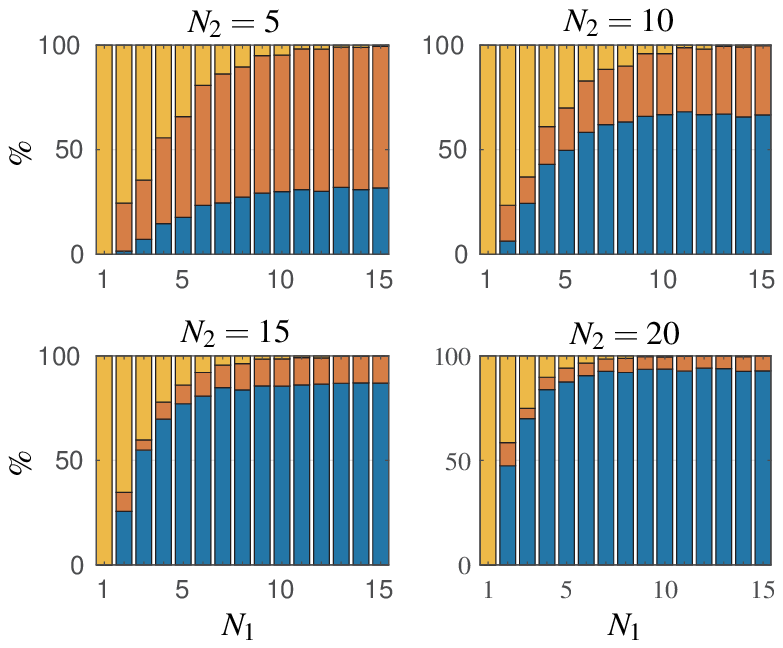}}
    {\caption{Model C, Experiment 5-2-3}\label{pict23-23}}
    \ffigbox[\FBwidth]{\includegraphics[scale=1.0]{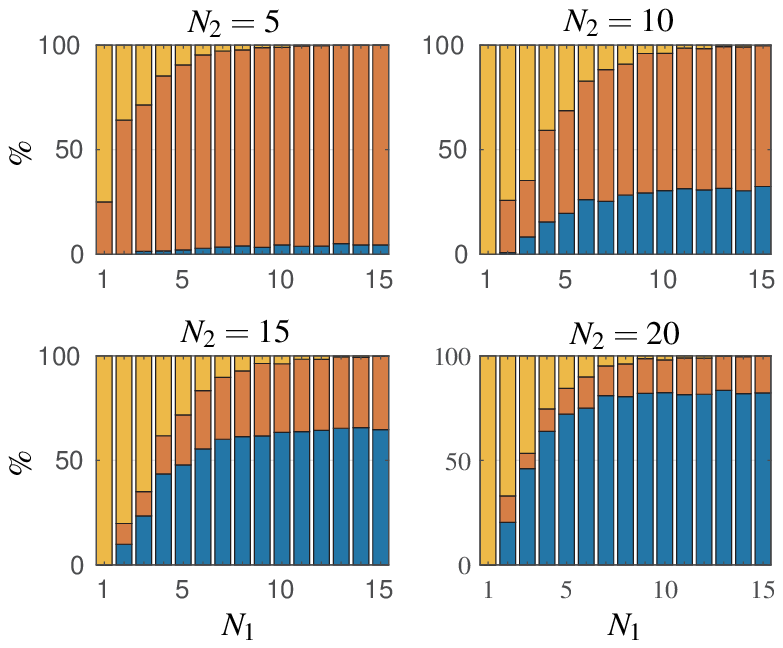}}
    {\caption{Model C, Experiment 5-2-5}\label{pict23-25}}
  \end{floatrow}
\end{figure}

\begin{figure}[p]
 \centering
  \begin{floatrow}
    \ffigbox[\FBwidth]{\includegraphics[scale=1.0]{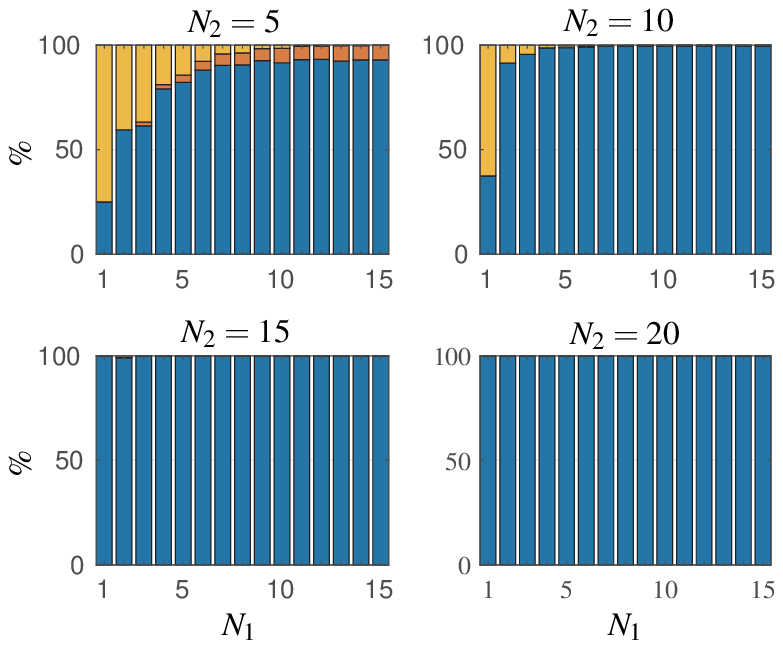}}
    {\caption{Model C, Experiment 5-3-1}\label{pict23-31}}
    \ffigbox[\FBwidth]{\includegraphics[scale=1.0]{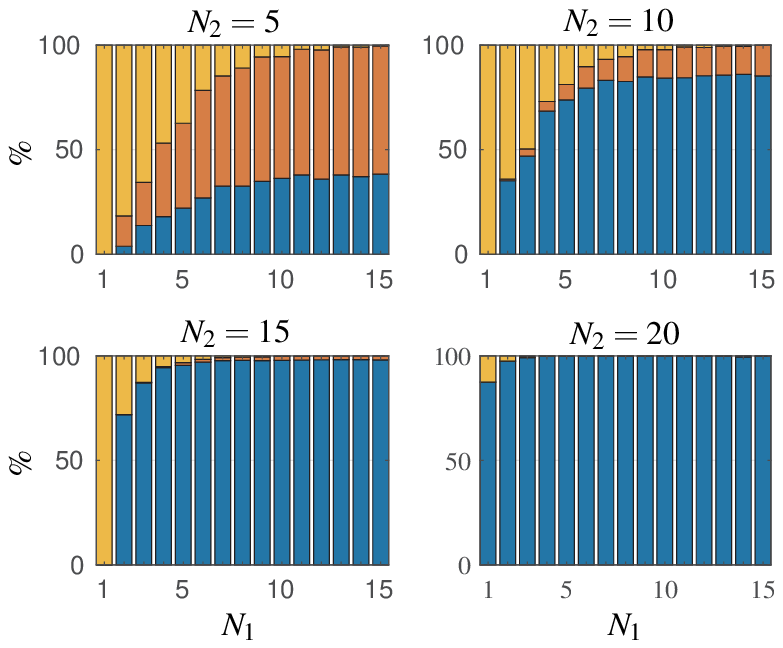}}
    {\caption{Model C, Experiment 5-3-3}\label{pict23-33}}
  \end{floatrow}
\end{figure}

\begin{figure}[p]
  \centering
  \begin{floatrow}
    \ffigbox[\FBwidth]{\includegraphics[scale=1.0]{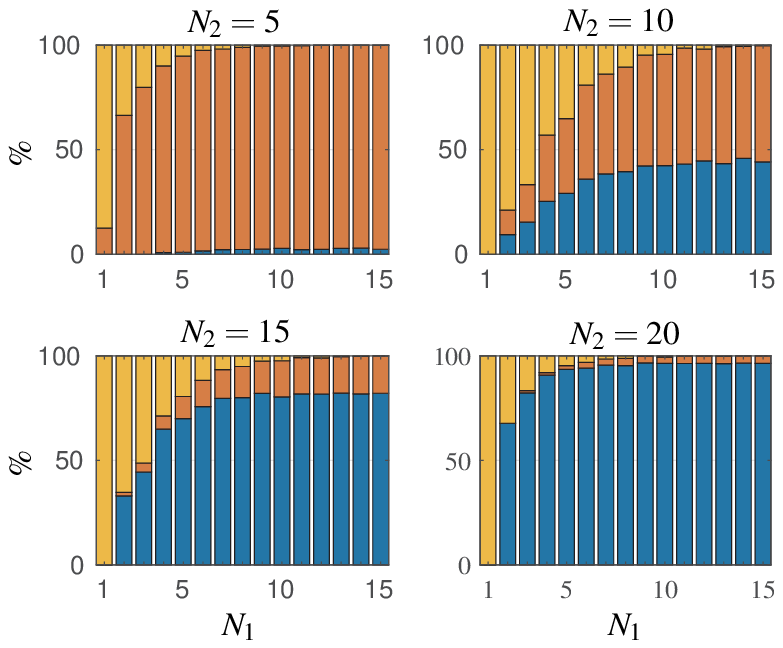}}
    {\caption{Model C, Experiment 5-3-5}\label{pict23-35}}
\end{floatrow}
\end{figure}

\begin{figure}[p]
 \centering
  \begin{floatrow}
    \ffigbox[\FBwidth]{\includegraphics[scale=1.0]{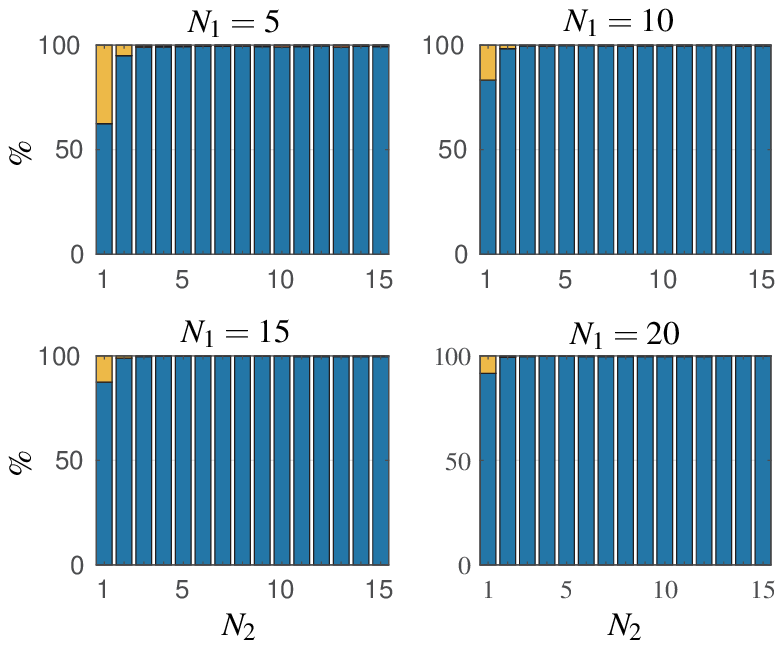}}
    {\caption{Model D, Experiment 4-1-1}\label{pict14-11}}
    \ffigbox[\FBwidth]{\includegraphics[scale=1.0]{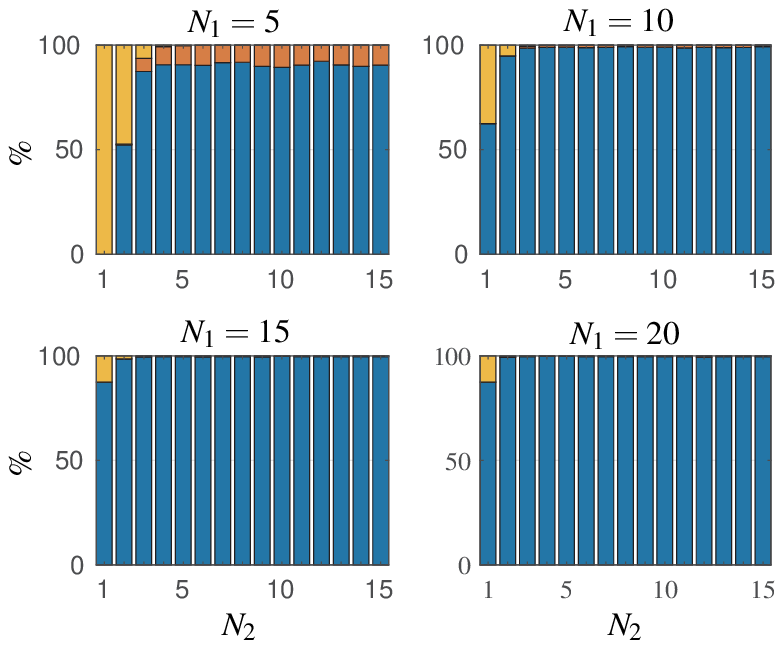}}
    {\caption{Model D, Experiment 4-1-3}\label{pict14-13}}
  \end{floatrow}
\end{figure}

\begin{figure}[p]
 \centering
  \begin{floatrow}
    \ffigbox[\FBwidth]{\includegraphics[scale=1.0]{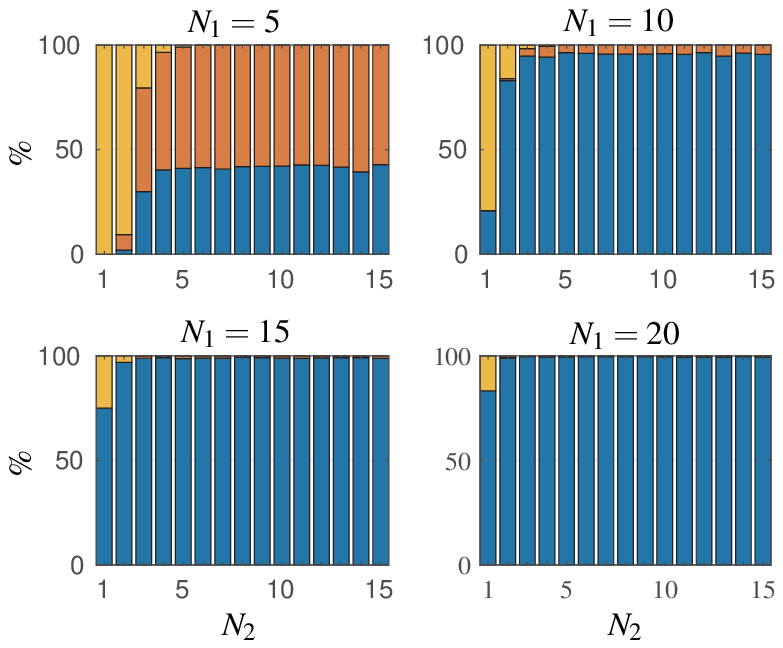}}
    {\caption{Model D, Experiment 4-1-5}\label{pict14-15}}
    \ffigbox[\FBwidth]{\includegraphics[scale=1.0]{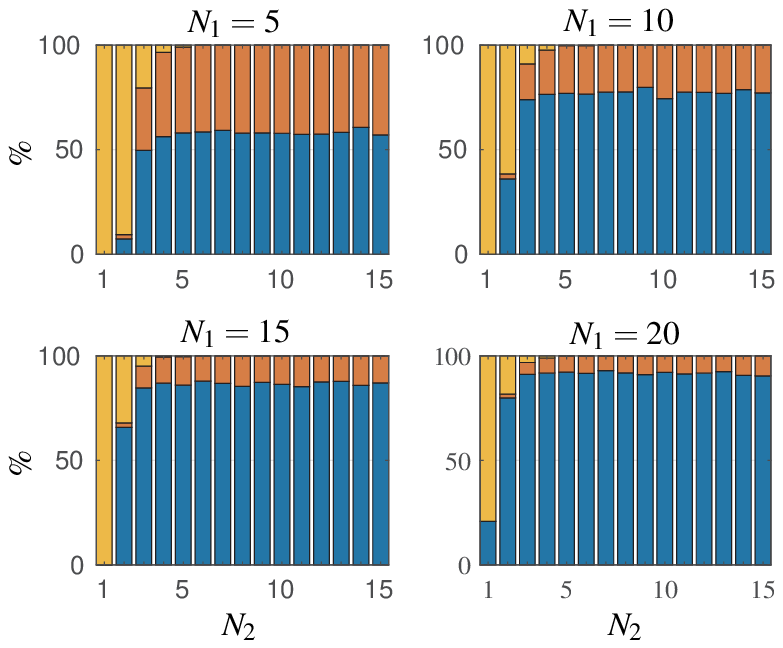}}
    {\caption{Model D, Experiment 4-2-1}\label{pict14-21}}
  \end{floatrow}
\end{figure}

\begin{figure}[p]
 \centering
  \begin{floatrow}
    \ffigbox[\FBwidth]{\includegraphics[scale=1.0]{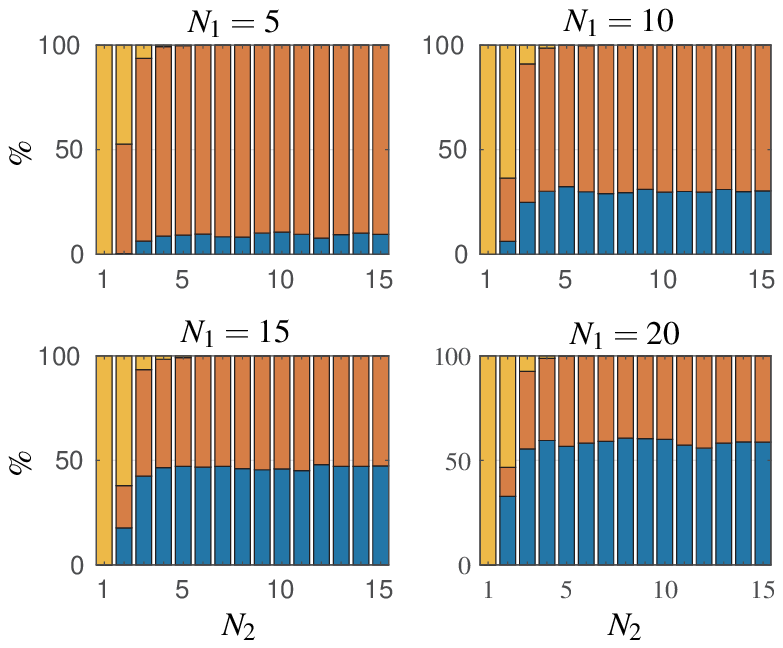}}
    {\caption{Model D, Experiment 4-2-3}\label{pict14-23}}
    \ffigbox[\FBwidth]{\includegraphics[scale=1.0]{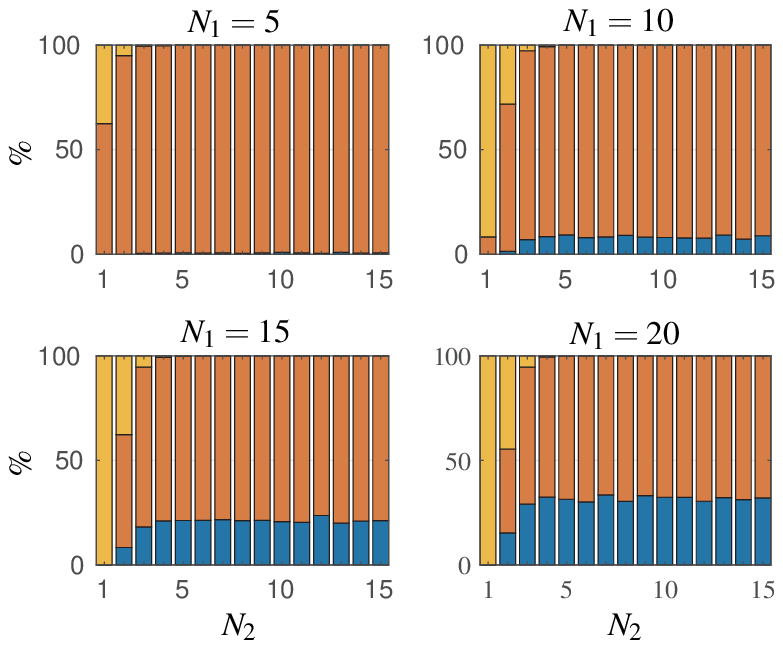}}
    {\caption{Model D, Experiment 4-2-5}\label{pict14-25}}
  \end{floatrow}
\end{figure}

\begin{figure}[p]
 \centering
  \begin{floatrow}
    \ffigbox[\FBwidth]{\includegraphics[scale=1.0]{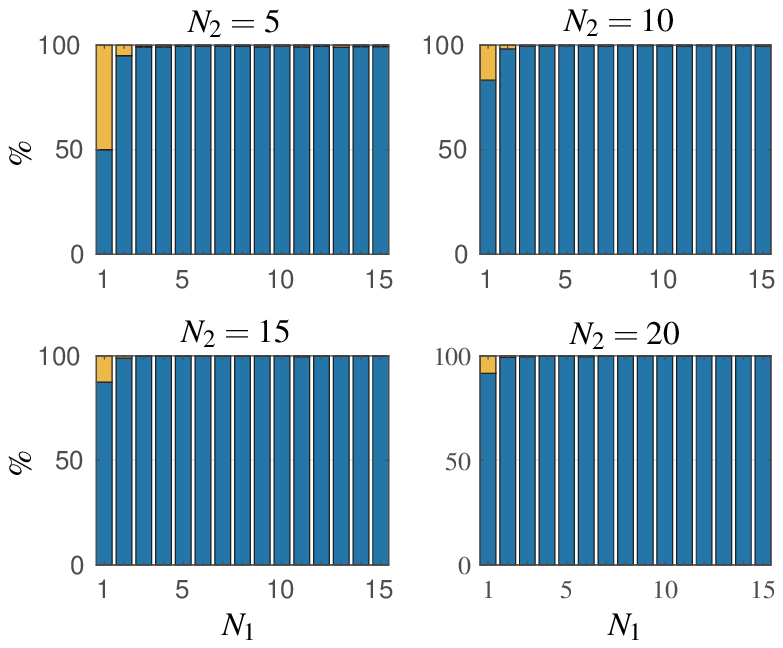}}
    {\caption{Model D, Experiment 5-1-1}\label{pict24-11}}
    \ffigbox[\FBwidth]{\includegraphics[scale=1.0]{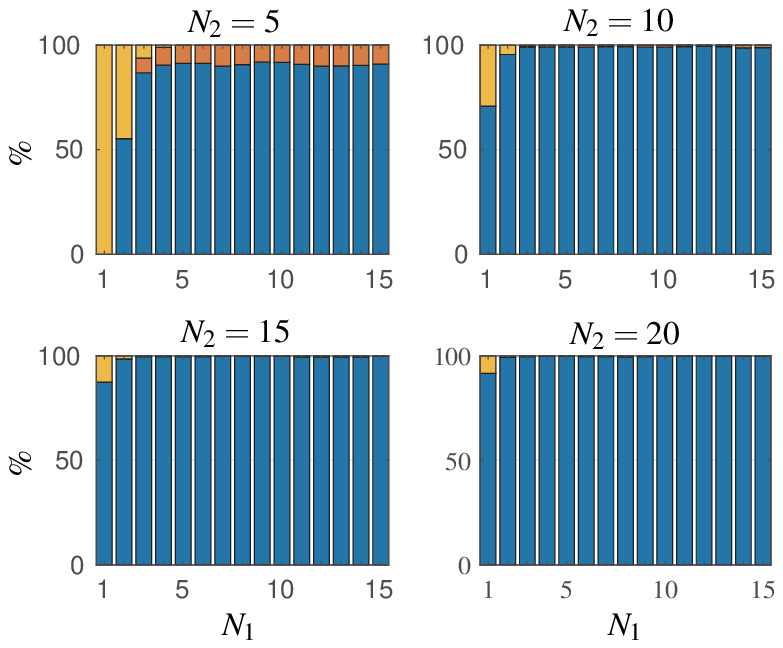}}
    {\caption{Model D, Experiment 5-1-3}\label{pict24-13}}
  \end{floatrow}
\end{figure}

\begin{figure}[p]
 \centering
  \begin{floatrow}
    \ffigbox[\FBwidth]{\includegraphics[scale=1.0]{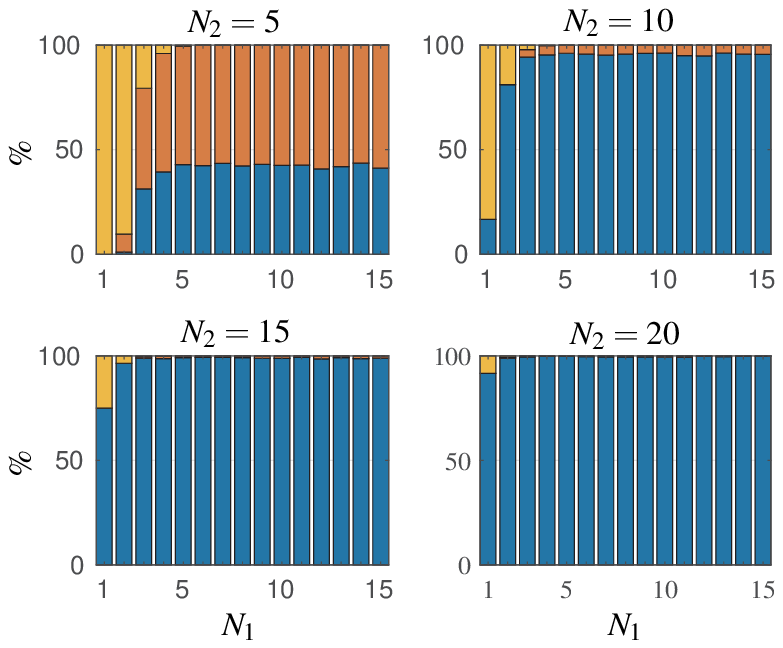}}
    {\caption{Model D, Experiment 5-1-5}\label{pict24-15}}
    \ffigbox[\FBwidth]{\includegraphics[scale=1.0]{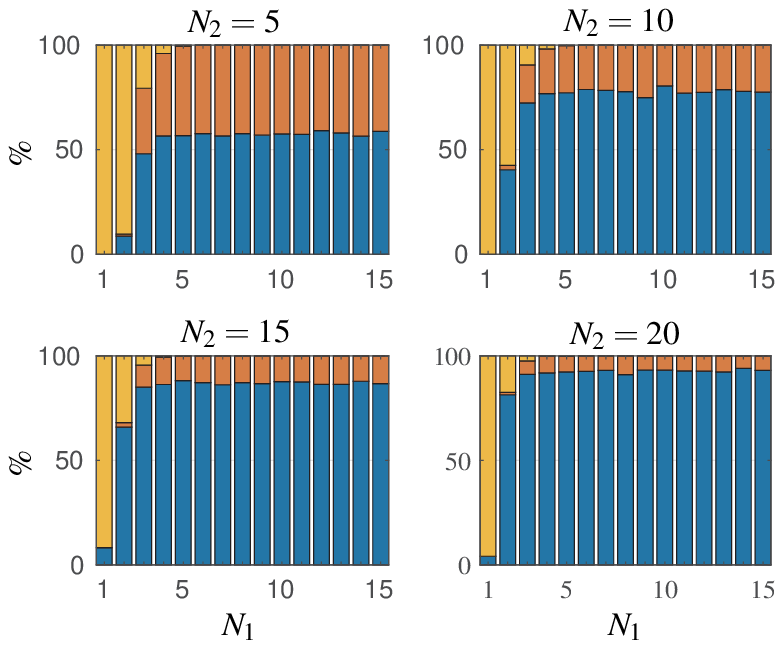}}
    {\caption{Model D, Experiment 5-2-1}\label{pict24-21}}
  \end{floatrow}
\end{figure}

\begin{figure}[p]
 \centering
  \begin{floatrow}
    \ffigbox[\FBwidth]{\includegraphics[scale=1.0]{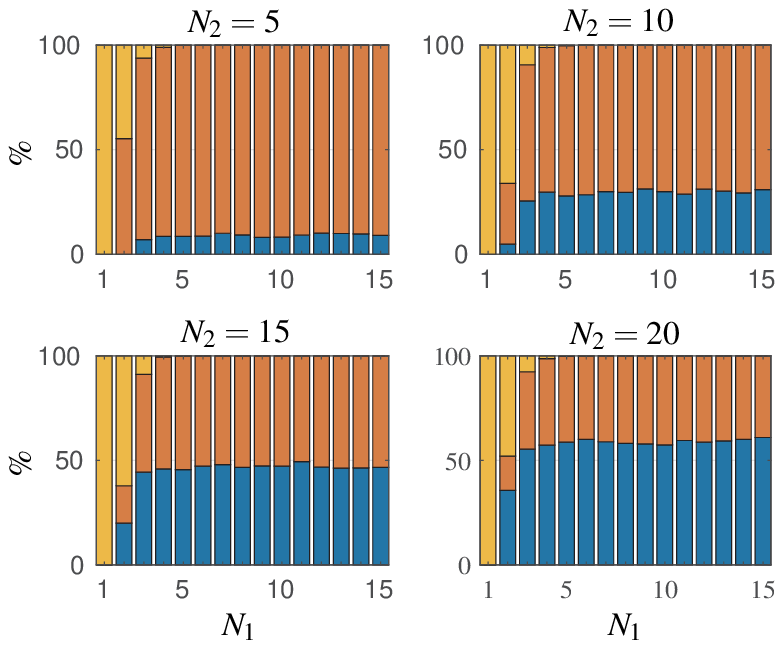}}
    {\caption{Model D, Experiment 5-2-3}\label{pict24-23}}
    \ffigbox[\FBwidth]{\includegraphics[scale=1.0]{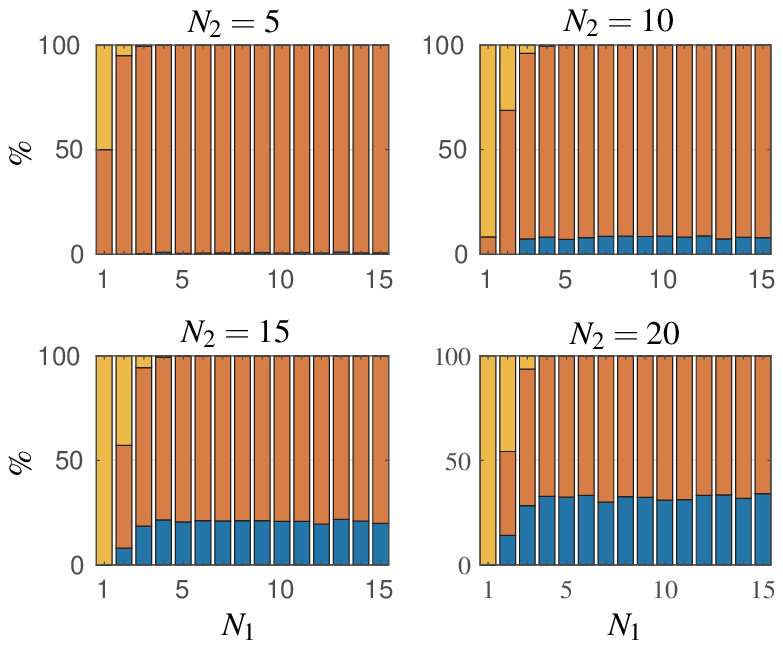}}
    {\caption{Model D, Experiment 5-2-5}\label{pict24-25}}
  \end{floatrow}
\end{figure}

\begin{figure}[p]
 \centering
  \begin{floatrow}
    \ffigbox[\FBwidth]{\includegraphics[scale=1.0]{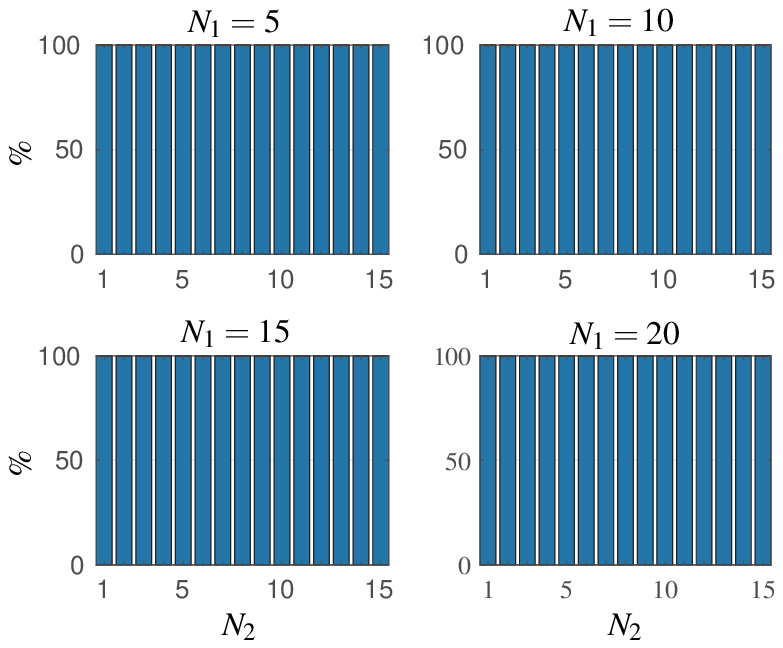}}
    {\caption{Model E, Experiment 4-1-1}\label{pict15-11}}
    \ffigbox[\FBwidth]{\includegraphics[scale=1.0]{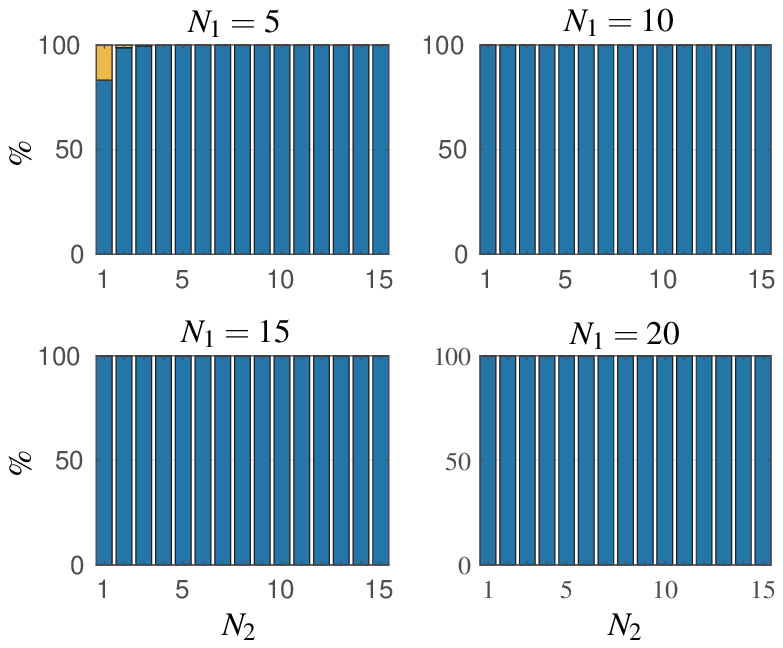}}
    {\caption{Model E, Experiment 4-1-3}\label{pict15-13}}
  \end{floatrow}
\end{figure}

\begin{figure}[p]
 \centering
  \begin{floatrow}
    \ffigbox[\FBwidth]{\includegraphics[scale=1.0]{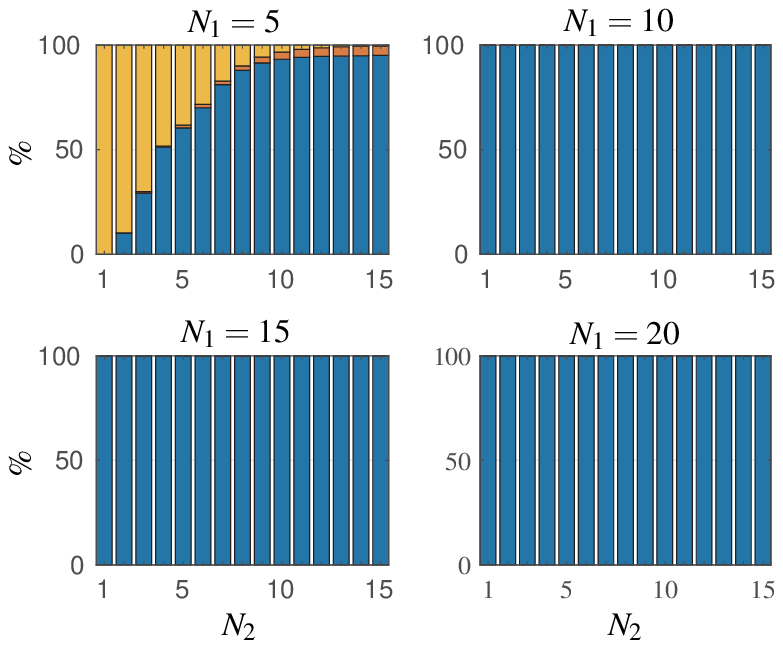}}
    {\caption{Model E, Experiment 4-1-5}\label{pict15-15}}
    \ffigbox[\FBwidth]{\includegraphics[scale=1.0]{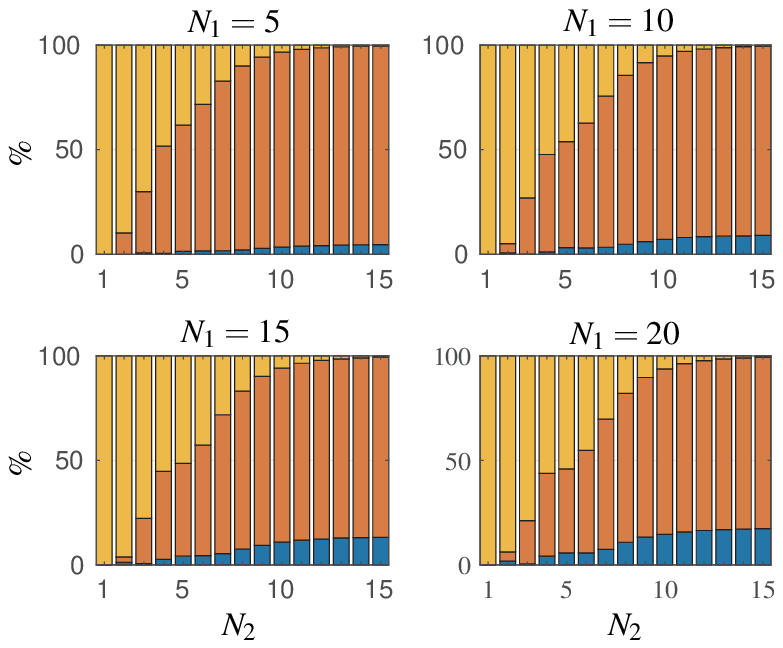}}
    {\caption{Model E, Experiment 4-2-1}\label{pict15-21}}
  \end{floatrow}
\end{figure}

\begin{figure}[p]
 \centering
  \begin{floatrow}
    \ffigbox[\FBwidth]{\includegraphics[scale=1.0]{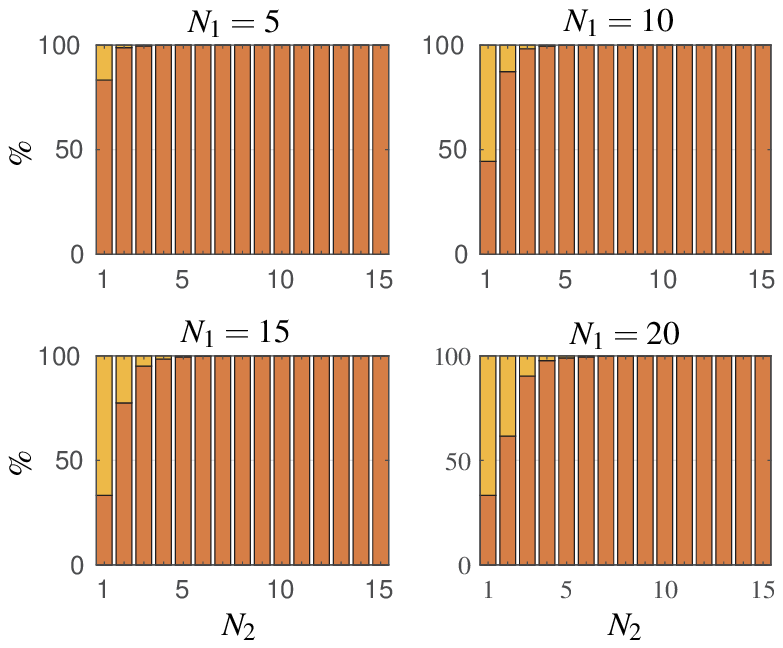}}
    {\caption{Model E, Experiment 4-2-3}\label{pict15-23}}
    \ffigbox[\FBwidth]{\includegraphics[scale=1.0]{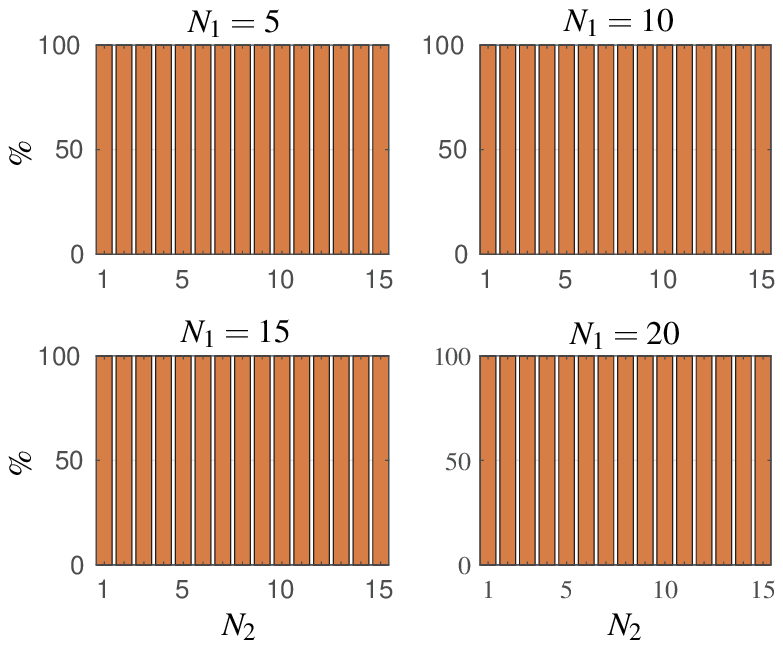}}
    {\caption{Model E, Experiment 4-2-5}\label{pict15-25}}
  \end{floatrow}
\end{figure}

\begin{figure}[p]
 \centering
  \begin{floatrow}
    \ffigbox[\FBwidth]{\includegraphics[scale=1.0]{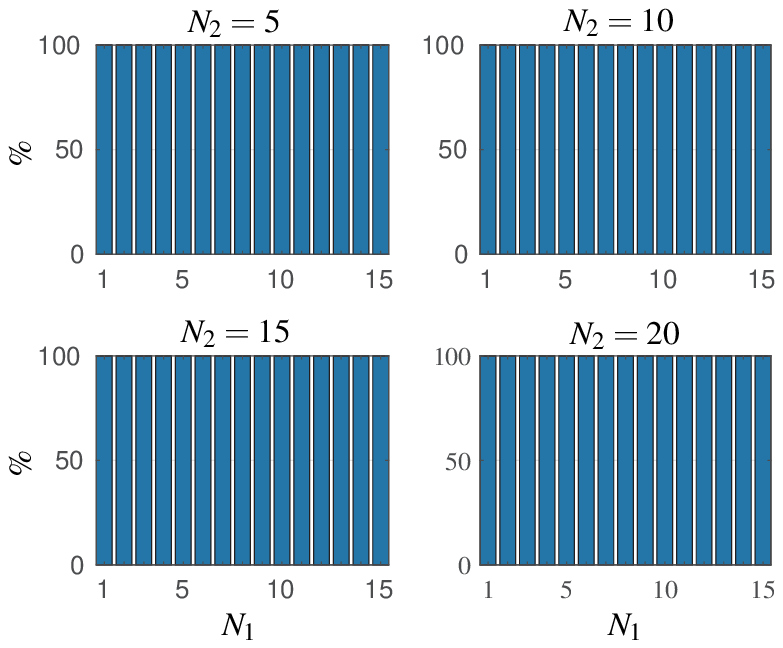}}
    {\caption{Model E, Experiment 5-1-1}\label{pict25-11}}
    \ffigbox[\FBwidth]{\includegraphics[scale=1.0]{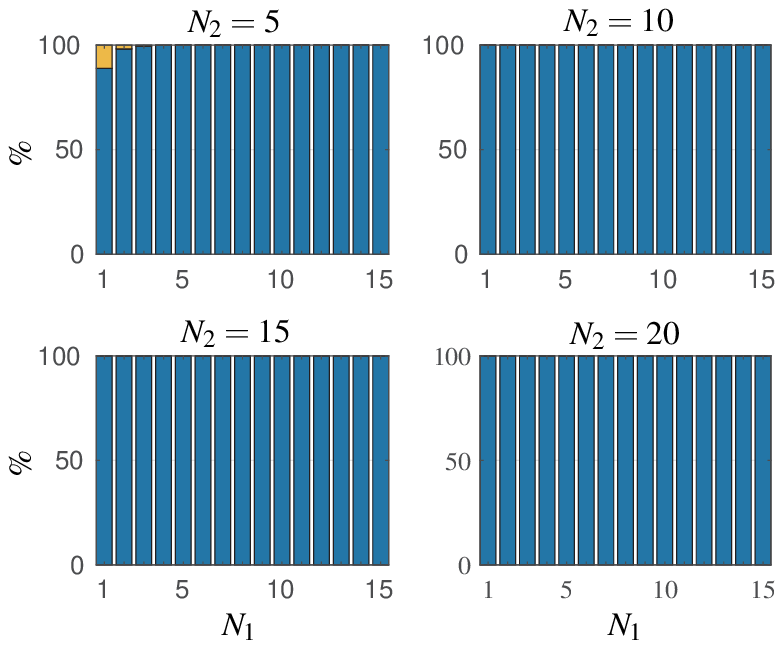}}
    {\caption{Model E, Experiment 5-1-3}\label{pict25-13}}
  \end{floatrow}
\end{figure}

\begin{figure}[p]
 \centering
  \begin{floatrow}
    \ffigbox[\FBwidth]{\includegraphics[scale=1.0]{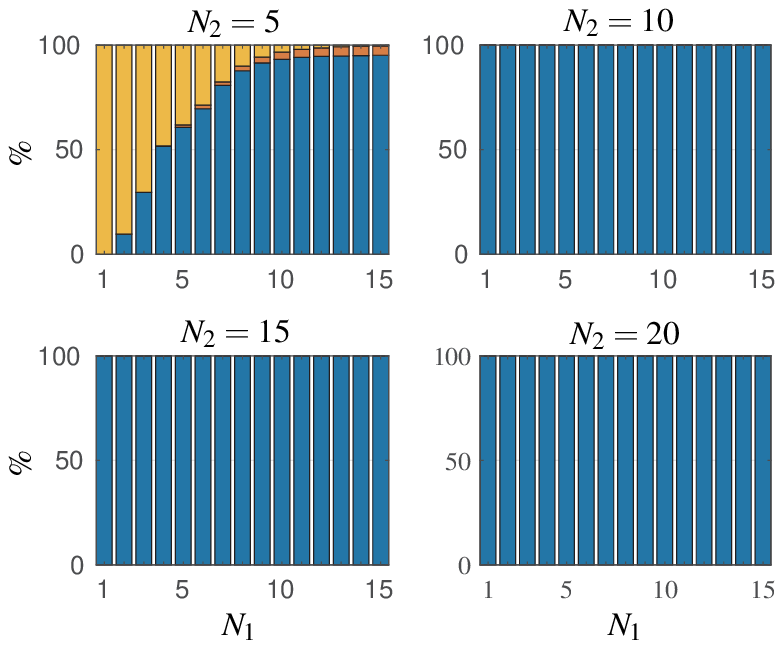}}
    {\caption{Model E, Experiment 5-1-5}\label{pict25-15}}
    \ffigbox[\FBwidth]{\includegraphics[scale=1.0]{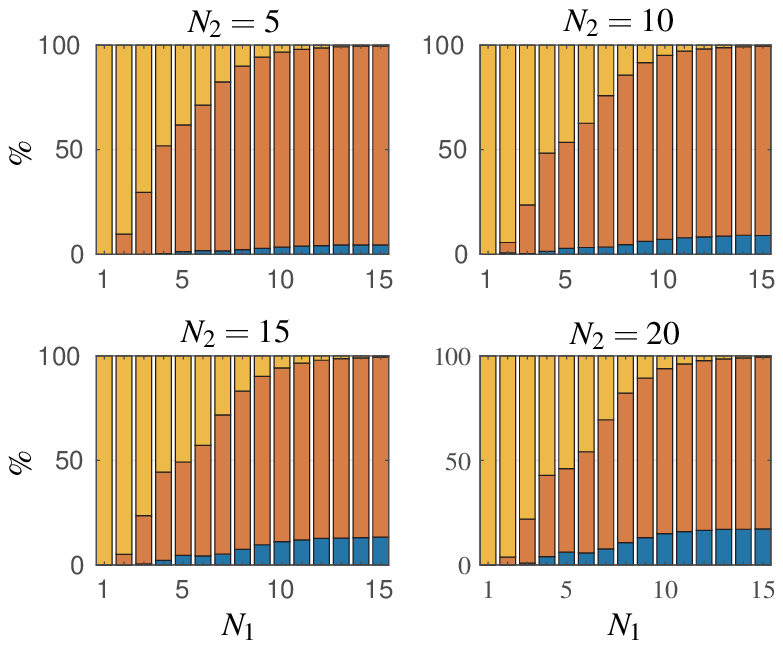}}
    {\caption{Model E, Experiment 5-2-1}\label{pict25-21}}
  \end{floatrow}
\end{figure}

\begin{figure}[p]
 \centering
  \begin{floatrow}
    \ffigbox[\FBwidth]{\includegraphics[scale=1.0]{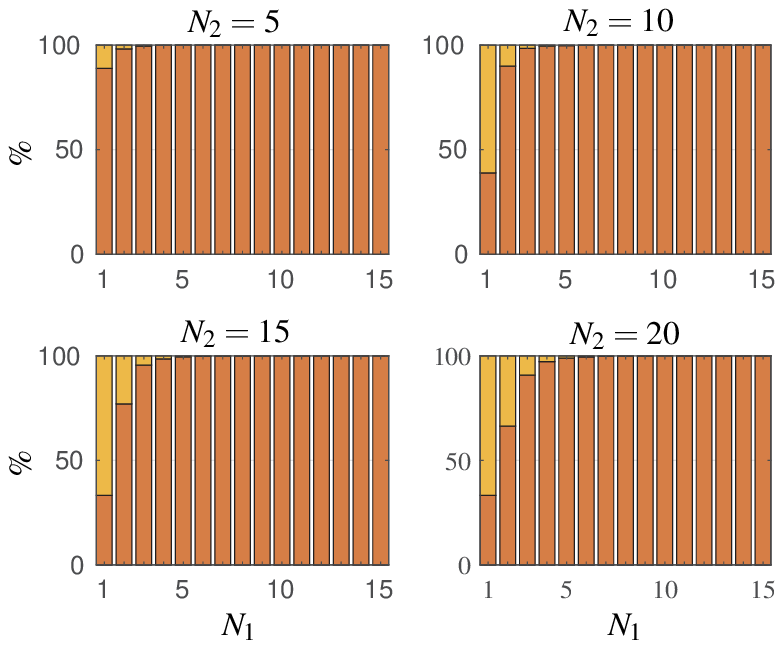}}
    {\caption{Model E, Experiment 5-2-3}\label{pict25-23}}
    \ffigbox[\FBwidth]{\includegraphics[scale=1.0]{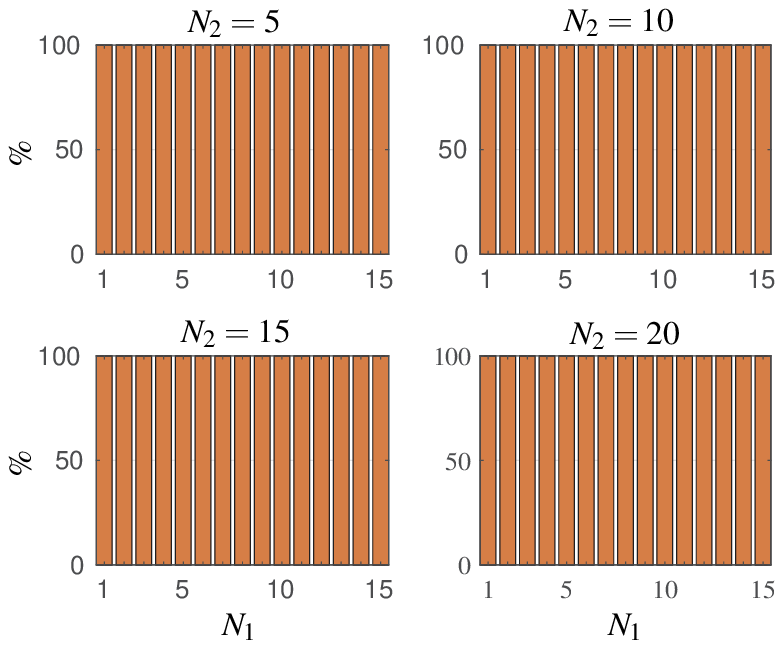}}
    {\caption{Model E, Experiment 5-2-5}\label{pict25-25}}
  \end{floatrow}
\end{figure}

\emph{Output:} Figures \ref{pict11-11}--\ref{pict25-25} 

\bigskip\model{Experiment 5:} \s{Matlab} script \code{mscript4.m}

\emph{Objective:} generate causal statements for Experiment 4

\emph{Output:} Table 3 in the main text

\bigskip\model{Experiment 6:} \s{Matlab} script \code{mscript5.m}

\emph{Objective:} assessment of reaction $N_1$-tuples as causes for the query
about $N_2$-tuples as effects; the experiments and figures are labeled by the
query, i.e., ``$5-C-F$'', where $F\geq 1$ is the required minimum number of
reactions from the reaction cluster $C\in\{1,2,\ldots\}$ occurring within
reaction $N_2$-tuples

\emph{Output:} Figures \ref{pict11-11}--\ref{pict25-25} 

\bigskip\model{Experiment 7:} \s{Matlab} script \code{mscript6.m}

\emph{Objective:} generate anti-causal statements for Experiment 6

\emph{Output:} Table 4 in the main text

\vfill

%
%

\begin{table}[ht]
  \centering
  \caption{Model A, clusters of reaction $N$-tuples}\label{tabS1}
  \begin{tabular}{ccccccc}
    \hline
    \xrot{$N$-tuple} & \xrot{Cluster} & \xrott{Uniq.}{tuples} &
    \xrott{Min}{freq.} & \xrott{Max}{freq.} &
    \xrott{Tuples}{total}\\
    \hline
    \multirow{6}{*}{$N=1$} & 1 & 1--2 & 457--5618 & 473--5618 & 930--5618 \\
    & 2 & 1--2 & 85--1281 & 92--1281 & 177--1281 \\
    & 3 & 1--2 & 5--473 & 7--473 & 8--473 \\
    & 4 & 1--1 & 1--126 & 1--126 & 1--126 \\
    & 5 & 1--1 & 13--13 & 13--13 & 13--13 \\
    & 6 & 1--1 & 1--1 & 1--1 & 1--1 \\
    \hline
    \multirow{7}{*}{$N=3$} & 1 & 1--8 & 55--3355 & 74--3355 & 559--3355 \\
    & 2 & 3--62 & 3--759 & 20--764 & 370--2284 \\
    & 3 & 4--32 & 2--154 & 2--188 & 8--1026 \\
    & 4 & 5--39 & 1--57 & 1--64 & 14--303 \\
    & 5 & 16--44 & 1--3 & 1--38 & 20--505 \\
    & 6 & 8--27 & 1--2 & 1--2 & 16--27 \\
    & 7 & 21--21 & 1--1 & 1--1 & 21--21 \\
    \hline
    \multirow{7}{*}{$N=5$} & 1 & 1--153 & 3--2090 & 15--2090 & 32--2090 \\
    & 2 & 5--109 & 2--457 & 2--486 & 114--2349 \\
    & 3 & 10--346 & 1--104 & 1--130 & 142--1153 \\
    & 4 & 5--304 & 1--58 & 1--72 & 188--331 \\
    & 5 & 138--266 & 1--3 & 1--40 & 266--1194 \\
    & 6 & 78--78 & 2--2 & 2--2 & 156--156 \\
    & 7 & 235--235 & 1--1 & 1--1 & 235--235 \\
    \hline
  \end{tabular}
\end{table}

\begin{table}[ht]
  \centering
  \caption{Model B, clusters of reaction $N$-tuples}\label{tabS2}
  \begin{tabular}{ccccccc}
    \hline
    \xrot{$N$-tuple} & \xrot{Cluster} & \xrott{Uniq.}{tuples} &
    \xrott{Min}{freq.} & \xrott{Max}{freq.} &
    \xrott{Tuples}{total}\\
    \hline
    \multirow{6}{*}{$N=1$} & 1 & 1--22 & 6--537 & 42--539 & 62--1076 \\
    & 2 & 1--19 & 3--64 & 14--123 & 25--1278 \\
    & 3 & 1--15 & 1--40 & 1--42 & 1--294 \\
    & 4 & 1--14 & 1--13 & 1--24 & 1--174 \\
    & 5 & 1--7 & 1--5 & 1--9 & 1--53 \\
    & 6 & 1--2 & 1--2 & 1--2 & 1--2 \\
    \hline
    \multirow{5}{*}{$N=3$} & 1 & 1--283 & 3--21 & 8--27 & 12--1604 \\
    & 2 & 1--186 & 2--12 & 2--14 & 12--372 \\
    & 3 & 46--540 & 1--3 & 1--8 & 92--540 \\
    & 4 & 59--259 & 1--2 & 1--2 & 118--259 \\
    & 5 & 218--218 & 1--1 & 1--1 & 218--218 \\
    \hline
    \multirow{4}{*}{$N=5$} & 1 & 1--29 & 2--5 & 2--5 & 3--83 \\
    & 2 & 4--577 & 1--3 & 1--3 & 12--577 \\
    & 3 & 15--2169 & 1--2 & 1--2 & 30--2169 \\
    & 4 & 517--541 & 1--1 & 1--1 & 517--541 \\
    \hline
  \end{tabular}
\end{table}

\begin{table}[ht]
  \centering
  \caption{Model C, clusters of reaction $N$-tuples}\label{tabS3}
  \begin{tabular}{ccccccc}
    \hline
    \xrot{$N$-tuple} & \xrot{Cluster} & \xrott{Uniq.}{tuples} &
    \xrott{Min}{freq.} & \xrott{Max}{freq.} &
    \xrott{Tuples}{total}\\
    \hline
    \multirow{5}{*}{$N=1$} & 1 & 2--16 & 7--538 & 40--548 & 155--1086 \\
    & 2 & 1--13 & 5--148 & 5--180 & 5--960 \\
    & 3 & 1--2 & 7--85 & 7--98 & 7--183 \\
    & 4 & 1--5 & 3--38 & 3--51 & 3--218 \\
    & 5 & 1--1 & 17--17 & 17--17 & 17--17 \\
    \hline
    \multirow{4}{*}{$N=3$} & 1 & 1--267 & 3--17 & 8--43 & 12--1895 \\
    & 2 & 38--127 & 2--3 & 2--11 & 76--276 \\
    & 3 & 37--313 & 1--2 & 1--2 & 74--313 \\
    & 4 & 132--192 & 1--1 & 1--1 & 132--192 \\
    \hline
    \multirow{4}{*}{$N=5$} & 1 & 1--43 & 2--6 & 2--6 & 3--141 \\
    & 2 & 2--552 & 1--3 & 1--3 & 6--552 \\
    & 3 & 18--1931 & 1--2 & 1--2 & 36--1931 \\
    & 4 & 377--487 & 1--1 & 1--1 & 377--487 \\
    \hline
  \end{tabular}
\end{table}

\begin{table}[ht]
  \centering
  \caption{Model D, clusters of reaction $N$-tuples}\label{tabS4}
  \begin{tabular}{ccccccc}
    \hline
    \xrot{$N$-tuple} & \xrot{Cluster} & \xrott{Uniq.}{tuples} &
    \xrott{Min}{freq.} & \xrott{Max}{freq.} &
    \xrott{Tuples}{total}\\
    \hline
    \multirow{5}{*}{$N=1$} & 1 & 1--24 & 3--21 & 10--31 & 14--330 \\
    & 2 & 1--22 & 1--9 & 1--11 & 1--146 \\
    & 3 & 1--6 & 1--6 & 1--6 & 1--13 \\
    & 4 & 1--4 & 1--3 & 1--3 & 1--6 \\
    & 5 & 1--4 & 1--1 & 1--1 & 1--4 \\
    \hline
    \multirow{3}{*}{$N=3$} & 1 & 1--206 & 1--3 & 1--3 & 2--206 \\
    & 2 & 6--319 & 1--2 & 1--2 & 12--319 \\
    & 3 & 146--321 & 1--1 & 1--1 & 146--321 \\
    \hline
    \multirow{2}{*}{$N=5$} & 1 & 1--334 & 1--2 & 1--2 & 2--334 \\
    & 2 & 332--332 & 1--1 & 1--1 & 332--332 \\
    \hline
  \end{tabular}
\end{table}

\begin{table}[ht]
  \centering
  \caption{Model E, clusters of reaction $N$-tuples}\label{tabS5}
  \begin{tabular}{ccccccc}
    \hline
    \xrot{$N$-tuple} & \xrot{Cluster} & \xrott{Uniq.}{tuples} &
    \xrott{Min}{freq.} & \xrott{Max}{freq.} &
    \xrott{Tuples}{total}\\
    \hline
    \multirow{4}{*}{$N=1$} & 1 & 6--12 & 1775--6391 & 1889--7173 & 22221--40805
    \\ 
    & 2 & 1--6 & 21--1674 & 36--2144 & 47--11491 \\
    & 3 & 1--5 & 21--116 & 21--137 & 21--373 \\
    & 4 & 3--3 & 29--29 & 33--33 & 94--94 \\
    \hline
    \multirow{4}{*}{$N=3$} & 1 & 27--1741 & 3--222 & 24--280 & 6725--23459 \\
    & 2 & 5--1767 & 2--3 & 2--148 & 15--44984 \\
    & 3 & 46--596 & 1--2 & 1--2 & 92--596 \\
    & 4 & 472--588 & 1--1 & 1--1 & 472--588 \\
    \hline
    \multirow{3}{*}{$N=5$} & 1 & 12--2695 & 3--3 & 3--33 & 36--13087 \\
    & 2 & 816--5051 & 2--2 & 2--2 & 1632--10102 \\
    & 3 & 20738--29570 & 1--1 & 1--1 & 20738--29570 \\
    \hline
  \end{tabular}
\end{table}

\clearpage

%
%

\end{document}